\documentclass[apj,appendixfloats]{emulateapj}
\usepackage{amsmath}
\usepackage{natbib}
\usepackage{graphicx}
\usepackage{appendix}
\usepackage{color}
\usepackage{relsize}
\DeclareGraphicsExtensions{.eps}
\begin{document}
\title{Changing Phases of Alien Worlds: Probing Atmospheres of $Kepler$ Planets with High-Precision Photometry}
\author{Lisa J. Esteves\altaffilmark{1}, Ernst J. W. De Mooij\altaffilmark{2} \& Ray Jayawardhana\altaffilmark{3}}
\altaffiltext{1}{Astronomy \& Astrophysics, University of Toronto, 50 St. George Street, Toronto, Ontario M5S 3H4, Canada}
\altaffiltext{2}{School of Mathematics and Physics, Queen's University Belfast, University Road Belfast, Northern Ireland, BT7 1NN, United Kingdom}
\altaffiltext{3}{Physics \& Astronomy, York University, Toronto, Ontario L3T 3R1, Canada}
\email{esteves@astro.utoronto.ca; e.demooij@qub.ac.uk; rayjay@yorku.ca}
\begin{abstract}
\indent We present a comprehensive analysis of planetary phase variations, including possible planetary light offsets, using eighteen quarters of data from the {\it Kepler} space telescope. After correcting for systematics, we found fourteen systems with significant detections in each of the phase curve components: planet's phase function, secondary eclipse, Doppler boosting and ellipsoidal variations. We model the full phase curve simultaneously, including primary and secondary transits, and derive albedos, day- and night-side temperatures and planet masses. Most planets manifest low optical geometric albedos ($<$0.25), with the exception of Kepler-10b, Kepler-91b and KOI-13b. We find that KOI-13b, with a small eccentricity of 0.0006$\pm$0.0001, is the only planet for which an eccentric orbit is favored. We detect a third harmonic with an amplitude of $1.9\pm0.2$ ppm for HAT-P-7b for the first time, and confirm the third harmonic for KOI-13b reported in Esteves et al. : both could be due to their spin-orbit misalignments. For six planets, we report a planetary brightness peak offset from the substellar point: of those, the hottest two (Kepler-76b and HAT-P-7b) exhibit pre-eclipse shifts or to the evening-side, while the cooler four (Kepler-7b, Kepler-8b, Kepler-12b and Kepler-41b) peak post-eclipse or on the morning-side. Our findings dramatically increase the number of {\it Kepler} planets with detected planetary light offsets, and provide the first evidence in the {\it Kepler} data for a correlation between the peak offset direction and the planet's temperature. Such a correlation could arise if thermal emission dominates light from hotter planets that harbor hot spots shifted toward the evening-side, as theoretically predicted, while reflected light dominates cooler planets with clouds on the planet's morning-side.
\end{abstract}
\maketitle
\bibpunct[; ]{(}{)}{;}{}{}{}
\section{Introduction}
\label{sec:intro}
\indent The {\it Kepler Space Telescope} monitored over $\sim$150,000 stars for nearly four years. Even though the main goal of the {\it Kepler} mission was to find planets using the transit method, the high precision, long baseline and continuous nature of its observations make the resulting photometry ideal for characterizing exoplanets through studies of their phase variations. \\ 
\indent From transit measurements, it is possible to derive both the planet's orbital parameters (period, scaled semi-major axis and impact parameter) as well as the planet-to-star radius ratio. Meanwhile, the out-of-transit light curve, known as the phase curve, can yield constraints on the planetary atmosphere (e.g. albedo, brightness temperature) and the planet's mass. \\
\indent A planet's phase curve is a composite of brightness variations caused by three independent phenomena: ellipsoidal variations stemming from changes in the starlight due to tides raised by the planet, Doppler boosting resulting from the reflex motion of the star, and a combination of reflected light and thermal emission from the planet. The secondary eclipse, when the planet passes behind the star, provides additional constraints on the light from the planet. \\
\indent In recent years, several groups have presented light curve data from {\it Kepler} for a few individual planets ~\citep[e.g.][]{HATP7_Welsh2010,KOI13_Shporer2011,TrES2_Barclay2012} and for small samples of planets (\citealp{Esteves2013}, hereafter as E13;~\citealp{Angerhausen2014}). With the exception of two ultra-short-period Earth-sized planets, Kepler-10b ~\citep{Kep10_Batalha2011} and Kepler-78b ~\citep{Kep78_SanchisOjeda2013}, all of these studies focus on hot Jupiters. \\
\indent Since {\it Kepler} observes in a single broad optical band, it is not possible to disentangle the individual contributions from reflected light and thermal emission. The {\it Spitzer Space Telescope}, on the other hand, primarily probes the thermal emission from hot Jupiters at infrared wavelengths. For a number of planets, {\it Spitzer} observations have provided direct measurements of the day-night contrasts, due to temperature differences between the two hemispheres of these presumably tidally locked worlds, and have shown that often the hottest spot in the planet's atmosphere is offset from the sub-stellar point ~\citep[e.g.][]{Knutson2007}. \\
\indent \citet{Kep7_Demory2013} have shown, using {\it Kepler} phase curve observations and {\it Spitzer} secondary eclipse measurements of Kepler-7b, that a planet's brightness can be dominated by reflected light and that the albedo can vary between the morning and evening sides of the planet. They attribute this to inhomogeneous reflective clouds, whose properties change as a function of longitude, influenced by the planet's wind and thermal patterns. \\
\indent Here we present the results of our transit and phase curve analysis, with the inclusion of a planetary brightness offset, for 14 planets (Kepler-5b, Kepler-6b, Kepler-7b, Kepler-8b, Kepler-10b, Kepler-12b, Kepler-41b, Kepler-43b, Kepler-76b, Kepler-91b, Kepler-412b, TrES-2b, HAT-P-7b and KOI-13b) using all 18 quarters of {\it Kepler}'s long-cadence and short-cadence data. In Section~\ref{sec:red} we present the dataset and our analysis method, while in Section~\ref{sec:model} we present our model and in Section~\ref{sec:analysis} we describe our fit to the data. The results are presented in Section~\ref{sec:results}, discussed in Section~\ref{sec:source} and finally we outline our conclusions in Section~\ref{sec:conclu}.
\section{Data Reduction}
\label{sec:red}
\indent After correcting for systematics (see Section~\ref{sec:cbv}), we visually inspected the phase-folded light curve of all publicly released {\it Kepler} planets and planet candidates with periods $<$10 days. Of these we found 14 planets (Kepler-5b, -6b, -7b, -8b, -10b, -12b, -41b, -43b, -76b, -91b, -412b, TrES-2b, HAT-P-7b and KOI-13b) that exhibited orbital phase variations resembling a planetary signal and had significant detections in each of our measured phase curve components (see Section~\ref{sec:model}). \\
\subsection{Removal of Systematics}
\label{sec:cbv}
\begin{table}[h]
\centering
\caption{{\it Kepler} Quarters of Data Used in Analysis}
\setlength{\tabcolsep}{2pt}
\begin{tabular*}{8.5cm}{@{\extracolsep{\fill}}l|cccccccccccccccccc}
System & \multicolumn{17}{c}{Quarters} \\
   & 0 & 1 & 2 & 3 & 4 & 5 & 6 & 7 & 8 & 9 & 10 & 11 & 12 & 13 & 14 & 15 & 16 & 17 \\
\hline
\hline
Kepler-5  & l & l & s & s & s & s & s & s & s & s & s & s & s & l & l & l & l & l \\
Kepler-6  & l & l & s & s & s & s & s &   & l & s & s &   & s & l & l &   & l & l \\
Kepler-7  & l & l & l & s & s & s & s & s & s & l & l & l & l & l & l & l & l & l \\
Kepler-8  & l & l & s & s & s & s & s & s & l & s & s & s & s & s & l & l & l & l \\
Kepler-10 & l & l & s & s &   & s & s & s &   & s & s & s &   & s & s & s &   & s \\
Kepler-12 & l & l & s & s &   & s & s & s &   & s & s & s &   & l & l & l &   & l \\
Kepler-41 &   & l & l & l & s & s & s & s & l & l & l & l & l & l & l & l & l & l \\
Kepler-43 &   & l & l & s & s & s & s & s & l & l & s & s & s & s & s & s & l & l \\
Kepler-76 &   & l & l & l & l & l &   & l & l & l &   & l & l & l &   & l & l & l \\
Kepler-91 & l & l & l & l & l & l & l & l & l & l & l & l & l & l & l & l & l & l \\
Kepler-412&   & l & l & l & s & s & s & s & l & l & l & l & l & l & l & l & l & l \\
KOI-13    & l & l & s & s & l & l & l & s & s & s & s & s & s & s & s & s & s & s \\
TrES-2    & s & s & s & s &   & s & s & s &   & s & s & s &   & s & s & s &   & s \\
HAT-P-7   & s & s & s & s & s & s & s & s & s & s & s & s & s & s & s & s & s & s \\
\hline
\end{tabular*}
\begin{tabular}{p{8.5cm}}
Note. Long and short cadence quarters are denoted by ``l'' and ``s'', respectively, while an empty space indicates quarters with no data.
\end{tabular}
\label{tab:quarters}
\end{table}
\indent In our analysis, we used both the {\it Kepler} LC and short cadence (SC) simple aperture data (see Table~\ref{tab:quarters}). In the only multiplanet system, Kepler-10b, we removed the second planet's transit. Instrumental signals were removed by performing a linear least squares fit\footnote{Using custom IDL procedures.} of the first eight cotrending basis vectors (CBVs) to the time-series of each quarter individually. Before cotrending, we removed any bad data points flagged by {\it Kepler} in the SAP or CBV files and to prevent contamination we only fit the CBVs to the out-of-transit time-series. The fitted basis vectors were then divided out of the quarter, in order to preserve the amplitude of the physical signals of interest. Since CBVs are only provided for the LC data, we interpolated onto the SC time-stamps using cubic splines. \\
\indent The CBVs are unique to each quarter and channel of the detector. They are ordered such that the first CBV is the dominant systematic signal of the channel and the subsequent CBVs decrease in signal strength. However, every target within a given channel is not similarly affected by instrumental systematics and therefore it is possible for the relative strengths of each CBV to vary. Determining which CBVs to fit in order to best remove systematics and preserve real astronomical signals is out of the scope of this paper. However, we do investigate the effect of varying the number of CBVs and find that there is little variation in the strength of the first four CBVs when additional CBVs are added. For each target and quarter we opt to fit the first eight CBVs, the maximum recommended by the {\it Kepler Data Processing Handbook}, as we cannot fully assess the affect that excluding CBVs would have on our analysis. \\
\indent In order to remove quarter-to-quarter discontinuities, we normalized each quarter to its out-of-transit median. After cotrending and combining all quarters, we removed outliers by calculating a running median and standard deviation of 21 measurements around each point and rejecting measurements that differed by more than 5$\sigma$. For each planet, the raw {\it Kepler} simple aperture photometry, the cotrended light curve and the cotrended out-of-transit light curve after outlier removal can be found in the Appendix Figs.~\ref{fig:A1}-\ref{fig:A7}.
\subsection{Companion Stars}
\label{sec:companion}
\begin{table}[h]
\caption{Detected Stellar Companions around Planet Host Stars}
\centering
\begin{tabular*}{8.5cm}{@{\extracolsep{\fill}}lcccc}
\hline
Host Star & Host Star                  & Comp.    & Comp. Est. & Comp. Est. \\
          & Kep Mag\textsuperscript{a} & Dist (") & Kep Mag    & Flux \%    \\
\hline
Kepler-5  & 13.369 &  0.9 \textsuperscript{b}     & 18.7 & $<$0.1\% \\
          &        & 3.39 \textsuperscript{b}     & 19.8 & $<$0.1\% \\
          &        & 4.94 \textsuperscript{b}     & 20.3 & $<$0.1\% \\
Kepler-6  & 13.303 & 4.01 \textsuperscript{b}     & 17.3 & $<$0.1\% \\
Kepler-7  & 12.885 & 1.9  \textsuperscript{b}     & 16.9 & $<$0.1\% \\
Kepler-8  & 13.563 & 3.04 \textsuperscript{b,c}   & 22.1 & $<$0.1\% \\
          &        & 3.74 \textsuperscript{b,c}   & 20.5 & $<$0.1\% \\
Kepler-10 & 10.961 & ...  \textsuperscript{b,d}   & ...  & ...      \\
Kepler-12 & 13.438 & 5.04 \textsuperscript{b,e}   & 22.2 & $<$0.1\% \\
Kepler-41 & 14.465 & ...  \textsuperscript{f}     & ...  & ...      \\
Kepler-43 & 13.958 & ...  \textsuperscript{g}     & ...  & ...      \\
Kepler-76 & 13.308 & ...  \textsuperscript{g}     & ...  & ...      \\
Kepler-91 & 12.495 & ...  \textsuperscript{h}     & ...  & ...      \\
Kepler-412& 14.309 & ...  \textsuperscript{g}     & ...  & ...      \\
KOI-13    &  9.958 & 1.12 \textsuperscript{i}     & ...  & 48\%  \\
TrES-2    & 11.338 & 1.09 \textsuperscript{j,k,l} & ...  & $<$0.1\% \\
HAT-P-7   & 10.463 & 3.8  \textsuperscript{k,m}   & ...  & $<$0.1\% \\
          &        & 3.9  \textsuperscript{k,l,m} & ...  & $<$0.1\% \\
\hline
\end{tabular*}
\begin{tabular}{p{3.8cm}}
\textsuperscript{a} From Kepler Input Catalog. \\
\textsuperscript{b} From~\citet{Companions_Adams2012}. \\
\textsuperscript{c} From~\citet{Kep8_Jenkins2010}. \\
\textsuperscript{d} From~\citet{Kep10_Batalha2011}. \\
\textsuperscript{e} From~\citet{Kep12_Fortney2011}. \\
\textsuperscript{f} No high resolution images available. However, ~\citet{Kep41_Quintana2013} test dilution scenarios and do not find a significant third light contribution. \\
\end{tabular}
\begin{tabular}{p{3.8cm}}
\textsuperscript{g} No data is available. \\
\textsuperscript{h} From~\citet{Kep91_LilloBox2014}. \\
\textsuperscript{i} From~\citet{KOI13_Shporer2014}. \\
\textsuperscript{j} From~\citet{TrES2_Daemgen2009}. \\
\textsuperscript{k} From~\citet{TrES2_HATP7_Bergfors2013}. \\
\textsuperscript{l} From~\citet{TrES2_HATP7_Faedi2013}. \\
\textsuperscript{m} From~\citet{HATP7_Narita2012}. \\
\end{tabular}
\label{tab:companion}
\end{table}
\indent {\it Kepler}'s large pixel size, with a width of 3.98", allows for the possibility of dilution from a background or foreground star or a nearby stellar companion. In the literature we found that several of our 14 systems have one or more nearby stellar companions (see Table~\ref{tab:companion}). However, only KOI-13 is significantly diluted by its companion. For KOI-13, studies find a large range of dilutions: 38-48\% ~\citep{KOI13_Szabo2011,Companions_Adams2012,KOI13_Shporer2014}. In our analysis, we adopt the~\citet{KOI13_Shporer2014} value of 48\%, corresponding to a dilution factor of 1.913$\pm$0.019. We also account for the quarter-to-quarter third light fraction provided by the {\it Kepler} Input Catalog, the average of which can be found in Tables~\ref{tab:res1}-\ref{tab:res4}.
\section{Light Curve Model}
\label{sec:model}
\begin{table}[t!]
\centering
\caption{Relevant Constants for Doppler Boosting and Ellipsoidal Modeling}
\setlength{\tabcolsep}{0.1cm}
\begin{tabular*}{8.5cm}{@{\extracolsep{\fill}}lccccccc}
\hline
& $\alpha_d$ & $\alpha_1$ & $\alpha_2$ & $f_1$ & $f_2$ & $u$ & $y$ \\
Kepler-5b &3.44&0.0649&1.23&0.194&0.324&0.545&0.290\\
Kepler-6b &3.84&0.0718&1.38&0.215&0.359&0.628&0.398\\
Kepler-7b &3.68&0.0679&1.30&0.197&0.338&0.582&0.344\\
Kepler-8b &3.51&0.0651&1.24&0.186&0.324&0.549&0.299\\
Kepler-10b &3.85&0.0692&1.36&0.201&0.345&0.603&0.391\\
Kepler-12b &3.66&0.0677&1.30&0.203&0.338&0.580&0.341\\
Kepler-41b &3.85&0.0718&1.39&0.197&0.356&0.629&0.402\\
Kepler-43b &3.61&0.0687&1.29&0.198&0.342&0.586&0.329\\
Kepler-76b &3.38&0.0637&1.20&0.150&0.311&0.532&0.273\\
Kepler-91b &4.72&0.0802&1.60&0.102&0.376&0.734&0.540\\
Kepler-412b &3.77&0.0706&1.35&0.186&0.348&0.613&0.379\\
TrES-2b &3.71&0.0674&1.31&0.192&0.335&0.580&0.354\\
HAT-P-7b &3.41&0.0657&1.22&0.184&0.326&0.551&0.282\\
KOI-13b\textsuperscript{a} &2.77&0.0550&1.29&0.163&0.274&0.477&0.406\\
KOI-13b\textsuperscript{b} &2.27&0.0492&1.39&0.146&0.246&0.443&0.539\\
KOI-13b\textsuperscript{c} &2.43&0.0517&1.49&0.153&0.258&0.476&0.624\\
\hline
\end{tabular*}
\begin{tabular}{p{8.5cm}}
Notes. 
Although $\alpha_1$, $u$ and $y$ are not explicitly mentioned in this paper their description and use in calculating $\alpha_2$, $f_1$ and $f_2$ can be found in~\citet{Esteves2013}. \\
Derived using stellar parameters from: \\
\textsuperscript{a} \citet{KOI13_Shporer2014}.\\
\textsuperscript{b} \citet{StellarParamsRevised_Huber2013}.\\
\textsuperscript{c} \citet{KOI13_Szabo2011}.\\
\end{tabular}
\label{tab:constants}
\end{table}
\begin{table}[t!]
\centering
\caption{$\Delta$BIC Values of Unfavored Models}
\setlength{\tabcolsep}{0.03cm}
\begin{tabular*}{8.5cm}{@{\extracolsep{\fill}}lcccccccc}
\hline
\hline
 & \multicolumn{8}{c}{Eccentricity} \\
 & \multicolumn{4}{c}{\scriptsize{Fixed}} & \multicolumn{4}{c}{\scriptsize{Free}} \\
\cline{2-9} 
 & \multicolumn{4}{c}{Planet Mass} & \multicolumn{4}{c}{Planet Mass} \\
 & \multicolumn{2}{c}{\scriptsize{Fixed}} & \multicolumn{2}{c}{\scriptsize{Free}}
 & \multicolumn{2}{c}{\scriptsize{Fixed}} & \multicolumn{2}{c}{\scriptsize{Free}} \\
\cline{2-9} 
 & \multicolumn{2}{c}{Offset} & \multicolumn{2}{c}{Offset}
 & \multicolumn{2}{c}{Offset} & \multicolumn{2}{c}{Offset} \\
 & \multicolumn{1}{c}{\scriptsize{Fixed}} & \multicolumn{1}{c}{\scriptsize{Free}}
 & \multicolumn{1}{c}{\scriptsize{Fixed}} & \multicolumn{1}{c}{\scriptsize{Free}}
 & \multicolumn{1}{c}{\scriptsize{Fixed}} & \multicolumn{1}{c}{\scriptsize{Free}}
 & \multicolumn{1}{c}{\scriptsize{Fixed}} & \multicolumn{1}{c}{\scriptsize{Free}} \\
\cline{2-9} 
Kepler-5b & 12 & 14 & * & 6 & 23 & 27 & 12 & 18 \\
Kepler-6b & * & 2 & 6 & 7 & 12 & 14 & 19 & 20 \\
Kepler-7b & 167 & * & 160 & 4 & 179 & 12 & 172 & 16 \\
Kepler-8b & 11 & * & 18 & 4 & 24 & 12 & 31 & 16 \\
Kepler-10b & * & 6 & 3 & 10 & 12 & 18 & 16 & 23 \\
Kepler-12b & 46 & * & 49 & 5 & 58 & 12 & 61 & 19 \\
Kepler-41b & 35 & * & 41 & 4 & 47 & 12 & 54 & 17 \\
Kepler-43b & 74 & 19 & 77 & * & 87 & 32 & 91 & 12 \\
Kepler-76b & 36 & * & 31 & 6 & 53 & 16 & 46 & 22 \\
Kepler-91b & 0.4 & * & 2 & 3 & 12 & 11 & 14 & 15 \\
Kepler-412b & * & 5 & 6 & 11 & 12 & 17 & 19 & 23 \\
TrES-2b & * & 5 & 5 & 11 & 13 & 16 & 17 & 23 \\
HAT-P-7b & 569 & 51 & 568 & * & 579 & 62 & 581 & 12 \\
KOI-13b\textsuperscript{a} & --- & --- & 45 & 51 & --- & --- & * & 9 \\
KOI-13b\textsuperscript{b} & --- & --- & 30 & 27 & --- & --- & * & 2 \\
KOI-13b\textsuperscript{c} & --- & --- & 66 & 48 & --- & --- & 35 & * \\
\hline
\end{tabular*}
\begin{tabular}{p{8.5cm}}
Derived using stellar parameters from: \\
\textsuperscript{a} \citet{KOI13_Shporer2014}.\\
\textsuperscript{b} \citet{StellarParamsRevised_Huber2013}.\\
\textsuperscript{c} \citet{KOI13_Szabo2011}.\\
When fixed, the eccentricity and brightness offset were set to zero, while the planet mass was set to its RV derived value. The favored model is indicated with an asterisk, while a dash indicates where models were not fit due to a lack of RV data. For HAT-P-7b and KOI-13b all models also include an independently fit cosine third harmonic. \\
\end{tabular}
\label{tab:bic}
\end{table}
\begin{figure*}[t!]
\begin{center}
\scalebox{0.8}{\includegraphics{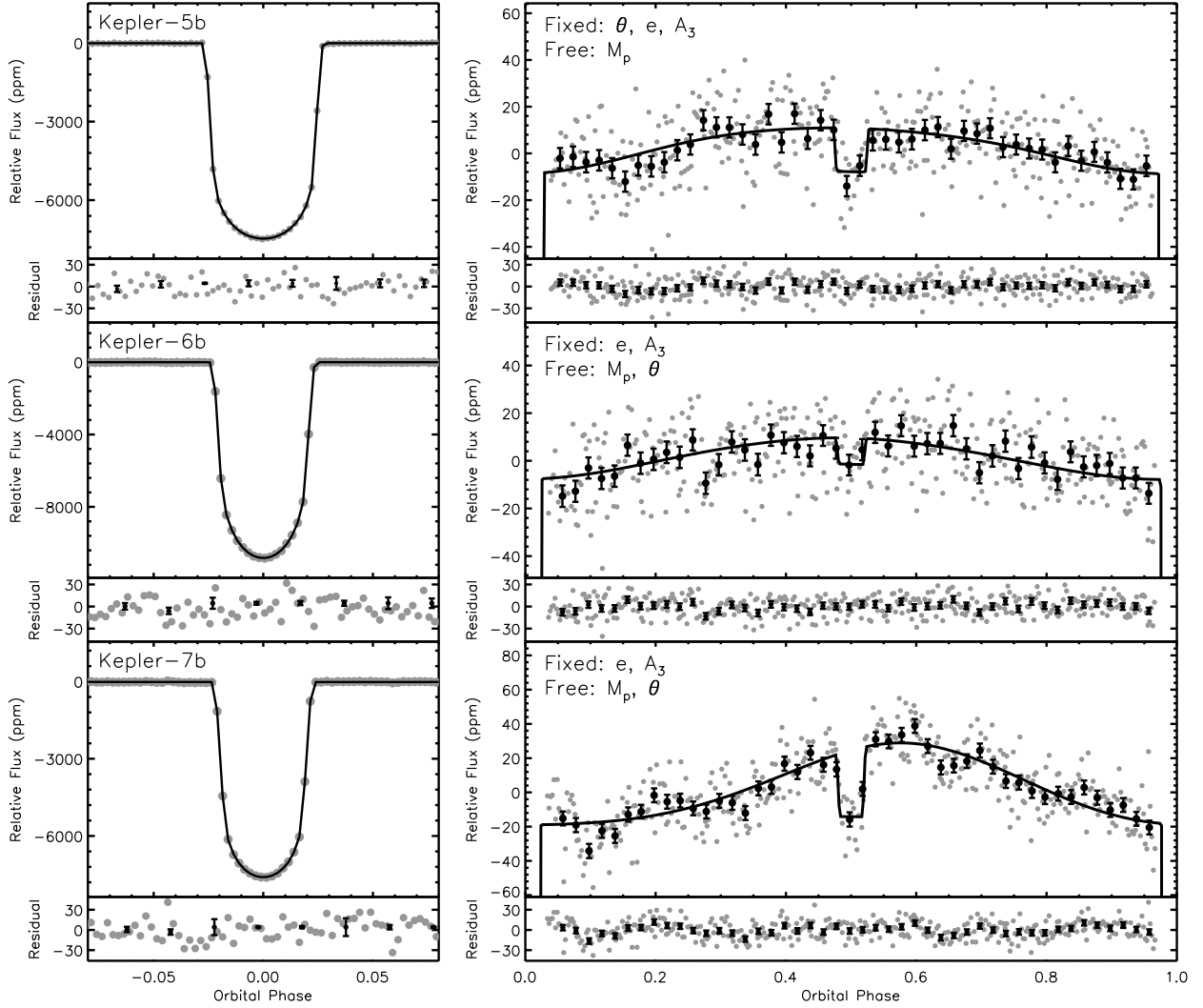}}
\end{center}
\caption{The left and right panels contain the binned and phase-folded transit and phase curves, respectively. For Kepler-5b, Kepler-6b and Kepler-7b the transit and phase curve was binned such that 400 points span the orbit, resulting in a bin size of 12.8, 11.6 and 17.6 minutes, respectively. Over-plotted on each is our best fit model and binned data, such that 50 points span the orbit (black).}
\label{fig:res1}
\end{figure*}
\begin{figure*}[t!]
\begin{center}
\scalebox{0.8}{\includegraphics{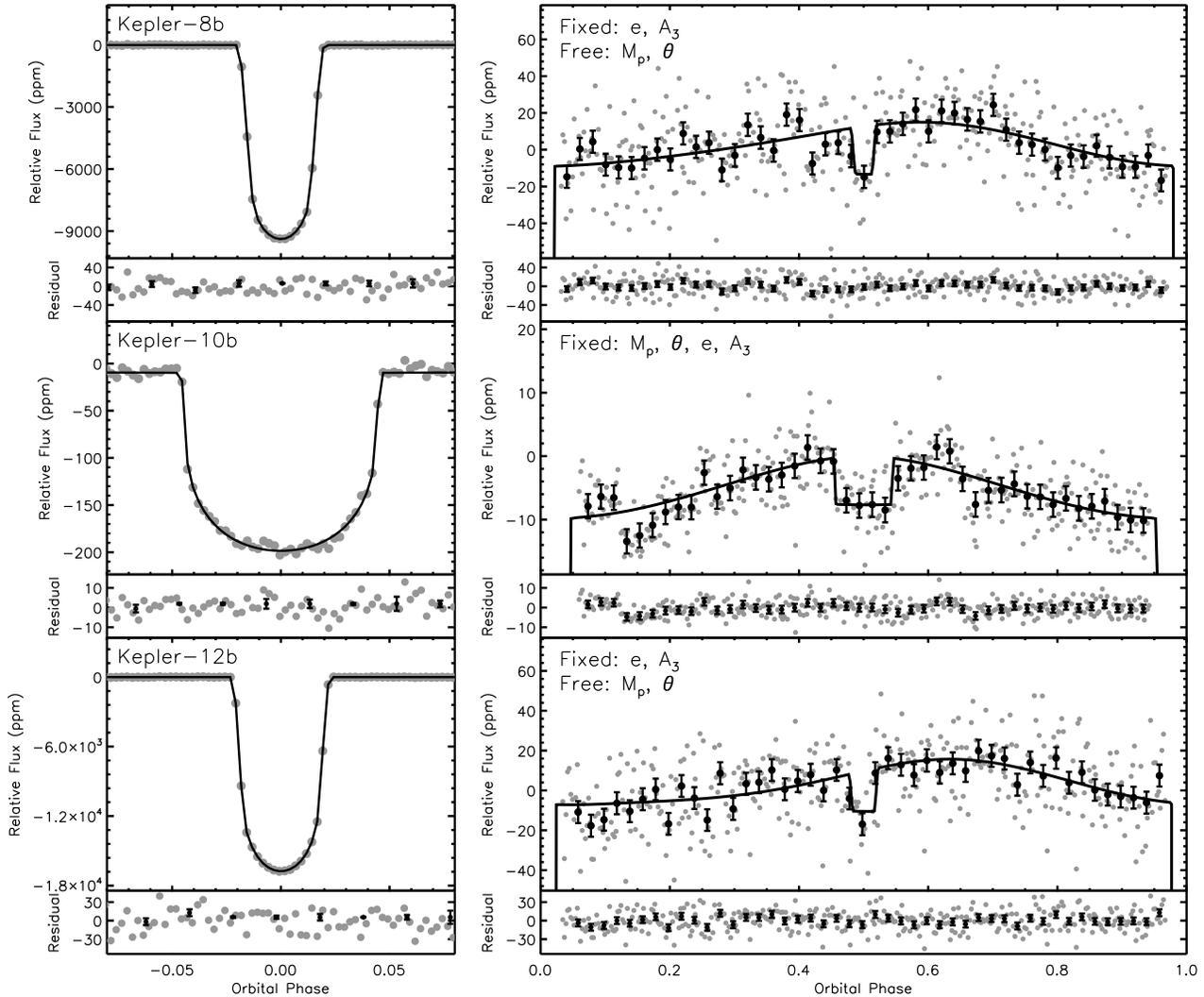}}
\end{center}
\caption{Same as Fig.~\ref{fig:res1}. However, for Kepler-8b, Kepler-10b and Kepler-12b the bin sizes are 12.7, 3.0 and 16.0 minutes, respectively.}
\label{fig:res2}
\end{figure*}
\begin{figure*}[t!]
\begin{center}
\scalebox{0.8}{\includegraphics{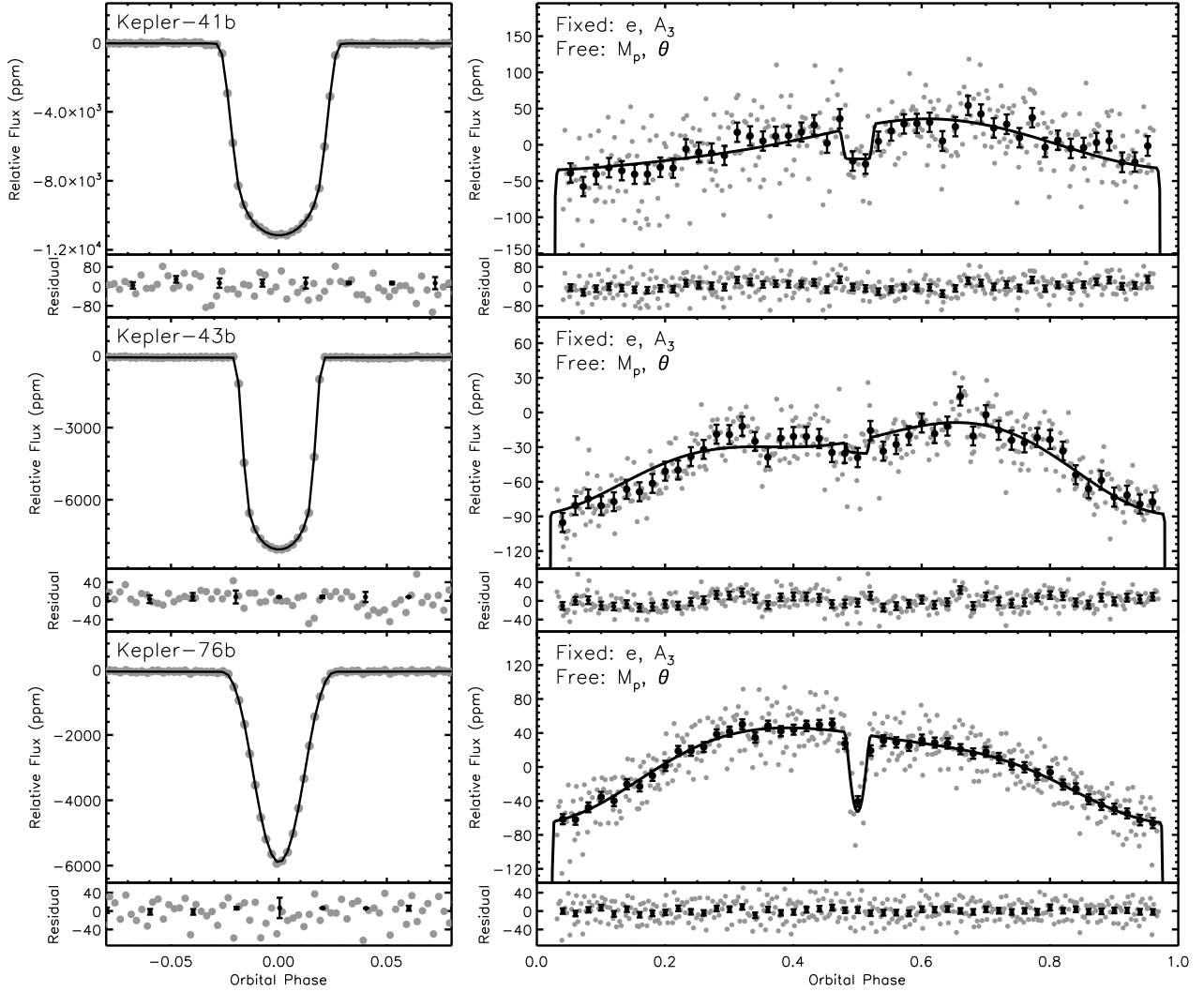}}
\end{center}
\caption{Same as Fig.~\ref{fig:res1}. However, for Kepler-41b, Kepler-43b and Kepler-76b the bin sizes are 6.7, 10.9 and 5.6 minutes, respectively.}
\label{fig:res3}
\end{figure*}
\begin{figure*}[t!]
\begin{center}
\scalebox{0.8}{\includegraphics{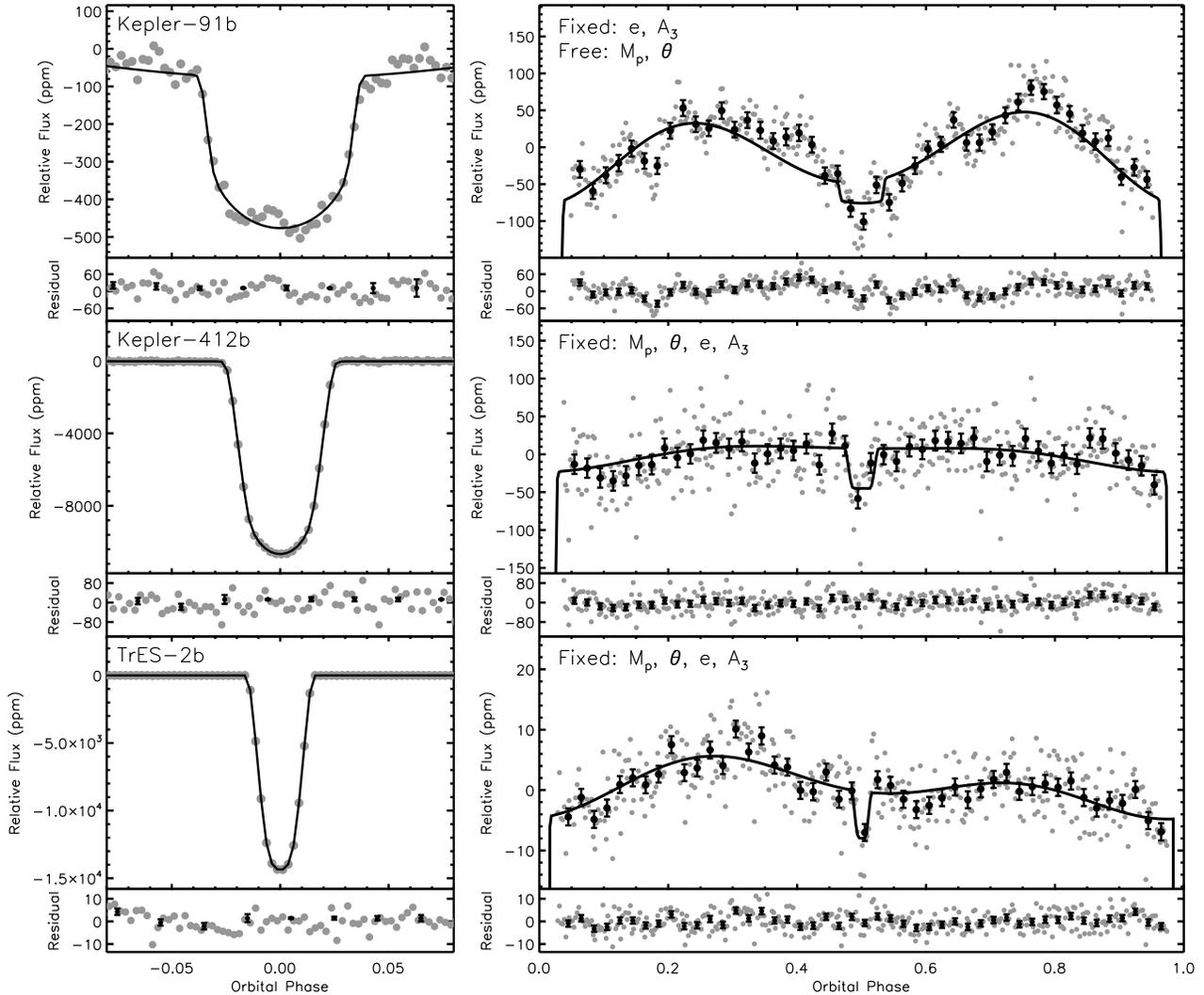}}
\end{center}
\caption{Same as Fig.~\ref{fig:res1}. However, for Kepler-91b, Kepler-412b and TrES-2b the bin sizes are 22.5, 6.2 and 8.9 minutes, respectively.}
\label{fig:res4}
\end{figure*}
\begin{figure*}[t!]
\begin{center}
\scalebox{0.8}{\includegraphics{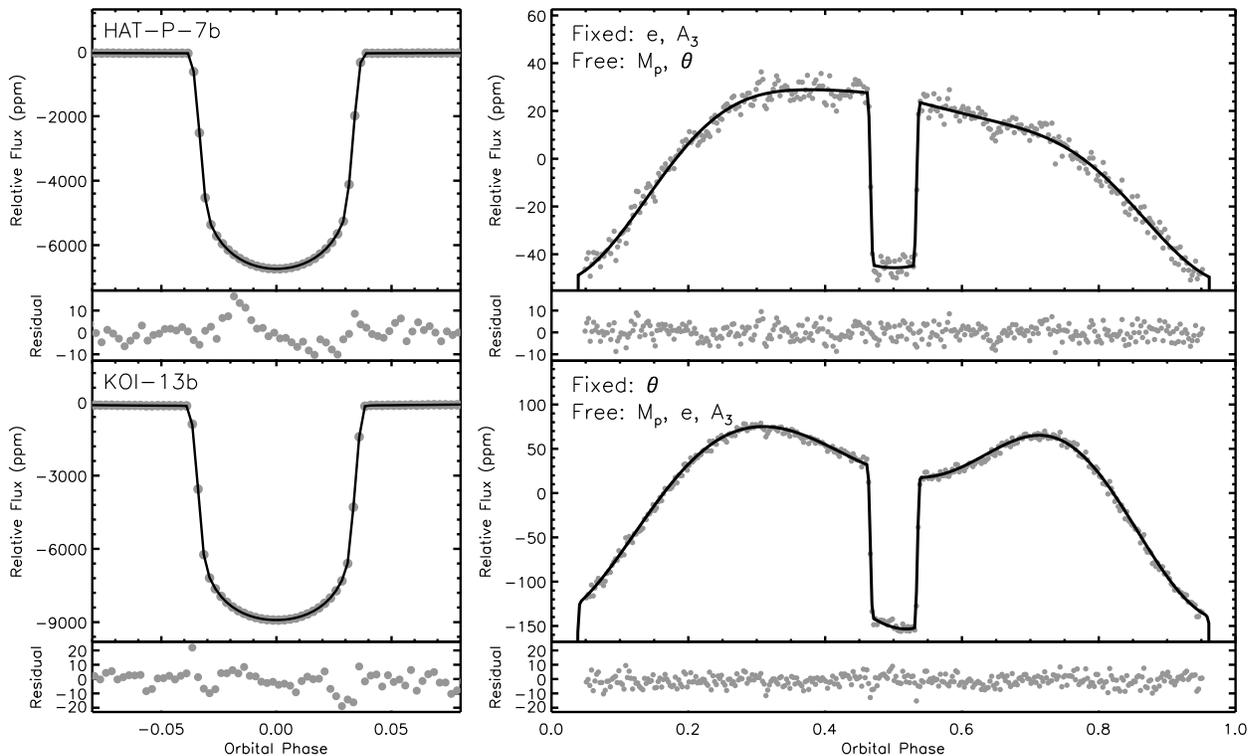}}
\end{center}
\caption{Same as Fig.~\ref{fig:res1}. However, for HAT-P-7b and KOI-13b the bin sizes are 7.9 and 6.4 minutes, respectively. For HAT-P-7b and KOI-13b we did not overplot the increasingly binned data as the error bars are smaller than the size of the data points.}
\label{fig:res5}
\end{figure*}
\begin{table*}[t!]
\centering
\caption{Stellar and Planetary Parameters}
\setlength{\tabcolsep}{0.1cm}
\begin{tabular*}{\textwidth}{@{\extracolsep{\fill}}lcccc}
\hline
\hline
Parameter & Kepler-5b & 
Kepler-6b & 
Kepler-7b & 
Kepler-8b \\
\hline \\ [-2.0ex]
KOI & 18.01 & 
17.01 & 
97.01 & 
10.01 \\
KIC & 8191672 & 
10874614 & 
5780885 & 
6922244 \\
Period (days)\textsuperscript{a} & 3.5484657$\pm$0.0000007 & 
3.2346996$\pm$0.0000004 & 
4.8854892$\pm$0.0000009 & 
3.5224991$\pm$0.0000007 \\
$T_{\star}$ (K) & 6297$\pm$60\textsuperscript{c} & 
5647$\pm$44\textsuperscript{d} & 
5933$\pm$44\textsuperscript{e} & 
6213$\pm$150\textsuperscript{g} \\
$\log g$ (cgs) & 3.96$\pm$0.10\textsuperscript{c} & 
4.236$\pm$0.011\textsuperscript{d} & 
3.98$\pm$0.1\textsuperscript{e} & 
4.174$\pm$0.026\textsuperscript{g} \\
$\mathrm{[Fe/H]}$ & 0.04$\pm$0.06\textsuperscript{c} & 
0.34$\pm$0.04\textsuperscript{d} & 
0.11$\pm$0.03\textsuperscript{e} & 
-0.055$\pm$0.03\textsuperscript{g} \\
$R_{\star}/R_{\sun}$ & 1.793$^{+0.043}_{-0.062}$\textsuperscript{c} & 
1.391$^{+0.017}_{-0.034}$\textsuperscript{d} & 
1.966$\pm$0.013\textsuperscript{f} & 
1.486$^{+0.053}_{-0.062}$\textsuperscript{g} \\
$M_{\star}/M_{\sun}$ & 1.374$^{+0.040}_{-0.059}$\textsuperscript{c} & 
1.209$^{+0.044}_{-0.038}$\textsuperscript{d} & 
1.359$\pm$0.031\textsuperscript{f} & 
1.213$^{+0.067}_{-0.062}$\textsuperscript{g} \\
$K$ (m s$^{-1}$) & 227.5$\pm$2.8\textsuperscript{c} & 
80.9$\pm$2.6\textsuperscript{d} & 
42.9$\pm$3.5\textsuperscript{e} & 
68.4$\pm$12.0\textsuperscript{g} \\
$M_{\mathrm{p}}$ from RV $({\mathrm{M_J}})$ & 2.111$^{+0.067}_{-0.086}$ & 
0.668$^{+0.038}_{-0.035}$ & 
0.441$^{+0.043}_{-0.042}$ & 
0.59$^{+0.13}_{-0.12}$ \\
Avg 3rd light (\%)\textsuperscript{b} & 3.14 & 
2.34 & 
1.42 & 
2.49 \\
[0.5ex] \hline
\multicolumn{5}{l}{Transit fit} \\
\hline \\ [-2.0ex]
$T_0$ (BJD-2454833) & 122.90144$\pm$0.00002 & 
121.486528$^{+0.000009}_{-0.000012}$ & 
134.27687$^{+0.00003}_{-0.00002}$ & 
121.11931$\pm$0.00002 \\
$b$ &0.0968$^{+0.035}_{-0.049}$ & 
0.141$^{+0.022}_{-0.025}$ & 
0.5599$^{+0.0045}_{-0.0046}$ & 
0.7191$^{+0.0021}_{-0.0023}$ \\
$R_{\mathrm{p}}/R_{\star}$ & 0.079965$^{+0.000087}_{-0.000071}$ & 
0.09424$^{+0.00012}_{-0.00011}$ & 
0.08294$\pm$0.00011 & 
0.095751$^{+0.00019}_{-0.00023}$ \\
$a/R_{\star}$ & 6.450$^{+0.021}_{-0.025}$ & 
7.503$\pm$0.022 & 
6.637$\pm$0.021 & 
6.854$^{+0.018}_{-0.017}$ \\
$i$ (degrees) & 89.14$^{+0.44}_{-0.32}$ & 
88.93$^{+0.19}_{-0.17}$ & 
85.161$^{+0.055}_{-0.054}$ & 
83.978$^{+0.036}_{-0.033}$ \\
$u_1$ & 0.3591$^{+0.0073}_{-0.0072}$ & 
0.4855$^{+0.0061}_{-0.0060}$ & 
0.368$^{+0.021}_{-0.016}$ & 
0.371$^{+0.039}_{-0.056}$ \\
$u_2$ & 0.152$^{+0.014}_{-0.015}$ & 
0.135$^{+0.013}_{-0.014}$ & 
0.206$^{+0.025}_{-0.033}$ & 
0.161$^{+0.071}_{-0.051}$ \\
[0.5ex] \hline
\multicolumn{5}{l}{Phase curve fit} \\
\hline \\ [-2.0ex]
$F_{\mathrm{ecl}}$ (ppm) & 18.6$^{+5.1}_{-5.3}$ & 
11.1$^{+4.8}_{-5.3}$ & 
39$\pm$11 & 
26$\pm$10 \\
$A_{\mathrm{p}}$ (ppm) & 19.3$^{+9.5}_{-7.6}$ & 
17.2$^{+6.3}_{-5.1}$ & 
48$^{+16}_{-17}$ & 
24$^{+16}_{-15}$ \\
$\theta$ (in phase)& 0\textsuperscript{*} & 
0\textsuperscript{*} & 
-0.0696$\pm$0.0052 & 
-0.069$^{+0.015}_{-0.016}$ \\
$F_{\mathrm{n}}$ (ppm) & -0.6$\pm$10 & 
-6$\pm$10 & 
-5$\pm$30 & 
4$\pm$20 \\
$M_{\mathrm{p}}$ $({\mathrm{M_J}})$ & 0.92$^{+0.93}_{-1.02}$ & 
0.668\textsuperscript{*} & 
0.441\textsuperscript{*} & 
0.59\textsuperscript{*} \\
[0.5ex] \hline
\multicolumn{5}{l}{Derived parameters} \\
\hline \\ [-2.0ex]
$R_{\mathrm{p}}$ $({\mathrm{R_J}})$ & 1.426$^{+0.036}_{-0.051}$ & 
1.304$^{+0.018}_{-0.033}$ & 
1.622$\pm$0.013 & 
1.416$^{+0.053}_{-0.062}$ \\
$a$ (Au) & 0.0538$^{+0.0015}_{-0.0021}$ & 
0.04852$^{+0.00074}_{-0.00133}$ & 
0.06067$\pm$0.00059 & 
0.0474$^{+0.0018}_{-0.0021}$ \\
$A_{\mathrm{g,ecl}}$ & 0.121$^{+0.034}_{-0.036}$ & 
0.070$^{+0.031}_{-0.034}$ & 
0.248$^{+0.073}_{-0.071}$ & 
0.133$\pm$0.053 \\
$T_{\mathrm{eq,max}}$ (K) & 2240$^{+30}_{-20}$ & 
1860$\pm$20 & 
2080$\pm$20 & 
2140$\pm$50 \\
$T_{\mathrm{eq,hom}}$ (K) & 1750$\pm$20 & 
1460$\pm$10 & 
1630$\pm$10 & 
1680$\pm$40 \\
$T_{\mathrm{B,day}}$ (K) & 2390$^{+80}_{-100}$ & 
2060$^{+90}_{-140}$ & 
2510$^{+90}_{-110}$ & 
2400$^{+100}_{-200}$ \\
$T_{\mathrm{B,night}}$ (K) & $^{<2300(1\sigma)}_{<2500(2\sigma)}$ & 
$^{<1900(1\sigma)}_{<2100(2\sigma)}$ & 
$^{<2400(1\sigma)}_{<2600(2\sigma)}$ & 
$^{<2400(1\sigma)}_{<2600(2\sigma)}$ \\
$\rho_p$ (g/cm$^3$) & 0.4$\pm$0.5 & 
0.40$^{+0.06}_{-0.04}$ & 
0.14$\pm$0.02 & 
0.27$^{+0.11}_{-0.08}$ \\
$\log g_p$ (cgs) & 3.1$^{+0.3}_{-3.1}$ & 
3.01$^{+0.05}_{-0.04}$ & 
2.64$\pm$0.05 & 
2.9$\pm$0.1 \\
\hline \\ [-2.0ex]
\end{tabular*}
\begin{tabular}{p{8.5cm}}
\textsuperscript{*} Best-fit model favored fixed values for these parameters. \\
\textsuperscript{a} From the NASA Exoplanet Archive \citep{archive2013}.\\
\textsuperscript{b} From the Kepler Input Catalog.\\
\textsuperscript{c} From \citet{Kep5_Koch2010}.\\
\end{tabular}
\begin{tabular}{p{8.5cm}}
\textsuperscript{d} From \citet{Kep6_Dunham2010}.\\
\textsuperscript{e} From \citet{Kep7_Latham2010}.\\
\textsuperscript{f} From \citet{Kep7_Demory2011}.\\
\textsuperscript{g} From \citet{Kep8_Jenkins2010}.\\
\end{tabular}
\label{tab:res1}
\end{table*}
\begin{table*}[t!]
\centering
\caption{Stellar and Planetary Parameters}
\setlength{\tabcolsep}{0.1cm}
\begin{tabular*}{\textwidth}{@{\extracolsep{\fill}}lcccc}
\hline
\hline
Parameter & Kepler-10b & 
Kepler-12b & 
Kepler-41b & 
Kepler-43b \\
\hline \\ [-2.0ex]
KOI & 72.01 & 
20.01 & 
196.01 & 
135.01 \\
KIC & 11904151 & 
11804465 & 
9410930 & 
9818381 \\
Period (days)\textsuperscript{a} & 0.837491$\pm$0.000002 & 
4.4379629$\pm$0.0000006 & 
1.8555577$\pm$0.0000003 & 
3.0240949$\pm$0.0000006 \\
$T_{\star}$ (K) & 5627$\pm$44\textsuperscript{c} & 
5947$\pm$100\textsuperscript{d} & 
5620$\pm$140\textsuperscript{e} & 
6041$\pm$143\textsuperscript{f} \\
$\log g$ (cgs) & 4.35$\pm$0.06\textsuperscript{c} & 
4.175$^{+0.015}_{-0.011}$\textsuperscript{d} & 
4.47$\pm$0.12\textsuperscript{e} & 
4.26$\pm$0.05\textsuperscript{f} \\
$\mathrm{[Fe/H]}$ & -0.15$\pm$0.04\textsuperscript{c} & 
0.07$\pm$0.04\textsuperscript{d} & 
0.29$\pm$0.16\textsuperscript{e} & 
0.33$\pm$0.11\textsuperscript{f} \\
$R_{\star}/R_{\sun}$ & 1.065$\pm$0.009\textsuperscript{c} & 
1.483$^{+0.025}_{-0.029}$\textsuperscript{d} & 
1.02$\pm$0.03\textsuperscript{e} & 
1.42$\pm$0.07\textsuperscript{f} \\
$M_{\star}/M_{\sun}$ & 0.913$\pm$0.022\textsuperscript{c} & 
1.166$^{+0.051}_{-0.054}$\textsuperscript{d} & 
1.12$\pm$0.07\textsuperscript{e} & 
1.32$\pm$0.09\textsuperscript{f} \\
$K$ (m s$^{-1}$) & 3.3$^{+0.8}_{-1.0}$\textsuperscript{c} & 
48.2$^{+4.4}_{-4.3}$\textsuperscript{d} & 
85$\pm$11\textsuperscript{e} & 
375$\pm$13\textsuperscript{f} \\
$M_{\mathrm{p}}$ from RV $({\mathrm{M_J}})$ & 0.0145$^{+0.0040}_{-0.0046}$ & 
0.432$^{+0.053}_{-0.051}$ & 
0.560$^{+0.0986}_{-0.093}$ & 
3.23$\pm$0.26 \\
Avg 3rd light (\%)\textsuperscript{b} & 1.13 & 
1.87 & 
4.54 & 
2.97 \\
[0.5ex] \hline
\multicolumn{5}{l}{Transit fit} \\
\hline \\ [-2.0ex]
$T_0$ (BJD-2454833) & 131.57513$\pm$0.00005 & 
171.00915$\pm$0.00001 & 
137.18104$\pm$0.00002 & 
132.41624$\pm$0.00002 \\
$b$ &0.30$^{+0.21}_{-0.20}$ & 
0.168$^{+0.010}_{-0.012}$ & 
0.6846$^{+0.0046}_{-0.0050}$ & 
0.6508$^{+0.0071}_{-0.0087}$ \\
$R_{\mathrm{p}}/R_{\star}$ & 0.01247$^{+0.00031}_{-0.00014}$ & 
0.118867$^{+0.000085}_{-0.000094}$ & 
0.10253$^{+0.00043}_{-0.00045}$ & 
0.08628$^{+0.00036}_{-0.00033}$ \\
$a/R_{\star}$ & 3.47$^{+0.13}_{-0.31}$ & 
8.019$^{+0.014}_{-0.013}$ & 
5.053$\pm$0.021 & 
6.975$^{+0.047}_{-0.041}$ \\
$i$ (degrees) & 85.1$^{+3.4}_{-4.3}$ & 
88.796$^{+0.088}_{-0.074}$ & 
82.214$^{+0.090}_{-0.085}$ & 
84.646$^{+0.107}_{-0.091}$ \\
$u_1$ & 0.459$^{+0.069}_{-0.051}$ & 
0.4205$^{+0.0060}_{-0.0058}$ & 
0.369$^{+0.085}_{-0.063}$ & 
0.290$^{+0.069}_{-0.039}$ \\
$u_2$ & 0.242$^{+0.070}_{-0.105}$ & 
0.137$\pm$0.013 & 
0.31$^{+0.095}_{-0.12}$ & 
0.338$^{+0.060}_{-0.104}$ \\
[0.5ex] \hline
\multicolumn{5}{l}{Phase curve fit} \\
\hline \\ [-2.0ex]
$F_{\mathrm{ecl}}$ (ppm) & 7.5$^{+2.0}_{-2.1}$ & 
20.2$^{+8.3}_{-7.6}$ & 
44$^{+15}_{-16}$ & 
11$^{+40}_{-37}$ \textsuperscript{$\dagger$} \\
$A_{\mathrm{p}}$ (ppm) & 9.79$^{+3.4}_{-3.2}$ & 
22.9$^{+6.2}_{-6.0}$ & 
69$\pm$25 & 
71$^{+56}_{-60}$ \textsuperscript{$\dagger$} \\
$\theta$ (in phase)& 0\textsuperscript{*} & 
-0.123$\pm$0.016 & 
-0.087$\pm$0.014 & 
-0.0747$^{+0.0074}_{-0.0078}$ \textsuperscript{$\dagger$} \\
$F_{\mathrm{n}}$ (ppm) & -2.3$^{+5.2}_{-5.6}$ & 
3$\pm$10 & 
-15$^{+37}_{-38}$ & 
-53$^{+94}_{-87}$ \textsuperscript{$\dagger$} \\
$M_{\mathrm{p}}$ $({\mathrm{M_J}})$ & 0.0145\textsuperscript{*} & 
0.432\textsuperscript{*} & 
0.560\textsuperscript{*} & 
6.3$^{+7.7}_{-7.0}$ \textsuperscript{$\dagger$} \\
[0.5ex] \hline
\multicolumn{5}{l}{Derived parameters} \\
\hline \\ [-2.0ex]
$R_{\mathrm{p}}$ $({\mathrm{R_J}})$ & 0.1321$^{+0.0044}_{-0.0026}$ & 
1.754$^{+0.031}_{-0.036}$ & 
1.040$\pm$0.035 & 
1.219$^{+0.065}_{-0.064}$ \textsuperscript{$\dagger$} \\
$a$ (Au) & 0.01720$^{+0.00081}_{-0.00168}$ & 
0.0553$^{+0.0010}_{-0.0012}$ & 
0.02396$^{+0.00081}_{-0.00080}$ & 
0.0460$^{+0.0026}_{-0.0025}$ \textsuperscript{$\dagger$} \\
$A_{\mathrm{g,ecl}}$ & 0.58$^{+0.23}_{-0.25}$ & 
0.092$^{+0.039}_{-0.035}$ & 
0.108$\pm$0.040 & 
0.071$^{+0.267}_{-0.071}$ \textsuperscript{$\dagger$} \\
$T_{\mathrm{eq,max}}$ (K) & 2730$^{+150}_{-70}$ & 
1900$\pm$30 & 
2260$\pm$60 & 
2070$\pm$60 \textsuperscript{$\dagger$} \\
$T_{\mathrm{eq,hom}}$ (K) & 2130$^{+120}_{-60}$ & 
1480$\pm$30 & 
1770$\pm$50 & 
1620$\pm$40 \textsuperscript{$\dagger$} \\
$T_{\mathrm{B,day}}$ (K) & 3300$^{+100}_{-200}$ & 
2100$\pm$100 & 
2400$\pm$100 & 
2200$^{+500}_{-2200}$ \textsuperscript{$\dagger$} \\
$T_{\mathrm{B,night}}$ (K) & $^{<2900(1\sigma)}_{<3300(2\sigma)}$ & 
$^{<2100(1\sigma)}_{<2200(2\sigma)}$ & 
$^{<2200(1\sigma)}_{<2500(2\sigma)}$ & 
$^{<2600(1\sigma)}_{<3000(2\sigma)}$ \textsuperscript{$\dagger$} \\
$\rho_p$ (g/cm$^3$) & 8$\pm$3 & 
0.11$\pm$0.02 & 
0.7$\pm$0.2 & 
5$^{+7}_{-5}$ \textsuperscript{$\dagger$} \\
$\log g_p$ (cgs) & 3.3$^{+0.1}_{-0.2}$ & 
2.56$\pm$0.07 & 
3.1$\pm$0.1 & 
4.0$^{+0.4}_{-4.0}$ \textsuperscript{$\dagger$} \\
\hline \\ [-2.0ex]
\end{tabular*}
\begin{tabular}{p{8.5cm}}
\textsuperscript{*} Best-fit model favored fixed values for these parameters. \\
\textsuperscript{$\dagger$} Planet's with evidence for non-planetary phase curve modulations (See Section~\ref{sec:k43_k91}). \\
\textsuperscript{a} From the NASA Exoplanet Archive \citep{archive2013}.\\
\textsuperscript{b} From the Kepler Input Catalog.\\
\end{tabular}
\begin{tabular}{p{8.5cm}}
\textsuperscript{c} From \citet{Kep10_Batalha2011}.\\
\textsuperscript{d} From \citet{Kep12_Fortney2011}.\\
\textsuperscript{e} From \citet{Kep41_Santerne2011}.\\
\textsuperscript{f} From \citet{Kep43_Bonomo2012}.\\
\end{tabular}
\label{tab:res2}
\end{table*}
\begin{table*}[t!]
\centering
\caption{Stellar and Planetary Parameters}
\setlength{\tabcolsep}{0.1cm}
\begin{tabular*}{\textwidth}{@{\extracolsep{\fill}}lcccc}
\hline
\hline
Parameter & Kepler-76b & 
Kepler-91b & 
Kepler-412b & 
TrES-2b \\
\hline \\ [-2.0ex]
KOI & 1658.01 & 
2133.01 & 
202.01 & 
1.01 \\
KIC & 4570949 & 
8219268 & 
7877496 & 
11446443 \\
Period (days)\textsuperscript{a} & 1.5449298$\pm$0.0000004 & 
6.24658$\pm$0.00008 & 
1.7208604$\pm$0.0000003 & 
2.47061317$\pm$0.00000009 \\
$T_{\star}$ (K) & 6409$\pm$95\textsuperscript{c} & 
4550$\pm$75\textsuperscript{d} & 
5750$\pm$90\textsuperscript{f} & 
5850$\pm$50\textsuperscript{g} \\
$\log g$ (cgs) & 4.2$\pm$0.3\textsuperscript{c} & 
2.953$\pm$0.007\textsuperscript{d} & 
4.30$\pm$0.07\textsuperscript{f} & 
4.426$^{+0.021}_{-0.023}$\textsuperscript{g} \\
$\mathrm{[Fe/H]}$ & -0.1$\pm$0.2\textsuperscript{c} & 
0.11$\pm$0.07\textsuperscript{d} & 
0.27$\pm$0.12\textsuperscript{f} & 
-0.15$\pm$0.10\textsuperscript{g} \\
$R_{\star}/R_{\sun}$ & 1.32$\pm$0.08\textsuperscript{c} & 
6.30$\pm$0.16\textsuperscript{d} & 
1.287$\pm$0.035\textsuperscript{f} & 
1.000$^{+0.036}_{-0.033}$\textsuperscript{g} \\
$M_{\star}/M_{\sun}$ & 1.2$\pm$0.2\textsuperscript{c} & 
1.31$\pm$0.10\textsuperscript{d} & 
1.167$\pm$0.091\textsuperscript{f} & 
0.980$\pm$0.062\textsuperscript{g} \\
$K$ (m s$^{-1}$) & 306$\pm$20\textsuperscript{c} & 
70$\pm$11\textsuperscript{e} & 
142$\pm$11\textsuperscript{f} & 
181.3$\pm$2.6\textsuperscript{h} \\
$M_{\mathrm{p}}$ from RV $({\mathrm{M_J}})$ & 2.01$^{+0.37}_{-0.35}$ & 
0.81$^{+0.18}_{-0.17}$ & 
0.941$^{+0.125}_{-0.019}$ & 
1.197$\pm$0.068 \\
Avg 3rd light (\%)\textsuperscript{b} & 5.55 & 
1.00 & 
5.96 & 
0.729 \\
[0.5ex] \hline
\multicolumn{5}{l}{Transit fit} \\
\hline \\ [-2.0ex]
$T_0$ (BJD-2454833) & 133.54841$^{+0.00001}_{-0.00002}$ & 
136.3958$\pm$0.0002 & 
133.02122$\pm$0.00002 & 
122.763360$^{+0.000003}_{-0.000002}$ \\
$b$ &0.96226$^{+0.0040}_{-0.0049}$ & 
0.8667$^{+0.0077}_{-0.0105}$ & 
0.7942$^{+0.0028}_{-0.0032}$ & 
0.84359$^{+0.00075}_{-0.00062}$ \\
$R_{\mathrm{p}}/R_{\star}$ & 0.1033$^{+0.0024}_{-0.0030}$ & 
0.02181$^{+0.00054}_{-0.00042}$ & 
0.10474$^{+0.00054}_{-0.00077}$ & 
0.12539$^{+0.00049}_{-0.00035}$ \\
$a/R_{\star}$ & 4.464$^{+0.049}_{-0.041}$ & 
2.496$^{+0.050}_{-0.043}$ & 
4.841$^{+0.024}_{-0.023}$ & 
7.903$^{+0.019}_{-0.016}$ \\
$i$ (degrees) & 77.55$^{+0.20}_{-0.17}$ & 
69.68$^{+0.67}_{-0.57}$ & 
80.559$^{+0.084}_{-0.079}$ & 
83.872$^{+0.020}_{-0.018}$ \\
$u_1$ & 0.762$^{+0.073}_{-0.139}$ & 
0.73$^{+0.13}_{-0.19}$ & 
0.36$^{+0.11}_{-0.15}$ & 
0.330$^{+0.080}_{-0.060}$ \\
$u_2$ & -0.346$^{+0.067}_{-0.047}$ & 
-0.002$^{+0.2}_{-0.1}$ & 
0.29$^{+0.18}_{-0.12}$ & 
0.285$^{+0.062}_{-0.087}$ \\
[0.5ex] \hline
\multicolumn{5}{l}{Phase curve fit} \\
\hline \\ [-2.0ex]
$F_{\mathrm{ecl}}$ (ppm) & 131.6$^{+8.7}_{-8.0}$ & 
35$\pm$18 \textsuperscript{$\dagger$} & 
53$^{+21}_{-24}$ & 
7.7$^{+2.4}_{-2.6}$ \\
$A_{\mathrm{p}}$ (ppm) & 106.9$^{+6.2}_{-5.6}$ & 
25$^{+18}_{-19}$ \textsuperscript{$\dagger$} & 
31$^{+20}_{-19}$ & 
4.1$^{+1.7}_{-1.8}$ \\
$\theta$ (in phase)& 0.0230$\pm$0.0034 & 
-0.132$^{+0.050}_{-0.047}$ \textsuperscript{$\dagger$} & 
0\textsuperscript{*} & 
0\textsuperscript{*} \\
$F_{\mathrm{n}}$ (ppm) & 28$\pm$14 & 
17$^{+31}_{-30}$ \textsuperscript{$\dagger$} & 
22$^{+40}_{-44}$ & 
3.6$^{+4.2}_{-4.3}$ \\
$M_{\mathrm{p}}$ $({\mathrm{M_J}})$ & 2.01\textsuperscript{*} & 
0.81\textsuperscript{* $\dagger$} & 
0.941\textsuperscript{*} & 
1.197\textsuperscript{*} \\
[0.5ex] \hline
\multicolumn{5}{l}{Derived parameters} \\
\hline \\ [-2.0ex]
$R_{\mathrm{p}}$ $({\mathrm{R_J}})$ & 1.36$\pm$0.12 & 
1.367$^{+0.069}_{-0.060}$ \textsuperscript{$\dagger$} & 
1.341$^{+0.044}_{-0.046}$ & 
1.247$^{+0.050}_{-0.045}$ \\
$a$ (Au) & 0.0274$^{+0.0020}_{-0.0019}$ & 
0.0731$^{+0.0034}_{-0.0031}$ \textsuperscript{$\dagger$} & 
0.02897$^{+0.00093}_{-0.00092}$ & 
0.0367$^{+0.0014}_{-0.0013}$ \\
$A_{\mathrm{g,ecl}}$ & 0.246$^{+0.038}_{-0.029}$ & 
0.46$^{+0.30}_{-0.26}$ \textsuperscript{$\dagger$} & 
0.113$^{+0.049}_{-0.053}$ & 
0.031$^{+0.0099}_{-0.010}$ \\
$T_{\mathrm{eq,max}}$ (K) & 2740$^{+50}_{-60}$ & 
2600$\pm$70 \textsuperscript{$\dagger$} & 
2360$\pm$40 & 
1880$\pm$20 \\
$T_{\mathrm{eq,hom}}$ (K) & 2140$\pm$40 & 
2040$\pm$50 \textsuperscript{$\dagger$} & 
1850$\pm$30 & 
1470$\pm$10 \\
$T_{\mathrm{B,day}}$ (K) & 2890$^{+70}_{-60}$ & 
3000$^{+200}_{-300}$ \textsuperscript{$\dagger$} & 
2400$^{+100}_{-200}$ & 
1910$^{+60}_{-80}$ \\
$T_{\mathrm{B,night}}$ (K) & 2380$^{+150}_{-200}$                 & 
$^{<3200(1\sigma)}_{<3400(2\sigma)}$ \textsuperscript{$\dagger$} & 
$^{<2500(1\sigma)}_{<2600(2\sigma)}$ & 
$^{<1900(1\sigma)}_{<2000(2\sigma)}$ \\
$\rho_p$ (g/cm$^3$) & 1.1$^{+0.6}_{-0.4}$ & 
0.4$^{+0.2}_{-0.1}$ \textsuperscript{$\dagger$} & 
0.52$^{+0.13}_{-0.04}$ & 
0.8$\pm$0.1 \\
$\log g_p$ (cgs) & 3.5$\pm$0.2 & 
3.0$\pm$0.1 \textsuperscript{$\dagger$} & 
3.13$^{+0.08}_{-0.02}$ & 
3.30$\pm$0.06 \\
\hline \\ [-2.0ex]
\end{tabular*}
\begin{tabular}{p{8.5cm}}
\textsuperscript{*} Best-fit model favored fixed values for these parameters. \\
\textsuperscript{$\dagger$} Planet's with evidence for non-planetary phase curve modulations (See Section~\ref{sec:k43_k91}). \\
\textsuperscript{$\S$} Depth has been scaled to account for the partial eclipse of Kepler-76b, $F_{\text{ecl,unscaled}} = 95.19$ ppm (see Section~\ref{sec:ec_ap_os}). \\
\textsuperscript{a} From the NASA Exoplanet Archive \citep{archive2013}.\\
\textsuperscript{b} From the Kepler Input Catalog.\\
\end{tabular}
\begin{tabular}{p{8.5cm}}
\textsuperscript{c} From \citet{Kep76_Faigler2013}.\\
\textsuperscript{d} From \citet{Kep91_LilloBox2014}.\\
\textsuperscript{e} From \citet{Kep91_Barclay2014}.\\
\textsuperscript{f} From \citet{Kep412_Deleuil2014}.\\
\textsuperscript{g} From \citet{TrES2_Sozzetti2007}.\\
\textsuperscript{h} From \citet{TrES2_ODonovan2006}.\\
\end{tabular}
\label{tab:res3}
\end{table*}
\begin{table*}[t!]
\centering
\caption{Stellar and Planetary Parameters}
\setlength{\tabcolsep}{0.1cm}
\begin{tabular*}{\textwidth}{@{\extracolsep{\fill}}lcccc}
\hline
\hline
Parameter & HAT-P-7b & 
 & 
KOI-13b & 
 \\
& & Model 1 & Model 2 & Model 3 \\
\hline \\ [-2.0ex]
KOI & 2.01 & 
 & 
13.01 & 
 \\
KIC & 10666592 & 
 & 
9941662 & 
 \\
Period (days)\textsuperscript{a} & 2.2047354$\pm$0.0000001 & 
 & 
1.763588$\pm$0.000001 & 
 \\
$T_{\star}$ (K) & 6350$\pm$80\textsuperscript{c} & 
7650$\pm$250\textsuperscript{d} & 
9107$^{+257}_{-425}$\textsuperscript{e} & 
8511$\pm$1\textsuperscript{f} \\
$\log g$ (cgs) & 4.07$^{+0.04}_{-0.08}$\textsuperscript{c} & 
4.2$\pm$0.5\textsuperscript{d} & 
3.867$^{+0.235}_{-0.148}$\textsuperscript{e} & 
3.9$\pm$0.1\textsuperscript{f} \\
$\mathrm{[Fe/H]}$ & 0.26$\pm$0.08\textsuperscript{c} & 
0.2$\pm$0.2\textsuperscript{d} & 
0.070$^{+0.140}_{-0.650}$\textsuperscript{e} & 
0.2$\pm$0.1\textsuperscript{f} \\
$R_{\star}/R_{\sun}$ & 1.84$^{+0.23}_{-0.11}$\textsuperscript{c} & 
1.74$\pm$0.04\textsuperscript{d} & 
3.031$^{+1.198}_{-0.944}$\textsuperscript{e} & 
2.55$\pm$0.1\textsuperscript{f} \\
$M_{\star}/M_{\sun}$ & 1.47$^{+0.08}_{-0.05}$\textsuperscript{c} & 
1.72$\pm$0.10\textsuperscript{d} & 
2.47$^{+0.45}_{-0.72}$\textsuperscript{e} & 
2.05$\pm$0.1\textsuperscript{f} \\
$K$ (m s$^{-1}$) & 213.5$\pm$1.9\textsuperscript{c} & 
 & 
--- & 
 \\
$M_{\mathrm{p}}$ from RV $({\mathrm{M_J}})$ & 1.781$^{+0.081}_{-0.056}$ & 
 & 
--- & 
 \\
Avg 3rd light (\%)\textsuperscript{b} & 0.184 & 
 & 
0.142 & 
 \\
[0.5ex] \hline
\multicolumn{5}{l}{Transit fit} \\
\hline \\ [-2.0ex]
$T_0$ (BJD-2454833) & 121.358572$\pm$0.000004 & 
 & 
120.56596$^{+0.00002}_{-0.00003}$ & 
 \\
$b$ &0.4960$^{+0.0011}_{-0.0013}$ & 
0.2536$^{+0.0038}_{-0.0035}$ & 
0.2537$^{+0.0037}_{-0.0034}$ & 
0.2535$^{+0.0033}_{-0.0034}$ \\
$R_{\mathrm{p}}/R_{\star}$ & 0.077524$^{+0.000017}_{-0.000022}$ & 
0.087373$^{+0.000023}_{-0.000024}$ & 
0.087373$^{+0.000025}_{-0.000022}$ & 
0.087371$^{+0.000022}_{-0.000023}$ \\
$a/R_{\star}$ & 4.1545$^{+0.0029}_{-0.0025}$ & 
4.5007$^{+0.0039}_{-0.0040}$ & 
4.5008$^{+0.0039}_{-0.0042}$ & 
4.4987$^{+0.0046}_{-0.0054}$ \\
$i$ (degrees) & 83.143$^{+0.023}_{-0.020}$ & 
86.770$^{+0.048}_{-0.052}$ & 
86.769$^{+0.046}_{-0.050}$ & 
86.770$^{+0.047}_{-0.046}$ \\
$u_1$ & 0.3497$^{+0.0026}_{-0.0035}$ & 
0.3183$\pm$0.0019 & 
0.3183$^{+0.0020}_{-0.0021}$ & 
0.3183$\pm$0.0019 \\
$u_2$ & 0.1741$^{+0.0057}_{-0.0044}$ & 
0.2024$\pm$0.0039 & 
0.2024$\pm$0.0042 & 
0.2024$^{+0.0039}_{-0.0037}$ \\
[0.5ex] \hline
\multicolumn{5}{l}{Phase curve fit} \\
\hline \\ [-2.0ex]
$F_{\mathrm{ecl}}$ (ppm) & 71.2$^{+1.9}_{-2.2}$ & 
172.0$^{+1.7}_{-1.6}$ & 
170.8$^{+1.6}_{-1.5}$ & 
170.8$^{+1.6}_{-1.5}$ \\
$A_{\mathrm{p}}$ (ppm) & 73.3$\pm$4.0 & 
151.9$^{+3.3}_{-3.2}$ & 
150.4$^{+3.1}_{-3.6}$ & 
149.9$^{+3.3}_{-3.2}$ \\
$\theta$ (in phase)& 0.01935$^{+0.00080}_{-0.00082}$ & 
0\textsuperscript{*} & 
0.00178$\pm$0.00052 & 
0.00280$\pm$0.00053 \\
$F_{\mathrm{n}}$ (ppm) & -1.1$^{+5.8}_{-6.2}$ & 
20.2$^{+4.8}_{-4.9}$ & 
20.6$^{+5.2}_{-4.6}$ & 
21.2$\pm$4.8 \\
$M_{\mathrm{p}}$ $({\mathrm{M_J}})$ & 1.63$\pm$0.13 & 
9.28$\pm$0.16 & 
12.83$^{+0.22}_{-0.23}$ & 
9.963$\pm$0.17 \\
$A_{\mathrm{3}}$ (ppm) & -1.93$\pm$0.23 & 
-7.03$\pm$0.27 & 
-7.14$\pm$0.26 & 
-7.33$\pm$0.27 \\
$\theta_{\mathrm{3}}$ (in phase)& 0.0163$^{+0.0054}_{-0.0052}$ & 
-0.0840$\pm$0.0024 & 
-0.09314$^{+0.00053}_{-0.00141}$ & 
-0.09880$^{+0.0025}_{-0.0026}$ \\
$e$ &0\textsuperscript{*} & 
0.00064$^{+0.00012}_{-0.00016}$ & 
0.00064$^{+0.00012}_{-0.000098}$ & 
0.00074$^{+0.00110}_{-0.00016}$ \\
$\omega$ (degrees) &0\textsuperscript{*} & 
5$^{+8}_{-10}$ & 
-1$^{+13}_{-7}$ & 
9$^{+26}_{-7}$ \\
[0.5ex] \hline
\multicolumn{5}{l}{Derived parameters} \\
\hline \\ [-2.0ex]
$R_{\mathrm{p}}$ $({\mathrm{R_J}})$ & 1.419$^{+0.178}_{-0.085}$ & 
1.512$\pm$0.035 & 
2.63$^{+1.04}_{-0.82}$ & 
2.216$\pm$0.087 \\
$a$ (Au) & 0.0355$^{+0.0045}_{-0.0021}$ & 
0.03641$\pm$0.00087 & 
0.063$^{+0.025}_{-0.020}$ & 
0.0533$^{+0.0021}_{-0.0022}$ \\
$A_{\mathrm{g,ecl}}$ & 0.2044$^{+0.0058}_{-0.0067}$ & 
0.4565$^{+0.0054}_{-0.0052}$ & 
0.4532$^{+0.0054}_{-0.0051}$ & 
0.4529$^{+0.0055}_{-0.0052}$ \\
$T_{\mathrm{eq,max}}$ (K) & 2820$\pm$40 & 
3300$\pm$100 & 
3900$^{+100}_{-200}$ & 
3626$^{+3}_{-2}$ \\
$T_{\mathrm{eq,hom}}$ (K) & 2200$\pm$30 & 
2550$\pm$80 & 
3040$^{+90}_{-140}$ & 
2837$\pm$2 \\
$T_{\mathrm{B,day}}$ (K) & 2860$\pm$30 & 
3490$^{+60}_{-70}$ & 
3830$^{+60}_{-90}$ & 
3715$\pm$7 \\
$T_{\mathrm{B,night}}$ (K) & $^{<2100(1\sigma)}_{<2300(2\sigma)}$ & 
2590$^{+110}_{-120}$                 & 
2790$^{+120}_{-140}$                 & 
2734$^{+76}_{-89}$                   \\
$\rho_p$ (g/cm$^3$) & 0.8$^{+0.2}_{-0.3}$ & 
3.6$\pm$0.3 & 
0.9$^{+2.0}_{-0.6}$ & 
1.2$\pm$0.2 \\
$\log g_p$ (cgs) & 3.32$^{+0.09}_{-0.14}$ & 
4.02$\pm$0.03 & 
3.7$\pm$0.3 & 
3.72$\pm$0.04 \\
\hline \\ [-2.0ex]
\end{tabular*}
\begin{tabular}{p{8.5cm}}
\textsuperscript{*} Best-fit model favored fixed values for these parameters. \\
\textsuperscript{a} From the NASA Exoplanet Archive \citep{archive2013}.\\
\textsuperscript{b} From the Kepler Input Catalog.\\
\end{tabular}
\begin{tabular}{p{8.5cm}}
\textsuperscript{c} From \citet{HATP7_Pal2008}.\\
\textsuperscript{d} From \citet{KOI13_Shporer2014}.\\
\textsuperscript{e} From \citet{StellarParamsRevised_Huber2013}.\\
\textsuperscript{f} From \citet{KOI13_Szabo2011}.\\
\end{tabular}
\begin{tabular}{p{17cm}}
Note: For KOI-13b values were reported for three sets of stellar parameters from the literature, where our chosen values from~\citet{KOI13_Shporer2014}. A stellar mass uncertainty of $\pm$0.1M$_{\odot}$ and a stellar radius uncertainity of $\pm$0.1R$_{\odot}$ was assumed when not given in the literature.
\end{tabular}
\label{tab:res4}
\end{table*}
\indent Our model is a combination of the four components: (i) $F_{\text{transit}}$, a~\citet{Mandel2002} transit model for a quadratically limb-darkened source; (ii) $F_{\text{p}}$, the planetary brightness; (iii) $F_{\text{ecl}}$, the secondary eclipse; (iv) $F_{\text{m}}$, a combination of Doppler boosting and ellipsoidal variations due to stellar variations induced by the planet's gravity; (v) $F_{3\phi}$, a cosine third harmonic of the planet's period. Each of these components is phase ($\phi$) dependent with $\phi$ running from 0 to 1 and mid-transit occurring at $\phi$=0. The relative flux of the planet-star system, as a function of phase, is then
\begin{eqnarray}
F_{\text{transit}}(\phi) \cdot F_{\text{m}}(\phi) + F_{\text{ecl}}(\phi) + F_{\text{p}}(\phi+\theta) +  F_{3}(\phi-\theta_{3}) \nonumber \\
\end{eqnarray}
where $\theta$ is an offset of the peak planetary brightness from $\phi=0.5$ and $\theta_{3}$ is the cosine third harmonic phase offset. The transit model includes the impact parameter of the transit ($b$), the ratio of the semi-major axis of the planet's orbit to the stellar radius ($a/R_{\star}$), the planet to star radius ratio ($R_{\text{p}}/R_{\star}$) and a linear combination of limb-darkening coefficients ($2u_1+u_2$, $u_1-2u_2$). While the phase curve model fits for the planet's mass ($M_{\text{p}}$), secondary eclipse depth ($F_{\text{ecl}}$) and planetary brightness amplitude and offset ($A_{\text{p}}$ and $\theta$). Our orbital period ($P$) was taken from the NASA Exoplanet Archive \citep{archive2013} and our time of mid-transit ($T_0$) was determined from a simple fit to speed up our Markov Chain Monte Carlo (MCMC) analysis. \\
%
\subsection{Eccentric Orbital Parameters}
\label{sec:ecc}
\indent Our model fits for $\sqrt{e}\cos\omega$ and $\sqrt{e}\sin\omega$ ~\citep[e.g.][]{Triaud2011}, where $e$ is the orbital eccentricity and $\omega$ is the argument of periapsis. \\
\indent For eccentric orbits, unlike for circular orbits, the true anomaly ($\nu$) does not change linearly in time and the star-planet separation ($d$) is time dependent. To adjust our phase curve model to account for this, the true anomaly and the separation were calculated in a manner similar to \citet{Bonavita2012,Bonavita2013}, derived using the ephemeris formulae of \citet[p. 115]{Heintz1978}, as
\begin{eqnarray}
E_0 &=& M + e \sin M + \frac{e^2}{2} \sin(2M) \\
E_{n+1} &=& E_n + \frac{ M - E_n + e \sin E_n }{1 - e \cos E_n } \\
\nu &=& \frac{1}{\pi}\arctan \left\{ \sqrt{ \frac{1+e}{1-e} } \tan \left( \frac{E}{2} \right ) \right\} \\
d &=& a \frac{1-e^2}{1+e\cos \nu}
\end{eqnarray}
where $M$ is mean anomaly, $E$ is the eccentric anomaly and $a$, the semi-major axis, can be calculated from $d$ and the planet-star separation during transit. \\
\indent The projected star-planet separation ($z_0$) can then be calculated by
\begin{eqnarray}
z_0 = \frac{d}{R_{\star}} \sqrt{ 1 - \sin^2(\nu+\omega) \sin^2 i }
\end{eqnarray}
where $i$ is the orbital inclination. \\
\indent Note that the mean anomaly is defined from periastron passage as
\begin{eqnarray}
M = 2\pi \frac{t - T_{\text{peri}}}{P}
\end{eqnarray}
where $T_{\text{peri}}$ is the time of periastron passage. This is important because the orbital phase used in our models is defined from mid-transit as
\begin{eqnarray}
\phi = 2\pi \frac{t - T_{\text{mid}}}{P}
\end{eqnarray}
where $T_{\text{mid}}$ is the time of mid-transit. As a result, there is a phase offset between $M$ and $\phi$. This phase offset is taken into account by requiring that $\nu+\omega$ at mid-transit be 90$^{\circ}$. Subsequently all parameters computed from $M$ ($\nu+\omega$, $d$ and $z_0$) must be similarly offset. \\
\indent In Sections~\ref{sec:ec_ap_os}-~\ref{sec:a3} we describe the circular phase curve models. The eccentric models can be obtained by substituting $\phi$ for $\nu+\omega$ and $a$ for $d$.
%
\subsection{Eclipse Depth, Planetary Brightness Amplitude and Offset}
\label{sec:ec_ap_os}
\indent We model the secondary eclipse using the formalism from \citet{Mandel2002} for a uniform source and normalize the model by $R_{\text{p}}^2/R_{\star}^2$. Here an eclipse model amplitude of one corresponds to a fully occulted planet while a grazing eclipse results in an amplitude less than one. The planetary brightness is modeled as a Lambert sphere described by
\begin{eqnarray}
F_{\text{p}} = A_{\text{p}} \frac{\sin z + (\pi-z)\cos z}{\pi}
\end{eqnarray}
where $A_{\text{p}}$ is the peak brightness amplitude, $\theta$ is a phase offset in peak brightness and $z$ is defined by
\begin{eqnarray}
\cos (z) = - \sin i \cos (2\pi[\phi+\theta])
\end{eqnarray}
Assuming only reflected light from the planet, the geometric albedo in the {\it Kepler} bandpass ($A_{\text{g}}$) can then be calculated by
\begin{eqnarray}
F_{\text{ecl}} = A_{\text{g}} \left(\frac{R_p}{a}\right)^2
\end{eqnarray}
%
\subsection{Planet Mass}
\label{sec:mp}
\indent Planet mass ($M_{\text{p}}$) was measured through the analysis of Doppler boosting and ellipsoidal variations. These signals, in-phase with planet's period, are stellar brightness fluctuations induced by the planet's gravity. \\
\indent Doppler boosting is a combination of a bolometric and a bandpass-dependent effect. The bolometric effect is the result of non-relativistic Doppler boosting of the stellar light in the direction of the star's radial velocity (RV). The observed periodic brightness change is proportional to the star's RV, which is a function of the planet's distance and mass \citep{TrES2_Barclay2012}. While the bandpass-dependent effect is a periodic red/blue shift of the star's spectrum, which results in a periodic change of the measured brightness as parts of the star's spectrum move in and out of the observed bandpass \citep{TrES2_Barclay2012}. \\
\indent Ellipsoidal variations, on the other hand, are periodic changes in observed stellar flux caused by fluctuations of the star's visible surface area as the stellar tide, created by the planet, rotates in and out of view of the observer \citep{HATP7_Mislis2012}. If there is no tidal lag, the star's visible surface area and ellipsoidal variations are at maximum when the direction of the tidal bulge is perpendicular to the observer's line of sight (i.e. $\phi$ of 0.25 and 0.75) and at minimum during the transit and the secondary eclipse.\\
\indent The Doppler (first term in Eq.~\ref{eqn:dop_ell_1}) and ellipsoidal (second term in Eq.~\ref{eqn:dop_ell_1}) contributions can be described by
\begin{eqnarray}\label{eqn:dop_ell_1}
F_{\text{m}} = M_{\text{p}} \cdot \bigg\{ &
\left(\frac{2\pi G}{P}\right)^{1/3} \frac{\alpha_{\text{d}} \sin i}{c \cdot M_{\star}^{2/3}}
\left( \frac{1 + e \cos \omega}{\sqrt{1-e^2}} \right) f_{\text{d}}
- \frac{ \alpha_2 \sin^2 i }{M_{\star}} f_e & \bigg\} \nonumber \\
&&
\end{eqnarray}
where  $M_{\star}$ is the host star mass, $G$ is the universal gravitational constant, $c$ is the speed of light and $\alpha_{\text{d}}$ is the photon-weighted bandpass-integrated beaming factor. Here we have assumed $M_{\mathrm{p}}<<M_{\star}$ and similar to E13 and~\citet{TrES2_Barclay2012} we calculate $\alpha_{\mathrm{d}}$ in the manner described by~\citet{Bloemen2011} and~\citet{Loeb2003}. \\
\indent In Eq.~\ref{eqn:dop_ell_1}, $f_d$ and $f_e$ are the phase dependent modulations of the Doppler boosting and ellipsoidal signals, respectively. They can be described by
\begin{eqnarray}
f_d &=& \sin ( 2 \pi \phi ) \\
f_e &=& \left(\frac{a}{R_{\star}}\right)^{-3} \cos(2 \cdot 2 \pi \phi) \nonumber \\
&& + \left( \frac{a}{R_{\star}} \right)^{-4} f_1\cos(2 \pi \phi) \nonumber \\
&& + \left( \frac{a}{R_{\star}} \right)^{-4} f_2\cos(3 \cdot 2 \pi \phi) 
\end{eqnarray}
where $f_1$ and $f_2$ are constants used to determine the amplitude of the higher-order ellipsoidal variations ~\citep{Morris1985} 
\begin{eqnarray}
f_1 &=& 3  \alpha_1 \frac{5 \sin^2 i - 4}{\sin i} \\
f_2 &=& 5  \alpha_1 \sin i \label{eqn:dop_ell_2}
\end{eqnarray}
In Eqs.~\ref{eqn:dop_ell_1}-\ref{eqn:dop_ell_2}, $\alpha_1$ and $\alpha_2$ are functions of the linear limb darkening and gravity darkening parameters ~\citep{Claret2011}, which we calculate in the same manner as E13 and~\citet{TrES2_Barclay2012}. The values of $f_1$, $f_2$, $\alpha_{\text{d}}$, $\alpha_1$ and $\alpha_2$ can be found in Table~\ref{tab:constants}. \\
\indent The phase curve is also fit with planet mass fixed to the radial velocity (RV) derived value calculated via
\begin{eqnarray}
K = \left( \frac{2 \pi G}{P} \right)^{1/3} \frac{M_{\text{p}} \sin i}{M_{\star}^{2/3} \sqrt{1-e^2}}
\end{eqnarray}
where $K$ is the published RV semi-amplitude (see Tables~\ref{tab:res1}-\ref{tab:res4}). 
%
\subsection{Cosine Third Harmonic}
\label{sec:a3}
\indent In E13, we discovered of a phase-shifted 6.7$\pm$0.3 ppm third harmonic in the residual of KOI-13b's phase curve. To investigate if this signal is present for other planets our model also includes a cosine third harmonic, which we allow to vary in amplitude ($A_{3}$) and phase ($\theta_{3}$) and is described by
\begin{eqnarray}
F_{3} = A_3 \cos(3 \cdot 2 \pi (\phi-\theta_{3}) ) 
\end{eqnarray}
%
\section{Analysis}
\label{sec:analysis}
\indent The light curves were phase-folded and binned such that 400 points spanned the orbit (i.e. a binsize of 0.0025 in phase), but with an increased sampling rate across the transit, such that it is spanned by 200 points. However, for Kepler-76b and Kepler-91b, which only have LC measurements, we fit to the unbinned light curve taking the sampling into account. This is important as sharp features in the light curve are smoothed significantly by the long exposures. \\
\indent KOI-13b's transit curve is asymmetric as a result of the planet's motion across a stellar surface temperature gradient during transit ~\citep{KOI13_Szabo2011}. To obtain a symmetric curve we combined the first and second half of the transit light curve, then interpolated onto our binned time-series to obtain a symmetric transit light curve. \\
\indent The transit and phase curve were simultaneous fit using a MCMC analysis. Five sequences of 400,000 steps were generated and the first 150,000 points were trimmed to avoid any contamination from the initial conditions. The chains were then combined after checking that they were well mixed ~\citep{Gelman1992}. \\
\indent For all planets we fit several permutations of our model. For each target we fit a total of 16 models, where either all parameters are free or one or more of the following parameters are fixed: i) planet mass fixed to RV derived value, ii) eccentricity fixed to zero, iii) brightness offset fixed to zero, iv) amplitude of the cosine third harmonic fixed to zero. Our best-fit model was chosen using the Bayesian Information Criterion (BIC, \citealt{BIC2007}) and the BIC values of our 8 most favored models can be found in Table~\ref{tab:bic}. HAT-P-7b and KOI-13b were the only planet's where a non-zero third harmonic amplitude was favored. In Table~\ref{tab:bic} each of the BIC values for HAT-P-7b and KOI-13b are for models which include a non-zero third harmonic, while for each of the other planets the reported BICs are for models without a third harmonic. \\
%
\section{Results}
\label{sec:results}
\indent Our best-fit model was chosen as the fit with the lowest BIC (see Table~\ref{tab:bic}). The results of our best-fits can be found in Tables~\ref{tab:res1}-\ref{tab:res4} and in Figs.~\ref{fig:res1}-\ref{fig:res5}. The uncertainties on phase curves parameters were increased to account for the influence of time-correlated noise (see Section~\ref{sec:corr_noise}). \\
\indent For Kepler-91b, we find low significance for the inclusion of a varying planetary brightness offset. Of our targets, KOI-13b, with a small eccentricity of 0.0006$\pm$0.0001, is the only planet where the BIC favored an eccentric orbit. To investigate the affect of stellar parameter uncertainty has on our derived parameters we repeat our fits of KOI-13b using three sets of stellar parameters ~\citep{KOI13_Shporer2014,StellarParamsRevised_Huber2013,KOI13_Szabo2011}. We do this only for KOI-13b because the reported stellar parameters differ greatly from one another and more importantly because the choice of stellar parameters strongly influences our derived equilibrium temperature (see Section~\ref{sec:k13} for discussion). \\
%
\subsection{Stellar Variability}
\label{sec:variability}
\indent For each system, we compute and analyze the light curve autocorrelation as described by~\citet{McQuillan2013}. For each target, the autocorrelation indicates the presence of correlated variability and/or evidence for systematic effects such as linear trends and jumps.  For seven planets (Kepler-7b, -8b, -41b, -43b, -76b, -412b and TrES-2b), the peak and shape of the autocorrelation shows evidence for sinusoidal light curve variability with an evolving phase and amplitude, most likely by stellar variability (e.g. starspots rotating in and out of view once per stellar rotation period). For TrES-2b and Kepler-412b we find a stellar rotation period at 9 and 10 times the planet’s period, respectively. The latter of which is in agreement with the stellar rotation period derived from~\citet{Kep412_Deleuil2014}. For the others we find that the rotation period is significantly offset from multiples of the orbital period. \\
\indent For Kepler-91b, the autocorrelation exhibits stochastic variability, with a dominant time-scale equal to the planet's period, and does not exhibit repeatability or resemble the signal expected from rotation. \citet{McQuillan2013} find these signal characteristics typical of red giant stars, like Kepler-91b's host star~\citet{Kep91_LilloBox2014}, as they are known to show significant correlated noise on timescales of hours to weeks due to stellar granulation ~\citep{Mathur2012,Kep91_Barclay2014}. The affect of Kepler-91b's orbital period phased stellar variability is discussed in Section~\ref{sec:k43_k91}. For the six other systems we did not find the stellar rotation period as the autocorrelation did not exhibit a clear dominant signal. \\
%
\subsection{Correlated Noise}
\label{sec:corr_noise}
\indent Since the phase and ampitude of this stellar variability evolves with time it is unclear how it will affect our fitted phase curve parameters. To assess this we performed a bootstrap, using a simple regression analysis, to measure $F_{\text{ecl}}$, $A_{\text{p}}$ and $M_{\text{p}}$ for subsets of the data. To simplify the model, the peak offset and third harmonic were fixed to the best-fit parameters from our MCMC analysis. For each fit we randomly selected half of the available individual orbits, allowing for orbits to be redrawn. However, to prevent skewing from incomplete phase curves we only selected orbits where at least 80\% of the data was available. For each planet we fit 5000 subsets for the amplitudes of $F_{\text{ecl}}$, $A_{\text{p}}$ and $M_{\text{p}}$. We then compared the peak and standard deviation of the bootstrap parameter distribution to the best-fit and 1-$\sigma$ values from our MCMC analysis. \\
\indent If only random noise is present we would expect bootstrap uncertainties $\sqrt2$ times larger than our MCMC uncertainties. However, we find that, although the peak value of the bootstrap distribution is in agreement with our best-fit MCMC values, the uncertainties derived from the bootstrap are consistently larger than thoses from the MCMC. Since the MCMC derived errors fail to reflect the time-variable noise characteristics of the data, we chose to adopt the larger uncertainty values derived from our bootstrap analysis. \\
\indent With the exception of Kepler-43b, the increase in uncertainty does not greatly alter the significance level of our fitted parameters. While for Kepler-43b the significance of our measurements of $M_p$, $A_p$ and $F_{ecl}$ is reduced to 1, 1 and 0 $\sigma$, respectively. The larger uncertainty in the bootstrap parameters reflect variations in the phase curve amplitude and/or shape between different subsection of the data. The effect of this increased phase curve variability on is discussed in Section~\ref{sec:k43_k91}. \\
%
\subsection{Non-Planetary Phase Curve Signals: Kepler-43b and Kepler-91b}
\label{sec:k43_k91}
\indent Previous studies of Kepler-91b's (KOI-2133) transit and phase curve have yielded differing orbital and planet parameters. The {\it Kepler} transit fit, performed by E13 yielded orbital parameters that are used, in conjunction with the eclipse depth, to conclude that, due to its large albedo ($>$1), Kepler-91b is a false positive. \citet{Kep91_Sliski2014} also use {\it Kepler} data to find a similar set of orbital parameters, but use an asteroseismic analysis to conclude that the stellar density is not consistent with their measured orbital parameters. \citet{Kep91_LilloBox2014} find a drastically different set of orbital parameters and confirm the planetary nature of Kepler-91b using high-resolution images and through the reanalysis of the {\it Kepler} data, including transit, phase curve and asteroseismic analysis.  Recently~\citet{Kep91_Barclay2014} simultaneously model RV and {\it Kepler} data using a Gaussian process to find a set of orbital parameters similar to those found by~\citet{Kep91_LilloBox2014}. \citet{Kep91_Barclay2014} hypothesize that the false positive conclusions found by previous studies are due to a temporally correlated stellar noise component, caused by stellar granulation on Kepler-91b's red giant host star. \\
\indent In this analysis we find orbital and planet parameters consistent with those found by~\citet{Kep91_Barclay2014} and with the planetary nature of Kepler-91b (see Table~\ref{tab:res3}). Possible causes for the differing results between our studies is the inclusion of an additional three quarters of data, leading to an increase in accuracy as variations are averaged over longer time-scales. In our current analysis we also simultaneously fit for the phase-curve and the transit, and fix the planetary mass to that derived from RV measurements by~\citet{Kep91_Barclay2014}. \\
\indent Similar to previous studies, our Kepler-91b phase curve exhibits dips and jumps with amplitudes similar to those of our measured eclipse depth. We find that the amplitudes of these variations are actually larger than the error estimates derived from our bootstrap, as correlated noise on time-scales of the Kepler-91b's planetary period is coherent over a significant number of transits. \\
\indent Kepler-43b's autocorrelation reveals light curve variations with a period of 12.8 days, which is also visible by eye in the unfolded time-series (see Fig.~\ref{fig:A4}). These variations, which exhibit a double-dip feature, are indicative of a bimodel brightness distribution, due to concentrations of active regions on opposite hemispheres~\citet{McQuillan2013}. \\
\indent Of our targets, Kepler-43b was the only planet where our bootstrap analysis considerably reduced the significance of our detection (see Section~\ref{sec:corr_noise}), which we attribute to large star spot induced variations. Further evidence for Kepler-43b's phase curve contamination can be found in residuals, where correlated noise, with an amplitude similar to the eclipse depth, can be seen. Furthermore, we find that the peak planetary brightness is several times larger than the eclipse depth. This is only possible if the planet does not fully pass behind the star, which would require a highly inclined and eccentric orbit both of which are not found by our phase curve analysis or the RV analysis from~\citet{Kep43_Bonomo2012}. \\
\indent Since Kepler-43b and Kepler-91b show compelling evidence for a non-planetary phase curve signal, we chose to exclude them from our discussion in Section~\ref{sec:source}.
%
\subsection{Impact of Different Stellar Parameters: KOI-13b/Kepler-13b}
\label{sec:k13}
\indent For KOI-13b, also referred to in the literature as Kepler-13b, we ran our fits using three sets of stellar parameters ~\citep{KOI13_Shporer2014,StellarParamsRevised_Huber2013,KOI13_Szabo2011}. We do this only for KOI-13b because the reported stellar parameters differ greatly from one another and more importantly because the choice of stellar parameters strongly influences our derived equilibrium temperature. This is of particular importance for KOI-13b as its blackbody peaks very close to the edge of the {\it Kepler} bandpass. Therefore a small increase in equilibrium temperature can lead to a significant increase in the contribution from thermal emission in the {\it Kepler} bandpass, while a small decrease in temperature will require a larger amount of reflected light to explain the observed secondary eclipse depth. Further discussion of KOI-13b's equilibrium temperature and albedo can be found in Section~\ref{sec:albedo}. \\
\indent The determination of KOI-13b's host star temperature is complicated by the fact that it has a stellar companion at 1.12" with almost the same apparent brightness. For the rest of our discussion we adopt the results obtained using the most recent spectroscopically derived values from~\citet{KOI13_Shporer2014} as our best-fit values, but we note that the results for this planet strongly depend on the assumed stellar parameters. \\
%
\subsection{Comparison to Previous Studies}
\label{sec:comparison}
\defcitealias{Angerhausen2014}{Ang14}
\defcitealias{TrES2_Barclay2012}{Bar12}
\defcitealias{Kep10_Batalha2011}{Bat11}
\defcitealias{HATP7_Borucki2009}{Bor09}
\defcitealias{Eclipses_Coughlin2012}{Cou12}
\defcitealias{Kep412_Deleuil2014}{Del14}
\defcitealias{Kep7_Demory2011}{Dem11}
\defcitealias{Kep7_Demory2013}{Dem13}
\defcitealias{Kep10_Demory2014}{Dem14}
\defcitealias{Kep5_Kep6_Desert2011}{D{\'e}s11}
\defcitealias{Esteves2013}{Est13}
\defcitealias{Kep76_Faigler2013}{Fai13}
\defcitealias{Kep10_FogtmannSchulz2014}{Fog14}
\defcitealias{Kep12_Fortney2011}{For11}
\defcitealias{HATP7_Jackson2012}{Jac12}
\defcitealias{K5_K6_K7_K8_KippingBakos2011}{Kip\&Bak11}
\defcitealias{TrES2_KippingSpiegel2011}{Kip\&Spi11}
\defcitealias{Kep91_LilloBox2014}{Lil14}
\defcitealias{KOI13_Mazeh2012}{Maz12}
\defcitealias{HATP7_Mislis2012}{Mis\&Hel12}
\defcitealias{KOI13_Mislis2012}{Mis\&Hod12}
\defcitealias{HATP7_Morris2013}{Mor13}
\defcitealias{Kep41_Quintana2013}{Qui13}
\defcitealias{Kep10_Rouan2011}{Rou11}
\defcitealias{Kep41_Santerne2011}{San11}
\defcitealias{Kep10_SanchisOjeda2014}{San14}
\defcitealias{KOI13_Shporer2011}{Shp11}
\defcitealias{KOI13_Shporer2014}{Shp14}
\defcitealias{KOI13_Szabo2011}{Sza11}
\defcitealias{HATP7_VanEylen2012}{Van12}
\defcitealias{HATP7_Welsh2010}{Wel10}
\defcitealias{Kep91_Barclay2014}{Bar14}
\begin{table*}[ht]
\scriptsize
\centering
\caption{Literature Phase Curve Parameters}
\setlength{\tabcolsep}{0.08cm}
\setlength{\tabcolsep}{0.05cm}
\begin{tabular*}{8.5cm}{@{\extracolsep{\fill}}lcccc}
\hline
\hline
Reference & $F_{\mathrm{ecl}}$ & $A_{\mathrm{p}}$ & $A_{\mathrm{e}}$ & $A_{\mathrm{d}}$ \\
\hline \\ [-2.0ex]
\multicolumn{5}{c}{KOI-13b} \\
This work & 172.0$^{+1.7}_{-1.6}$ & 151.9$^{+3.3}_{-3.2}$ & 72.9$^{+6.0}_{-5.4}$ & 10.02$^{+0.58}_{-0.54}$ \\
\citetalias{Angerhausen2014} & 162$\pm$10\textsuperscript{e} & 151$\pm$10\textsuperscript{e} & 71$\pm$10\textsuperscript{e} & 0 \\ 
\citetalias{KOI13_Shporer2014} & 173.7$\pm$1.8\textsuperscript{e} & 156.4$\pm$2.1\textsuperscript{f} & 60.1$\pm$1.0 & 9.45$\pm$0.49 \\ 
\citetalias{Esteves2013} & 169.6$^{+2.2}_{-2.3}$\textsuperscript{e} & 143.5$^{+1.7}_{-1.8}$\textsuperscript{e} & 72.31$\pm$0.81\textsuperscript{e} & 8.58$\pm$0.31\textsuperscript{e} \\ 
\citetalias{KOI13_Mazeh2012} & 163.8$\pm$3.8 & 142$\pm$3\textsuperscript{f} & 66.8$\pm$1.6 & 8.6$\pm$1.1 \\ 
\citetalias{KOI13_Mislis2012} & ...\textsuperscript{a} & ...\textsuperscript{a} & ...\textsuperscript{c} & ...\textsuperscript{c} \\ 
\citetalias{Eclipses_Coughlin2012} & 124.3$^{+6.9}_{-7.8}$ & ... & ... & ... \\ 
\citetalias{KOI13_Shporer2011} & ... & 152.2$\pm$2.5\textsuperscript{e,f} & 57.9$\pm$1.3\textsuperscript{e} & 10.1$\pm$0.8\textsuperscript{e} \\ 
\citetalias{KOI13_Szabo2011} & 120$\pm$10 & ... & ... & ... \\ 
\hline \\ [-2.0ex]
\multicolumn{5}{c}{Kepler-7b} \\
This work & 39$\pm$11 & 48$^{+16}_{-17}$ & 1\textsuperscript{$\dagger$} & 0.5\textsuperscript{$\dagger$} \\
\citetalias{Angerhausen2014} & 46.6$\pm$3.9 & 47.8$\pm$5.2 & 0 & 3.9$\pm$4.0 \\ 
\citetalias{Kep7_Demory2013} & 48$\pm$3 & 50$\pm$2 & 0 & 0 \\ 
\citetalias{Eclipses_Coughlin2012} & 53.2$^{+14.1}_{-13.0}$ & 0 & ... & ... \\ 
\citetalias{Kep7_Demory2011} & 44$\pm$5 & 42$\pm$4 & 0 & ... \\ 
\citetalias{K5_K6_K7_K8_KippingBakos2011} & 47$\pm$14 & 30$\pm$17\textsuperscript{g} & ... & ... \\ 
\hline \\ [-2.0ex]
\multicolumn{5}{c}{Kepler-5b} \\
This work & 18.6$^{+5.1}_{-5.3}$ & 19.3$^{+9.5}_{-7.6}$ & 2.9$^{+3.3}_{-3.2}$ & 1.1$^{+1.2}_{-1.3}$ \\
\citetalias{Angerhausen2014} & 19.8$\pm$3.6 & 8.3$\pm$2.6 & 3.1$\pm$1.4 & 0 \\ 
\citetalias{Esteves2013} & 18.8$\pm$3.7 & 16.5$\pm$2.0 & 4.7$^{+1.0}_{-1.1}$ & 0 \\ 
\citetalias{Kep5_Kep6_Desert2011} & 21$\pm$6 & 0 & ... & ... \\ 
\citetalias{K5_K6_K7_K8_KippingBakos2011} & 26$\pm$17 & 0 & ... & ... \\ 
\hline \\ [-2.0ex]
\multicolumn{5}{c}{Kepler-41b} \\
This work & 44$^{+15}_{-16}$ & 69$\pm$25 & 5\textsuperscript{$\dagger$} & 1\textsuperscript{$\dagger$} \\
\citetalias{Angerhausen2014} & 46.2$\pm$8.7 & 46.3$\pm$7.9 & 2.8$\pm$4.3 & 0 \\ 
\citetalias{Kep41_Quintana2013} & 60$\pm$11 & 37.4$\pm$6.3 & 4.5$\pm$3.3 & 0 \\ 
\citetalias{Kep41_Santerne2011} & 64$^{+10}_{-12}$ & 64$^{+10}_{-12}$ & 0 & 0 \\ 
\citetalias{Eclipses_Coughlin2012} & 77$\pm$24 & 0 & ... & ... \\ 
\hline \\ [-2.0ex]
\multicolumn{5}{c}{Kepler-412b} \\
This work & 53$^{+21}_{-24}$ & 31$^{+20}_{-19}$ & 9\textsuperscript{$\dagger$} & 2\textsuperscript{$\dagger$} \\
\citetalias{Angerhausen2014} & 40.2$\pm$9.0 & 17.3$\pm$7.4 & 6.3$\pm$3.9 & 2.7$\pm$2.2 \\ 
\citetalias{Kep412_Deleuil2014} & 47.4$\pm$7.4 & 28.8$\pm$3.2 & 10.2$\pm$2.5 & 1.81$\pm$0.14 \\ 
\citetalias{Eclipses_Coughlin2012} & 69$^{+31}_{-30}$ & 0 & ... & ... \\ 
\hline \\ [-2.0ex]
\multicolumn{5}{c}{Kepler-12b} \\
This work & 20.2$^{+8.3}_{-7.6}$ & 22.9$^{+6.2}_{-6.0}$ & 0.9\textsuperscript{$\dagger$} & 0.6\textsuperscript{$\dagger$} \\
\citetalias{Angerhausen2014} & 18.7$\pm$4.9 & 16.7$\pm$2.6 & 0 & 0 \\ 
\citetalias{Kep12_Fortney2011} & 31$\pm$8 & 0 & ... & ... \\ 
\hline \\ [-2.0ex]
\multicolumn{5}{c}{Kepler-43b} \\
This work & 11$^{+40}_{-37}$ & 71$^{+56}_{-60}$ & 17$^{+24}_{-19}$ & 9$\pm$10 \\
\citetalias{Angerhausen2014} & 17.0$\pm$5.3 & 51.7$\pm$3.3 & 16.9$\pm$2.2 & 0 \\ 
& & & & \\
\hline
\end{tabular*}
\quad
\setlength{\tabcolsep}{0.05cm}
\begin{tabular*}{8.5cm}{@{\extracolsep{\fill}}lcccc}
\hline
\hline
Reference & $F_{\mathrm{ecl}}$ & $A_{\mathrm{p}}$ & $A_{\mathrm{e}}$ & $A_{\mathrm{d}}$ \\
\hline \\ [-2.0ex]
\multicolumn{5}{c}{HAT-P-7b} \\
This work & 71.2$^{+1.9}_{-2.2}$ & 73.3$\pm$4.0 & 17.7$^{+2.1}_{-2.3}$ & 2.22$^{+0.23}_{-0.24}$ \\
\citetalias{Angerhausen2014} & 69.3$\pm$0.5 & 60.8$\pm$0.5 & 16.8$\pm$0.3 & 5.3$\pm$0.1 \\ 
\citetalias{Esteves2013} & 68.31$\pm$0.69 & 65.75$\pm$0.48 & 19.09$\pm$0.25 & 5.80$\pm$0.19 \\ 
\citetalias{HATP7_Morris2013} & 69.1$\pm$3.8 & ... & ... & ... \\ 
\citetalias{HATP7_VanEylen2012} & 71.85$\pm$0.23 & ...\textsuperscript{a} & 18.9$\pm$0.3\textsuperscript{b} & ... \\ 
\citetalias{HATP7_Jackson2012} & 61$\pm$3 & 61 & ...\textsuperscript{c} & ...\textsuperscript{c} \\ 
\citetalias{HATP7_Mislis2012} & ...\textsuperscript{a} & ...\textsuperscript{a} & 15.5\textsuperscript{d} & 3.0 \\ 
\citetalias{Eclipses_Coughlin2012} & 75.0$^{+9.5}_{-8.1}$ & 63.2$^{+13.6}_{-12.6}$ & ... & ... \\ 
\citetalias{HATP7_Welsh2010} & 85.8 & 63.7 & 18.65\textsuperscript{d} & ... \\ 
\citetalias{HATP7_Borucki2009} & 130$\pm$11 & 122 & ... & ... \\ 
\hline \\ [-2.0ex]
\multicolumn{5}{c}{TrES-2b} \\
This work & 7.7$^{+2.4}_{-2.6}$ & 4.1$^{+1.7}_{-1.8}$ & 3\textsuperscript{$\dagger$} & 2\textsuperscript{$\dagger$} \\
\citetalias{Angerhausen2014} & 10.9$\pm$2.2 & 3.0$\pm$0.8 & 2.9$\pm$0.5 & 2.0$\pm$0.2 \\ 
\citetalias{Esteves2013} & 7.5$\pm$1.7 & 4.77$^{+0.65}_{-0.63}$ & 3.67$\pm$0.33 & 2.40$\pm$0.30 \\ 
\citetalias{TrES2_Barclay2012} & 6.5$^{+1.7}_{-1.8}$ & 3.41$^{+0.55}_{-0.82}$ & 2.79$^{+0.44}_{0.62}$ & 3.44$^{+0.35}_{-0.33}$ \\ 
\citetalias{TrES2_KippingSpiegel2011} & 0 & 6.5$\pm$1.9 & 1.50$^{+0.92}_{-0.93}$ & 0.22$^{+0.88}_{-0.87}$ \\ 
\hline \\ [-2.0ex]
\multicolumn{5}{c}{Kepler-10b} \\
This work & 7.5$^{+2.0}_{-2.1}$ & 9.79$^{+3.4}_{-3.2}$ & 0.5\textsuperscript{$\dagger$} & 0.04\textsuperscript{$\dagger$} \\
\citetalias{Kep10_Demory2014} & 7.4$^{+1.1}_{-1.0}$ & 0 & 0 & 0 \\ 
\citetalias{Kep10_FogtmannSchulz2014} & 9.91$\pm$1.01 & 8.13$\pm$0.68 & 0 & 0 \\ 
\citetalias{Kep10_SanchisOjeda2014} & 7.5$\pm$1.4 & ... & ... & ... \\ 
\citetalias{Kep10_Rouan2011} & 5.6$\pm$2.0 & 5.6$\pm$2.0 & 0 & 0 \\ 
\citetalias{Kep10_Batalha2011} & 5.8$\pm$2.5 & 7.6$\pm$2.0 & 0 & 0 \\ 
\hline \\ [-2.0ex]
\multicolumn{5}{c}{Kepler-6b} \\
This work & 11.1$^{+4.8}_{-5.3}$ & 17.2$^{+6.3}_{-5.1}$ & 2\textsuperscript{$\dagger$} & 1\textsuperscript{$\dagger$} \\
\citetalias{Angerhausen2014} & 11.3$\pm$4.2 & 9.5$\pm$2.7 & 0 & 1.0$\pm$0.9 \\ 
\citetalias{Esteves2013} & 8.9$\pm$3.8 & 12.4$\pm$2.0 & 2.7$\pm$1.0 & 0 \\ 
\citetalias{Kep5_Kep6_Desert2011} & 22$\pm$7 & 0 & ... & ... \\ 
\hline \\ [-2.0ex]
\multicolumn{5}{c}{Kepler-8b} \\
This work & 26$\pm$10 & 24$^{+16}_{-15}$ & 2\textsuperscript{$\dagger$} & 0.8\textsuperscript{$\dagger$} \\
\citetalias{Angerhausen2014} & 16.5$\pm$4.4 & 13.8$\pm$3.7 & 3.7$\pm$2.0 & 0 \\ 
\citetalias{Esteves2013} & 26.2$\pm$5.6 & 25.3$^{+2.7}_{-2.6}$ & 2.5$\pm$1.2 & 4.0$\pm$1.4 \\ 
\hline \\ [-2.0ex]
\multicolumn{5}{c}{Kepler-76b} \\
This work & 131.6$^{+8.7}_{-8.0}$ & 106.9$^{+6.2}_{-5.6}$ & 21\textsuperscript{$\dagger$} & 3\textsuperscript{$\dagger$} \\
\citetalias{Angerhausen2014} & 75.6$\pm$5.6 & 101.3$\pm$3.6 & 22.6$\pm$1.9 & 11.4$\pm$1.0 \\ 
\citetalias{Kep76_Faigler2013} & 98.9$\pm$7.1 & 112$\pm$5.0\textsuperscript{f} & 21.5$\pm$1.7 & 15.6$\pm$2.2 \\ 
\hline \\ [-2.0ex]
\multicolumn{5}{c}{Kepler-91b} \\
This work & 35$\pm$18 & 25$^{+18}_{-19}$ & 50\textsuperscript{$\dagger$} & 1\textsuperscript{$\dagger$} \\
\citetalias{Kep91_LilloBox2014} & 0 & 25$\pm$15 & 60.5$^{+16}_{-17}$\textsuperscript{d} & 1.5$^{+0.5}_{-1.0}$\textsuperscript{d} \\ 
\citetalias{Esteves2013} & 38.7$\pm$8.2 & 13.1$^{+5.8}_{-6.0}$ & 45.2$^{+3.1}_{-2.1}$ & 0 \\ 
\citetalias{Kep91_Barclay2014} & 49$\pm$15 & 27.4$^{+21}_{-16}$ & 50.8$^{+4.8}_{-5.0}$ & ...\textsuperscript{h} \\ 
\hline
\end{tabular*}
\begin{tabular}{p{8.5cm}}
Zero values indicate non-detections. \\
$^{\dagger}$ A$_{\text{e}}$ and A$_{\text{d}}$ were fixed in our best-fit model. \\
\textsuperscript{a} A$_{\text{g}}$ reported instead of an amplitude. \\
\textsuperscript{b} Adjusted by 1/$\pi$. \\
\textsuperscript{c} M$_{\text{p}}$ reported instead of an amplitude. \\
\end{tabular}
\quad
\begin{tabular}{p{8.5cm}}
\textsuperscript{d} Converted to a semi-amplitude. \\
\textsuperscript{e} Corrected by a dilution factor of 1.913$\pm$0.019 (see Section \ref{sec:companion}). \\
\textsuperscript{f} Converted to a peak-to-peak amplitude. \\
\textsuperscript{g} A$_{\text{p}}$ derived from reported F$_{\text{ecl}}$ and F$_{\text{n}}$. \\
\textsuperscript{h} A$_{\text{d}}$ fixed to RV derived value. \\
\end{tabular}
\label{tab:comparison}
\end{table*}
\begin{figure}[ht]
\begin{center}
\scalebox{0.4}{\includegraphics{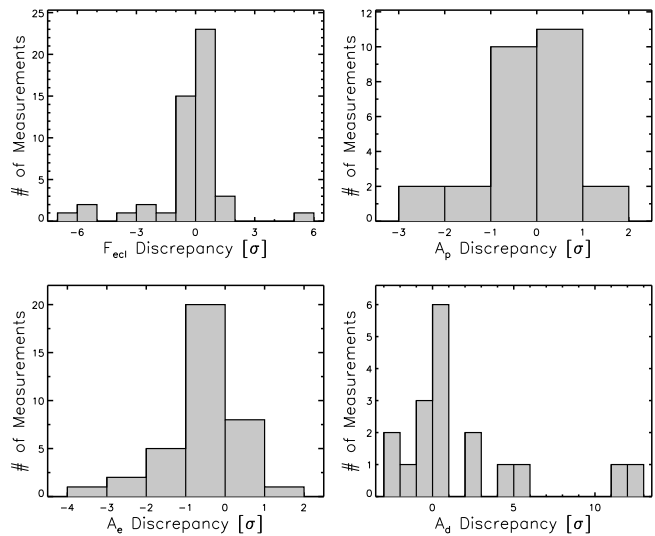}}
\end{center}
\caption{Histogram of the $\sigma$ deviation between published values and our results (see Table~\ref{tab:comparison}) for eclipse depths ($F_{\text{ecl}}$), phase function amplitudes ($A_{\text{p}}$), ellipsoidal variations ($A_{\text{e}}$) and Doppler boosting ($A_{\text{d}}$). Errors for fixed $A_{\text{e}}$ and $A_{\text{d}}$ values were derived using RV uncertainties.}
\label{fig:comparison}
\end{figure}
\begin{figure}[t]
\begin{center}
\scalebox{0.8}{\includegraphics{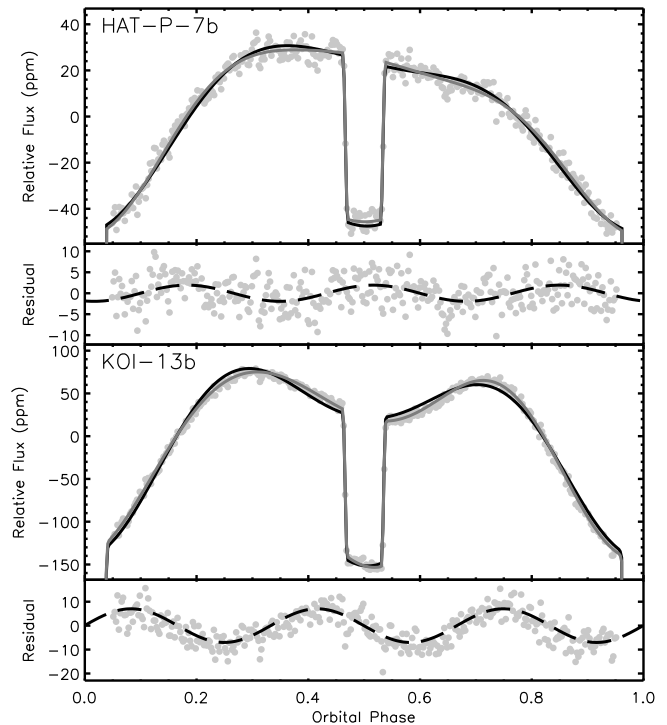}}
\end{center}
\caption{Phase curves of the two planets, HAT-P-7b (top) and KOI-13b (bottom), where we find a non-zero third cosine harmonic. Solid grey line is our best-fit model, which includes the third harmonic and the solid black line is a model without the harmonic. The residuals are for the no harmonic model and the dashed black line maps the third harmonic within the residuals. For KOI-13b the third harmonic is present during secondary eclipse, which indicates that this modulation is not of planetary origin.}
\label{fig:noA3}
\end{figure}
\begin{figure}[ht]
\begin{center}
\scalebox{0.4}{\includegraphics{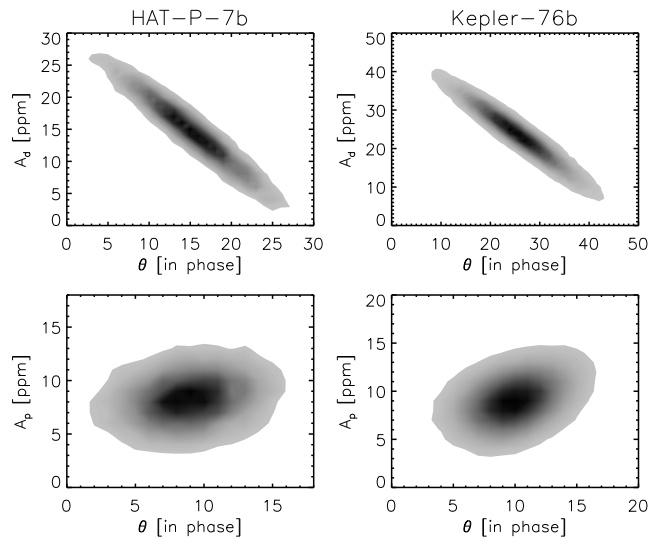}}
\end{center}
\caption{ Top: Contour plot of the posterior distributions of offset ($\theta$) vs Doppler boosting amplitude (A$_{\text{d}}$) for HAT-P-7b (left) and Kepler-76b (right). Bottom: Contour plot of the posterior distributions of offset ($\theta$) vs planetary brightness amplitude (A$_{\text{p}}$) for HAT-P-7b (left) and Kepler-76b (right). }
\label{fig:degen}
\end{figure}
\begin{table*}[t!]
\centering
\caption{Self-Consistent Albedos and Temperatures}
\begin{tabular*}{\textwidth}{@{\extracolsep{\fill}}l|cccc|cc}
\hline
\hline
& \multicolumn{4}{c}{Self-Consistent} & \multicolumn{2}{c}{Reflected Light Fraction} \\
& A$_{\text{g,max}}$ & T$_{\text{max}}$ & A$_{\text{g,hom}}$ & T$_{\text{hom}}$ & f$_{\text{ref,max}}$ & f$_{\text{ref,hom}}$ \\
\hline \\ [-2.0ex]
Kepler-5b &0.067$^{+0.050}_{-0.058}$&2182$^{+77}_{-70}$&0.118$^{+0.035}_{-0.037}$&1670$^{+47}_{-46}$&0.549$^{+0.451}_{-0.058}$&0.9706$^{+0.029}_{-0.037}$\\
Kepler-6b &0.049$^{+0.037}_{-0.041}$&1827$^{+47}_{-44}$&0.069$^{+0.032}_{-0.035}$&1419$^{+33}_{-32}$&0.706$^{+0.294}_{-0.041}$&0.9849$^{+0.015}_{-0.035}$\\
Kepler-7b &0.232$^{+0.081}_{-0.083}$&1870$^{+100}_{-110}$&0.247$^{+0.073}_{-0.071}$&1450$^{+72}_{-80}$&0.933$^{+0.067}_{-0.083}$&0.99643$^{+0.0036}_{-0.0714}$\\
Kepler-8b &0.094$^{+0.072}_{-0.090}$&2060$^{+130}_{-120}$&0.131$^{+0.054}_{-0.055}$&1589$\pm$81&0.708$^{+0.292}_{-0.090}$&0.9820$^{+0.018}_{-0.055}$\\
Kepler-10b &0.583$^{+0.083}_{-0.345}$&1600$^{+960}_{-1600}$&0.584$^{+0.082}_{-0.260}$&1270$^{+640}_{-1270}$&0.99794$^{+0.0021}_{-0.3452}$&1.00$^{+0.00}_{-0.26}$\\
Kepler-12b &0.071$\pm$0.046&1845$^{+68}_{-69}$&0.091$^{+0.039}_{-0.035}$&1432$^{+47}_{-50}$&0.775$^{+0.225}_{-0.046}$&0.9882$^{+0.012}_{-0.035}$\\
Kepler-41b &0.044$^{+0.070}_{-0.044}$&2220$^{+99}_{-120}$&0.104$^{+0.042}_{-0.043}$&1695$^{+78}_{-77}$&0.407$^{+0.593}_{-0.044}$&0.9611$^{+0.039}_{-0.043}$\\
Kepler-43b\textsuperscript{$\dagger$} &0.030$^{+0.303}_{-0.030}$&2043$^{+79}_{-352}$&0.069$^{+0.270}_{-0.069}$&1574$^{+86}_{-255}$&0.423$^{+0.577}_{-0.030}$&0.9649$^{+0.035}_{-0.069}$\\
Kepler-76b &0.144$^{+0.082}_{-0.103}$&2580$^{+170}_{-160}$&0.239$^{+0.041}_{-0.033}$&1920$^{+74}_{-86}$&0.58$^{+0.42}_{-0.10}$&0.9711$^{+0.029}_{-0.033}$\\
Kepler-91b\textsuperscript{$\dagger$} &0.44$^{+0.22}_{-0.44}$&1980$^{+690}_{-1980}$&0.46$^{+0.21}_{-0.27}$&1520$^{+390}_{-1520}$&0.9688$^{+0.031}_{-0.445}$&0.99825$^{+0.0018}_{-0.2655}$\\
Kepler-412b &0.031$^{+0.082}_{-0.031}$&2334$^{+70}_{-120}$&0.107$^{+0.051}_{-0.057}$&1769$^{+76}_{-73}$&0.271$^{+0.729}_{-0.031}$&0.9506$^{+0.049}_{-0.057}$\\
TrES-2b &0.0049$^{+0.0144}_{-0.0049}$&1877$^{+21}_{-28}$&0.029$^{+0.010}_{-0.011}$&1455$\pm$20&0.1591$^{+0.7918}_{-0.0049}$&0.9548$^{+0.045}_{-0.011}$\\
HAT-P-7b &0.047$^{+0.046}_{-0.047}$&2764$^{+87}_{-88}$&0.1937$^{+0.0078}_{-0.0093}$&2022$^{+36}_{-34}$&0.228$^{+0.243}_{-0.047}$&0.9478$^{+0.0522}_{-0.0093}$\\
KOI-13b\textsuperscript{a} &0.404$^{+0.030}_{-0.078}$&2580$^{+260}_{-160}$&0.4529$^{+0.0069}_{-0.0077}$&1919$^{+81}_{-79}$&0.884$^{+0.077}_{-0.078}$&0.99206$^{+0.0079}_{-0.0077}$\\
KOI-13b\textsuperscript{b} &0.00$^{+0.37}_{-0.00}$&3880$^{+110}_{-850}$&0.441$^{+0.011}_{-0.013}$&2310$^{+99}_{-140}$&0.00$^{+0.82}_{-0.00}$&0.9736$^{+0.026}_{-0.013}$\\
KOI-13b\textsuperscript{c} &0.316$^{+0.018}_{-0.019}$&3088$\pm$43&0.4453$^{+0.0060}_{-0.0057}$&2154$^{+15}_{-16}$&0.698$^{+0.049}_{-0.019}$&0.98332$^{+0.0167}_{-0.0057}$\\
\hline
\end{tabular*}
\begin{tabular}{p{17cm}}
\textsuperscript{$\dagger$} Planet's with evidence for non-planetary phase curve modulations (See Section~\ref{sec:k43_k91}). \\
Derived using stellar parameters from: \\
\textsuperscript{a} \citet{KOI13_Shporer2014}.\\
\textsuperscript{b} \citet{StellarParamsRevised_Huber2013}.\\
\textsuperscript{c} \citet{KOI13_Szabo2011}.\\
\end{tabular}
\label{tab:albedo}
\end{table*}
\begin{figure*}[t]
\begin{center}
\scalebox{0.83}{\includegraphics{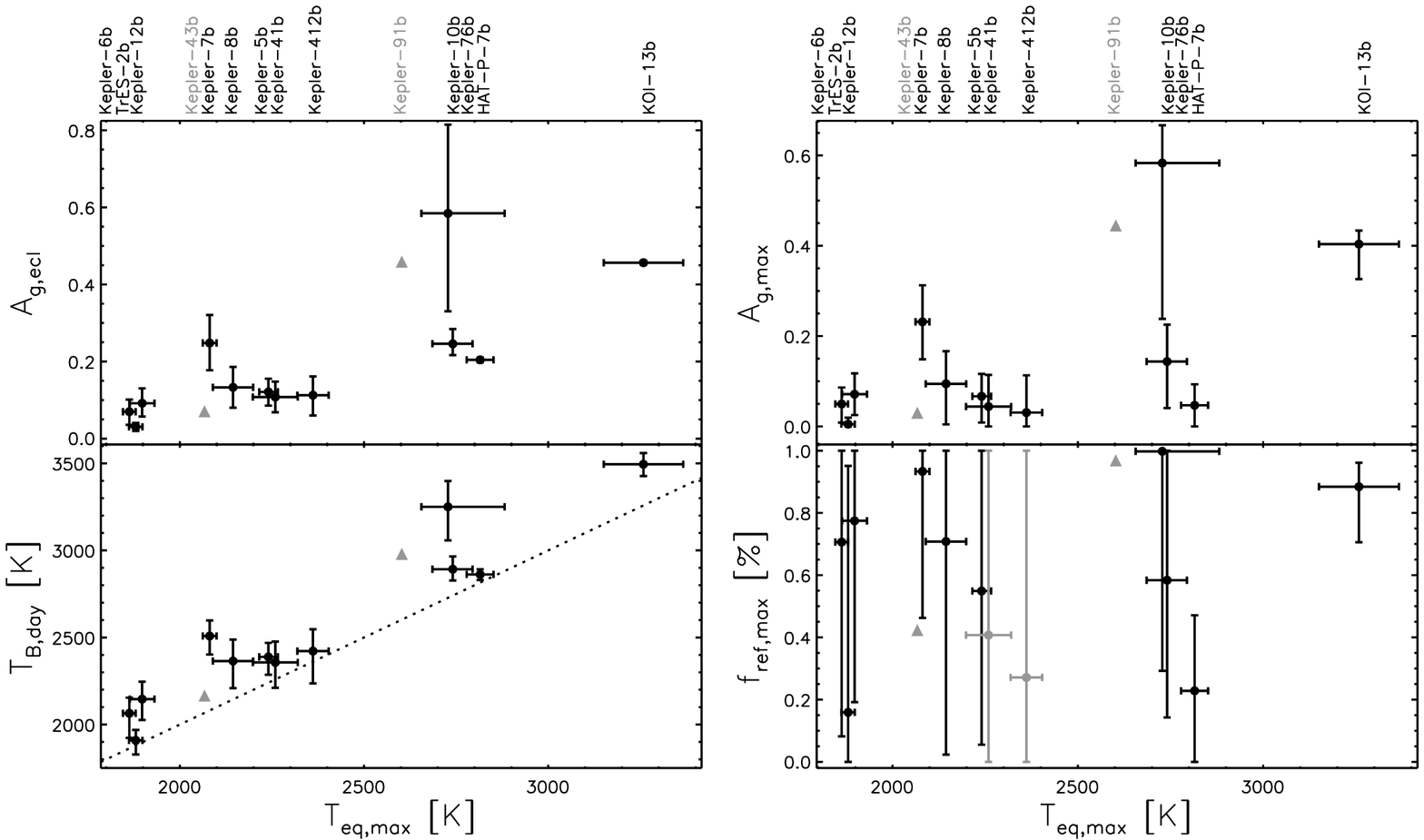}}
\end{center}
\caption{Secondary eclipse derived planet properties plotted as a function of the maximum equilibrium temperature: Top left, geometric albedo; Top right, self-consistent geometric albedo; Bottom left, day-side brightness temperature, where the dotted grey line indicates equal brightness and equilibrium temperatures; Bottom right, peak offset of the planet's phase function, where the dotted grey line indicates the boundary between eastward and westward shifts. Grey triangles are for targets where we find evidence for a non-planetary phase curve modulations and grey circles are for values where the we find no contraint on the reflected light fraction. Located directly above each point (i.e. on top of the plots) is the planet's name.}
\label{fig:tmax}
\end{figure*}
%
\indent Only three planet's in our sample (HAT-P-7b and Kepler-43b) did not favor a model with a fixed mass. However, for both planets our light curve derived mass is within 2$\sigma$ of the RV derived value. 
%
\indent In Table~\ref{tab:comparison} we present all $>$1$\sigma$ published {\it Kepler} eclipse depths, planetary brightness, ellipsoidal (A$_{\text{e}}$) and Doppler boosting (A$_{\text{d}}$) amplitudes for the 14 planets in our sample. In this table we have presented all planetary brightness measurements as peak-to-peak amplitudes and all ellipsoidal and Doppler measurements as semi-amplitudes. In addition we have adjusted the amplitudes for KOI-13b from E13, \citet{Angerhausen2014} and \citet{KOI13_Shporer2011}, to account for dilution due to a nearby companion (see Section~\ref{sec:companion}). \\
\indent For all published measurements, where errors were reported, we calculate the $\sigma$ deviation from our results. A histogram of the differences can be found in Fig.~\ref{fig:comparison}. For planets where a fixed mass was favored we report RV derived amplitudes and use their associated errors in our comparison to published values. \\ 
\indent In the majority of studies, the eclipse depths, planetary brightness and ellipsoidal amplitude are within 1$\sigma$ of our values, while for Doppler boosting most previous values lie within 2$\sigma$ and are skewed toward higher values. There are also some notable outliers. For HAT-P-7b and KOI-13b early phase curve studies, using significantly less data, differ by more than 5$\sigma$. While more recent studies of Kepler-76b and HAT-P-7b ~\citep[E13;][]{Kep76_Faigler2013, Angerhausen2014}, find significantly higher Doppler boosting amplitudes. We hypothesize that the inclusion of a planetary brightness offset is the source of this discrepancy between Doppler measurements and that pre-eclipse brightness shifts were previously measured as increased Doppler boosting. \\
\indent To test if a degeneracy exist, we refit HAT-P-7b and Kepler-76b with ellipsoidal variations and Doppler boosting as separate free parameters with no priors. From our posterior distributions (see top panel of Fig.~\ref{fig:degen}), we find that there is a correlation between the Doppler boosting and brightness offset, where a smaller positive offset (i.e. a pre-eclipse shift) results in a larger Doppler amplitude. We also find that also find that reducing the offset results in a lower planetary brightness amplitude (see bottom panel of Fig.~\ref{fig:degen}), which could account for the significantly larger brightness amplitude we find for HAT-P-7b (see Table~\ref{tab:comparison}). For the other phase curve parameters, eclipse depth and ellipsoidal amplitude, we do not find a correlation with brightness offset. \\
%
\subsection{Third Harmonic and Spin-Orbit Misalignment}
\label{sec:spin_orbit}
\indent For two of our targets, HAT-P-7b and KOI-13b, we found that the favored model included a phase-shifted cosine third harmonic signal (see Section~\ref{sec:a3} for description and Table~\ref{tab:res4} for values).
\indent For KOI-13b we measure a third harmonic amplitude in agreement with our previous value from E13 and $\sim$3 ppm less than the amplitude found by ~\citet{KOI13_Shporer2014}. This third harmonic could be due to the movement of the stellar tidal bulge raised by the planet (i.e. the source of ellipsoidal variations) across areas of the star with different surface brightnesses. The motivation for this reasoning is the asymmetry in KOI-13b's transit caused by a spin-orbit misalignment \citep{KOI13_Barnes2011,KOI13_Szabo2011} and significant gravity darkening due to rapid stellar rotation ($v\sin i$=65-70 km s$^{-1}$;~\citealt{KOI13_Szabo2011}). \\
\indent For HAT-P-7b we find that the BIC favors a model including a 8$\sigma$ detection of a 1.93 ppm third harmonic. Interestingly, similar to E13, \citet{HATP7_VanEylen2013} and \citet{HATP7_Morris2013}, we find an asymmetry in HAT-P-7b's transit. Analysis of this asymmetry is outside the scope of this paper, but, like KOI-13b, it could be related to the system's significant spin-orbit misalignment ~\citep{HATP7_Winn2009, HATP7_Narita2009}. Although, unlike KOI-13b's host star, HAT-P-7 has an unusually low, and somewhat disputed, $v\sin i$ of 2-6 km s$^{-1}$, indicating a nearly pole-on view ~\citep{HATP7_Winn2009, HATP7_Narita2009, HATP7_Albrecht2012, StellarParams_Torres2012}. \\
\indent Recently~\citet{Cowan2013} have shown that phase curves can contain contributions in odd harmonics due to the brightness map of the planet. However, we favor a stellar origin as the signal is still present when KOI-13b is obscured during secondary eclipse (see Figure~\ref{fig:noA3}). \\
%
\subsection{Planet Brightness, Temperature and Albedo}
\label{sec:albedo}
\indent With our eclipse depths we calculate each planet's geometric {\it Kepler} albedo ($A_{\text{g,ecl}}$), which assumes no thermal brightness contribution, and the brightness temperature of the day and night-side ($T_{\text{B,day}}$ and $T_{\text{B,night}}$), which assumes no contribution from reflected light and applies to the {\it Kepler} bandpass. We also calculate the equilibrium temperature, which assumes A$_{\text{B}}$=0, for the two limiting cases, instant re-radiation ($f$=2/3, $T_{\text{eq,max}}$) and homogeneous re-distribution ($f$=1/4, $T_{\text{eq,hom}}$). These values can be found in Tables~\ref{tab:res1}-\ref{tab:res4} and their calculation described in E13 and references therein. \\
\indent For Kepler-10b, Kepler-91b and KOI-13b we find $A_{\text{g,ecl}}$ between 0.4 and 0.6, while for all the other planets we find albedos $<$0.25. Kepler-10b's very high albedo is a clear outlier from our sample, as is its small radius, at only 1.4R$_{\oplus}$, ultra short period and rocky composition ~\citep{Kep10_Batalha2011}. Of our targets Kepler-10b is only planet where the presence of an atmosphere is not required by its mass and radius measurements. If this is the case then it is possible that Kepler-10b's high albedo, when compared to the hot-Jupiter albedos, is actually common of close-in rocky planets. \\
\indent For Kepler-91b our noise analysis finds correlated stellar variability that varies on time-scales equal to the planet period (see Sections~\ref{sec:variability} and~\ref{sec:k43_k91}). We suspect that this variability is contaminating our measurement of Kepler-91b's eclipse depth, resulting in an unusually high albedo. \\
\indent KOI-13b's high albedo is most likely the result of blackbody emission leaking into the optical as KOI-13b's equilibrium temperature, at 3300-3900 K, peaks very close to the edge of the {\it Kepler} bandpass (see Section~\ref{sec:k13} for further discussion).\\
\indent Since the eclipse depths at optical wavelengths are likely a combination of reflected light and thermal emission, we self-consistently solve for $A_{\text{g}}$ by assuming a Bond albedo of $\frac{3}{2} A_{\text{g}}$ and taking into account both contributions using
\begin{eqnarray}
F_{\text{ecl}} = & \left( \frac{R_{\text{p}}}{R_{\star}} \right)^2 \frac{\int B_{\lambda} \!
\mathsmaller{\left\{ T_{\star} \left( \frac{a}{R_{\star}} \right)^{-1/2} [f(1-\frac{3}{2}A_{\text{g}})]^{1/4}\right\}} \;
T_{\text{K}} \, \text{d}\lambda}{\int (T_{\text{K}} F_{\lambda,\star} \text{d}\lambda)} & \nonumber \\
& + \; A_{\text{g}} \left( \frac{R_{\text{p}}}{a} \right)^2 &
\end{eqnarray}
where $B_{\lambda}$ is the Planck function of the equilibrium temperature expression in parentheses, $T_{\text{K}}$ is the {\it Kepler} transmission function and $F_{\lambda,\star}$ is the stellar flux computed using the NEXTGEN model spectra \citep{Hauschildt1999}. \\
\indent We calculate each target's self-consistent albedo and corresponding temperature for the two limiting cases, denoted {\it max} and {\it hom}, and present the values in Table~\ref{tab:albedo}. For several planets we find that the self-consistent albedo, in the instant re-radiation limit, is marginally consistent with zero, meaning that thermal emission alone can explain their observed eclipse depths. This can be seen in the upper left panel of Fig.~\ref{fig:tmax} where we compare the maximum equilibrium temperature to the reflection only geometric albedo ($A_{\text{g,max}}$) and in the upper right panel of Fig.~\ref{fig:tmax} where we compare the equilibrium temperature to the self-consistent geometric albedo (i.e. considering both reflected and thermal emission). \\
\indent For KOI-13b, we calculate the albedos and temperatures using three sets of stellar parameters reported in the literature (see Table~\ref{tab:res4}). These three studies report a stellar temperature between 7650K and 9100K, corresponding to an equilibrium temperature from 3300K to 3900K and leading to a variation in the self-consistent albedo from 0.0 to 0.4 (see Table~\ref{tab:albedo}). \\
\indent In the bottom left panel of Fig.~\ref{fig:tmax} we compare the day-side brightness temperature ($T_{\text{B,day}}$) to the maximum equilibrium temperature ($T_{\text{eq,max}}$) and find that for several planets the brightness temperature is significantly higher than the equilibrium temperature. However, the majority of this excess brightness temperature can again be accounted for if we include a reflected light contribution as, at optical wavelengths, even low albedos can produce a planet brightness dominated by reflected light. For all planets we find self-consistent albedos consistent with the planetary nature of our targets (i.e. $<$1). \\
\indent There doesn't appear to be a correlation between the excess brightness temperature and equilibrium temperature as an excess is seen in both the hottest and coolest planets in our sample. It is possible is that this extra flux is seen because we are probing significantly hotter layers of the planet's atmosphere, which is supported by our finding that for Kepler-76b and KOI-13b we need an increase in brightness to explain both their observed day- and night-side flux (see Section~\ref{sec:nightside}). \\
%
\subsection{Night-side Emission}
\label{sec:nightside}
\indent It is possible to constrain the planet's night-side brightness using the day-side brightness provided by the eclipse depth and the planetary brightness amplitude, which measures the day-night flux difference. Note that for the planets with a peak brightness offset we calculate the difference between the eclipse depth and brightness amplitude at $\phi$=0.5, not the peak amplitude. For Kepler-76b and KOI-13b we find a 2 and 4 $\sigma$ night-side flux detection, respectively, while for all other planets we find a night-side flux consistent with zero. For Kepler-76b the eclipse appears shallower than the brightness amplitude (see Fig.~\ref{fig:res4}). However, after compensating for Kepler-76b's partial secondary eclipse we find that the eclipse depth is a significantly larger than the planet's brightness. \\
\indent By combining the day and night side measurements it is possible to place a constraint on the planet's re-distribution factor ($f'$) by simultaneously solving for $f'$ and $A_{\text{g,max}}$ in the following equations
\begin{eqnarray}
F_{\text{ecl}} = & \left( \frac{R_{\text{p}}}{R_{\star}} \right)^2 \frac{\int B_{\lambda} \!
\mathsmaller{\left\{ T_{\star} \left( \frac{a}{R_{\star}} \right)^{-1/2} 
[(\frac{1}{4}+f') (1-\frac{3}{2}A_{\text{g}})]^{1/4}\right\}} \;
T_{\text{K}} \, \text{d}\lambda}{\int (T_{\text{K}} F_{\lambda,\star} \text{d}\lambda)} & \nonumber \\
& + \; A_{\text{g}} \left( \frac{R_{\text{p}}}{a} \right)^2 & \\
F_{\text{n}} = & \left( \frac{R_{\text{p}}}{R_{\star}} \right)^2 \frac{\int B_{\lambda} \!
\mathsmaller{\left\{ T_{\star} \left( \frac{a}{R_{\star}} \right)^{-1/2} 
[(\frac{1}{4}-f') (1-\frac{3}{2}A_{\text{g}})]^{1/4}\right\}} \;
T_{\text{K}} \, \text{d}\lambda}{\int (T_{\text{K}} F_{\lambda,\star} \text{d}\lambda)} & \nonumber \\
\end{eqnarray}
where $f'$=1/4 corresponds maximum day-side and zero night-side temperature, while $f'$=0 is an equal day-night temperature. For planets with a marginal night-side detection, Kepler-8b, Kepler-12b, Kepler-412b and TrES-2b, we find albedos of 0.13$\pm$0.06, 0.09$\pm$0.04, 0.11$\pm$0.06 and 0.03$\pm$0.01, respectively, but do not find any constraint on $f'$. \\
\indent For the planets with a significant night-side flux (Kepler-76b and KOI-13b) our simple model fails, and we do not find any solution. This is most likely due to the assumption that the atmosphere is isothermal, with two temperatures, one for the day-side and one for the night-side. However, this can be accounted for if we are probing a significantly hotter layer of the planet's atmosphere in the {\it Kepler} bandpass. \\
%
\subsection{Fraction of Planetary Brightness from Reflected Light}
\label{sec:reflected}
\indent The self-consistent albedo measurements give strong evidence that, for most planets, the light from the planet in the {\it Kepler} band is a combination of both reflected light and thermal emission. Since {\it Kepler} observes in a single broad optical band, it is difficult to constrain the relative contributions from each source. However, using our self-consistent albedo equation we can estimate the reflected light fraction (f$_{\text{ref}}$) by dividing the observed geometric albedo by our self-consistent albedo. The values for the two limiting, instant re-radiation or homogeneous re-distribution of absorbed stellar energy, cases can be found in Table~\ref{tab:albedo}. \\
\indent In the bottom right panel of Fig.~\ref{fig:tmax} we compare the reflected light fraction and maximum equilibrium temperature and do not find a significant correlation. This is not surprising as most, if not all, values are not well constrained and several are completely unconstrained. \\ 
%
%
\subsection{Peak Offset of Planetary Light}
\label{sec:offset}
\begin{figure*}[t]
\begin{center}
\scalebox{0.83}{\includegraphics{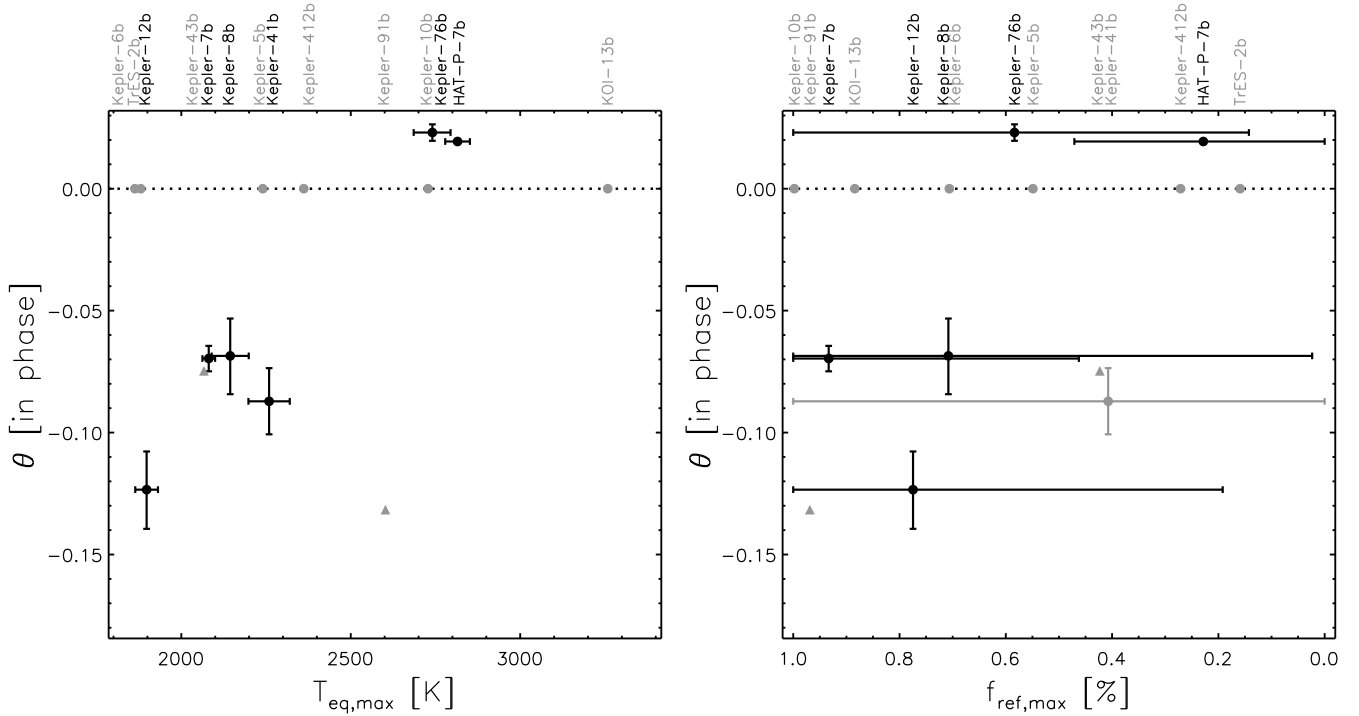}}
\end{center}
\caption{Left: Maximum equilibrium temperature vs. peak offset of the planet's phase function. Right: The fraction of the planet's brightness due to reflected light vs. peak offset of the planet's phase function. Black circles are for offsets of the planetary phase curve signal, grey triangles are for offsets that are due to non-planetary phase curve modulations, grey circles without error bars are for planet's where an offset was not favoured and grey circles with error bars are for planet's where there is no constraint on the reflected light fraction. Located directly above each point (i.e. on top of the plots) is the planet's name.}
\label{fig:offset}
\end{figure*}
\indent For eight of our fourteen targets, our model favored the inclusion of an offset of the peak planetary bright from mid-eclipse. Kepler-7b, -8b, -12b, -41b, -43b and -91b exhibit brightness peaks after eclipse, referred to as a negative offset in our formalism, while the Kepler-76b and HAT-P-7b reach peak brightness before eclipse (positive offset). \\
\indent Of our sample, only Kepler-7b, -12b and -43b have published offset measurements, all with the peak after the eclipse ~\citep{Kep7_Demory2013,Angerhausen2014}. For Kepler-7b, studies, including this work, find an offset ranging from -0.07 to -0.11 ~\citep{Kep7_Demory2013,Angerhausen2014}. While for the two other planets, both this work and~\citet{Angerhausen2014} find significant negatives offsets, ranging from -0.08 to -0.10 for Kepler-12b and -0.12 to -0.19 for Kepler-43b. 

The other planets in our sample for which we find an offset, Kepler-8b, Kepler-41b, Kepler-76b and HAT-P-7b, are not found in other work ~\citep{Angerhausen2014}. The latter two of which are possibly be due to a model degeneracy between Doppler boosting and a pre-eclipse shift in planetary brightness (see Section~\ref{sec:comparison}). \\
\indent In the left panel of Fig.~\ref{fig:offset}, we plot peak offset as a function of equilibrium temperature ($T_{\text{eq,max}}$) for all planet's, including those where a zero fixed offset was favored. We find that when all planets are considered there appears to be no correlation between offset and temperature. However, if we only consider planets where a peak offset was found, we find that the six cooler planet's exhibit negative offsets ranging from -0.7 to -0.12, while the hotter two have smaller positive offsets of $\sim$0.02. \\
\indent When we exclude planet's where our analysis shows compelling evidence that correlated noise, most likely due to the star, is dominating the phase curve signal (see Section~\ref{sec:k43_k91}). The remaining planets (Kepler-6b, -7b, -8b, -41b, -76b and HAT-P-7b) fall into two regions, cooler planets (T$_{\text{eq,max}}$$<$2300 K) with large post-eclipse shifts and hotter planets (T$_{\text{eq,max}}>$2300K) with small pre-eclipse shifts (see black circles in Fig.~\ref{fig:offset}). This bimodal distribution in peak offset and temperature could be a result of two distinct mechanisms, which we discuss in the following section.  \\
%
\section{Discussion: Cloudy Mornings and Hot Afternoons}
\label{sec:source}
\indent Close-in planets, such as those in our sample, are expected to be tidally locked and will therefore rotate prograde, with the planet's night-side moving in the direction of orbital motion. With this assumption, the planet's evening-side will be visible during the first half of the orbit (pre-eclipse), while the morning-side will be more visible in the second (post-eclipse), regardless of on-sky orientation. \\
\indent Therefore, if atmospheric circulation moves in the direction of rotation ~\citep[e.g.][]{PreShift_Showman2011}, winds would transport energy from the substellar point, towards the planet's evening-side and then onwards towards the morning-side. This shift away from the substellar point, where incident flux is largest, then results in a longitudinally asymmetric thermal day-side, as hot material moves towards the evening-side. Considering only the thermal emission of a planet with prograde rotation, this would result in increased brightness before eclipse, when the hot evening-side is more visible. Evidence of this can be seen in circulation models of tidally locked hot-Jupiters ~\citep[e.g.][]{PreShift_Cooper2005,PreShift_Heng2011,PreShift_Showman2011} and thermal phase curve observations of the hot-Jupiters such as WASP-43b ~\citep{Stevenson2014}, HD209458 ~\citep{Zellem2014} and HD189733 ~\citep{Knutson2012}. \\
\indent We hypothesize that an analogous mechanism is affecting the planet's reflected day-side brightness. The same atmospheric circulation patterns result in the movement of reflective cloud particulates from the night-side, where cooler temperatures allow for condensation, to the morning-side. As these clouds move across the planet's day-side, they quickly heat-up and dissipate from increased irradiation. Resulting in a longitudinally asymmetric reflective day-side brightness ~\citep{Kep7_Demory2013}. Considering only the reflected light from a planet with prograde rotation, this would result in an increased brightness after eclipse, when the cloudy, reflective morning-side is more visible. \\
\indent With the exclusion Kepler-43b and Kepler-91b, which show compelling evidence for a non-planetary phase curve signal (see Section~\ref{sec:k43_k91}), we find that these two mechanisms can be used to explain why the four planets with large shifts to after eclipse are all under 2300 K, while the two with smaller pre-eclipse shifts are over 2700 K (see black circles in Fig.~\ref{fig:offset}). \\
\indent It is possible that both these mechanisms operate simultaneously, with thermal emission mostly in the infrared and reflected light mostly in the optical. Since the thermal emission from the cooler planets (T$_{\text{eq,max}}$$<$2300K) is not expected to contribute significantly in the Kepler bandpass, if a thermal shift did exist, it would most likely not be detectable by {\it Kepler}. We suspect that for Kepler-6b, -7b, -8b and -41b we are observing their longitudinally asymmetric reflected day-side brightness, caused by clouds on the morning-side. Supporting evidence includes: i) Cloud formation is possible as T$_{\text{eq,max}}$$<$2500K ~\citep[e.g.][]{Condensation_Fortney2008a,Condensation_Fortney2008b,Condensation_Morley2013}. ii) The negative offset measurements are large ($>$25$^{\text{o}}$), which could result from the brightest longitude being far from the substellar point, as clouds would be thickest, and therefore most reflective, close to the morning-side terminator. iii) Our reflected light fraction (see left panel of Fig.~\ref{fig:offset}) is consistent with a completely reflective eclipse depth. iv) This is further supported by the Spitzer observations from~\citet{Kep7_Demory2013}, who use the lack of significant thermal emission in the {\it Spitzer} 3.6 and 4.5 $\mu$m bandpasses and the presence of a post-eclipse shift to conclude that Kepler-7b's phase curve is dominated by reflected light. Furthermore they also state that the most likely cause of the post-eclipse shift is the presence of inhomogeneous reflective clouds, whose properties change as a function of longitude and are influenced by the planet's wind and thermal patterns. \\ 
\indent For the hotter planets (T$_{\text{eq,max}}$$>$2700K) it is likely that thermal emission contributes significantly in the {\it Kepler} band. If for Kepler-76b and HAT-P-7b the thermal emission dominates over the reflected light, the phase curve offset will be dominated by the transport of the hot-spot away from the sub-stellar point and towards the evening-side. Supporting evidence includes: i) Their high temperatures make it difficult for cloud formation, resulting low albedos and reflection. ii) The positive offset measurements are small ($<$8$^{\text{o}}$), which could be attributed to the brightest longitude being close to the substellar point. Possibly due to the rapid re-emission of absorb stellar energy once material has left the substellar point. iii) For HAT-P-7b our reflected light fraction (see left panel of Fig.~\ref{fig:offset}) is consistent with a completely thermal eclipse depth, while for Kepler-76 the reflected light fraction is not well constrained by current measurements. \\
\indent If this interpretation is correct, it appears these processes are not ubiquitous as we do not find detections for half of our sample (Kepler-5b, -6b, -10b, -13b, -41b, -412b, TrES-2b). If large offsets were present for these planets they should be detectable as their phase curve signal-to-noise is similar to those with detections. When comparing planet properties of those with an offset to those without we do not find a clear trend. The two populations both span a similar range in temperature and reflectivity (see Fig.~\ref{fig:offset}) as well as in surface gravity and density (see Tables~\ref{tab:res1}-\ref{tab:res4}). These non-detections could be attributed to: i) longitudinally symmetric coverage of reflective particles and/or a symmetric thermal surface brightness about the substellar point. ii) A combination of both a reflected and thermal offset, that, when measured in the {\it Kepler} band, mimics the phase curve expected for a planet with no offset. However, this scenario is unlikely for the cooler planets (Kepler-5b, -6b, -41b, -412b, TrES-2b) as their thermal emission is not expected to contribute significantly in the optical. iii) The {\it Kepler} data is not precise enough to allow us to determine the offset. \\
%
\section{Conclusions}
\label{sec:conclu}
\indent Our re-analysis of Kepler-91b's transit found very different results than in E13, where we previously derived a geometric albedo $>$1 and therefore concluded that Kepler-91b, previously referred to as KOI-2133, was a self-luminous object, not a planet. However, with our new set of transit parameters, we derive a geometric albedo of 0.46, fully consistent with it being a planet. \\
\indent From our light curve analysis we conclude that the phase curves of Kepler-43b and Kepler-91b's are strongly affected by stellar variability and therefore their phase curve measurements are not well constrained with current data. \\
\indent Of the fourteen {\it Kepler} planets, with significant detections in each of the phase curve components, we find that most have low geometric albedos $<$0.25 dominated by reflected light, with the exception of Kepler-10b, Kepler-91b and KOI-13b where we derive values of 0.4-0.6. Of our targets, KOI-13b, with a small eccentricity of 0.0006$\pm$0.0001, is the only planet where an eccentric orbit is favored. \\
\indent For both HAT-P-7b and KOI-13b, we identify a third harmonic in the lightcurve with an amplitude of $1.9\pm0.2$ ppm and $7.0\pm0.3$ ppm, respectively. For KOI-13b the harmonic has been seen before, but for HAT-P-7b this is the first detection. The cause of this additional third harmonic is unknown, but could be from the movement of the stellar tidal bulge raised by the planet, the source of ellipsoidal variations, across areas of the star with different surface brightnesses. \\
\indent For seven planets (Kepler-5b, Kepler-6b, Kepler-10b, Kepler-412b, TrES-2b and KOI-13b) our analysis did not favor an offset in the peak planetary brightness, while for Kepler-7b, Kepler-8b, Kepler-12b, Kepler-41b, Kepler-43b, Kepler-76b and HAT-P-7b we find both eastward and westward offsets. For the two hottest planets, Kepler-76b and HAT-P-7b, with offsets of 0.023$\pm$0.003 and 0.0194$\pm$0.0008, respectively, we find that the planetary light peaks before the eclipse, corresponding to a peak brightness eastward of the substellar point or on the evening-side of the planet. While for the cooler planets, Kepler-7b, Kepler-8b, Kepler-12b, Kepler-41b and Kepler-43b, with offsets between -0.07 and -0.12, the planetary light peaks after the eclipse (i.e. westward or on the morning-side). \\
\indent These results have drastically increased the number of {\it Kepler} planets with detected planetary light offsets and provided the first evidence, in the {\it Kepler} data, for a correlation between the direction of the peak offset and the planet's temperature. This correlation could possibly arise if hotter planets are dominated by thermal emission and therefore exhibit a hot spot shifted to the east, as theoretically predicted, whereas cooler planets are dominated by reflected light and have clouds westward of the substellar point (i.e. on the morning-side), as seen for Kepler-7b. However, with this study alone we are not able to determine whether the planets with peak offsets are seen predominantly in reflected light or thermal emission. \\
\acknowledgments
\indent This work was supported by grants from the Natural Sciences and Engineering Research Council (NSERC) of Canada to R.J. L.J.E. is supported in part by an NSERC CGS while E.d.M. received partial support from an Ontario Postdoctoral Fellowship.
%
%

%
\appendix
\begin{figure*}[ht]
\begin{center}
\scalebox{0.8}{\includegraphics{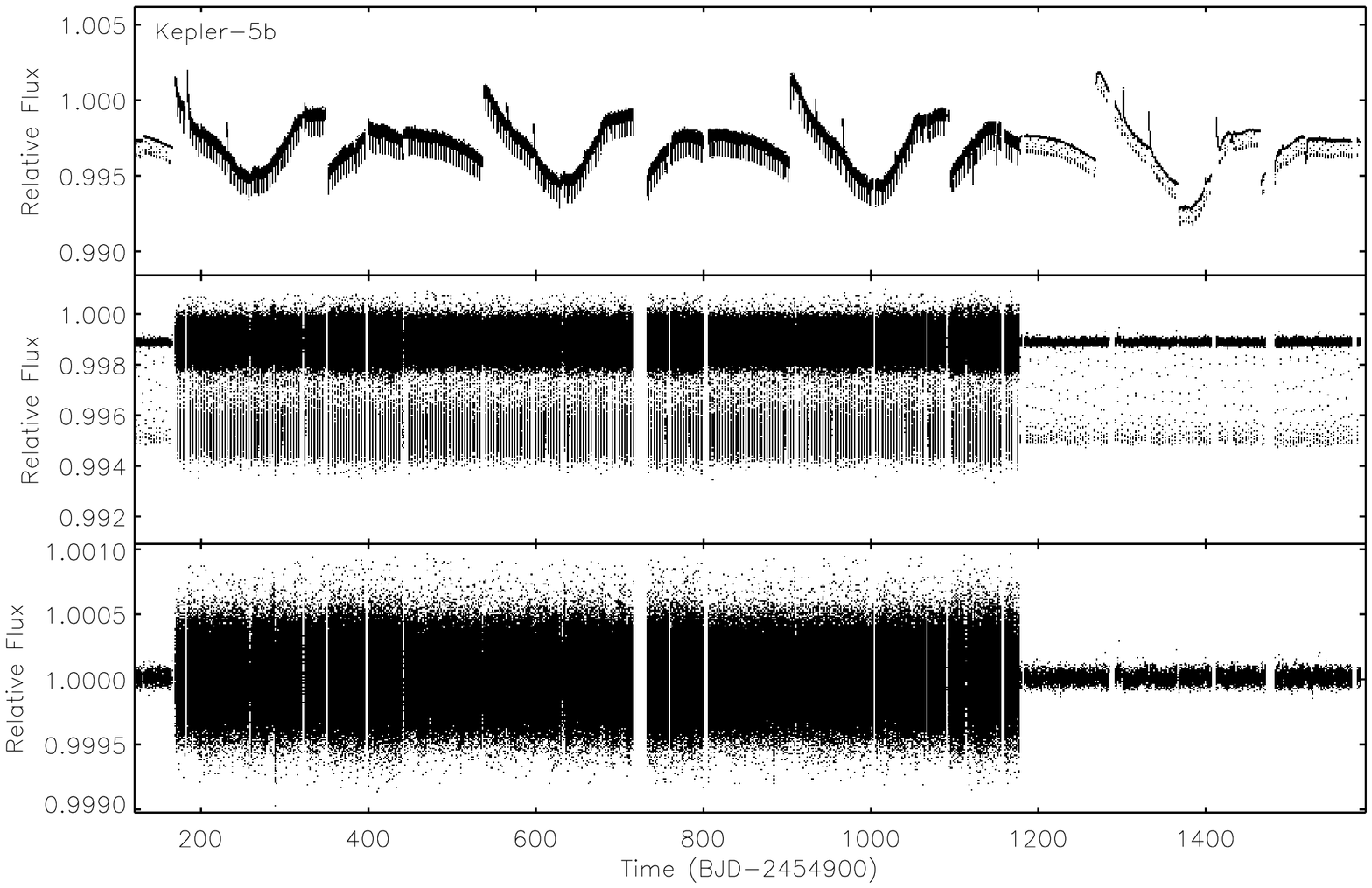}}
\scalebox{0.8}{\includegraphics{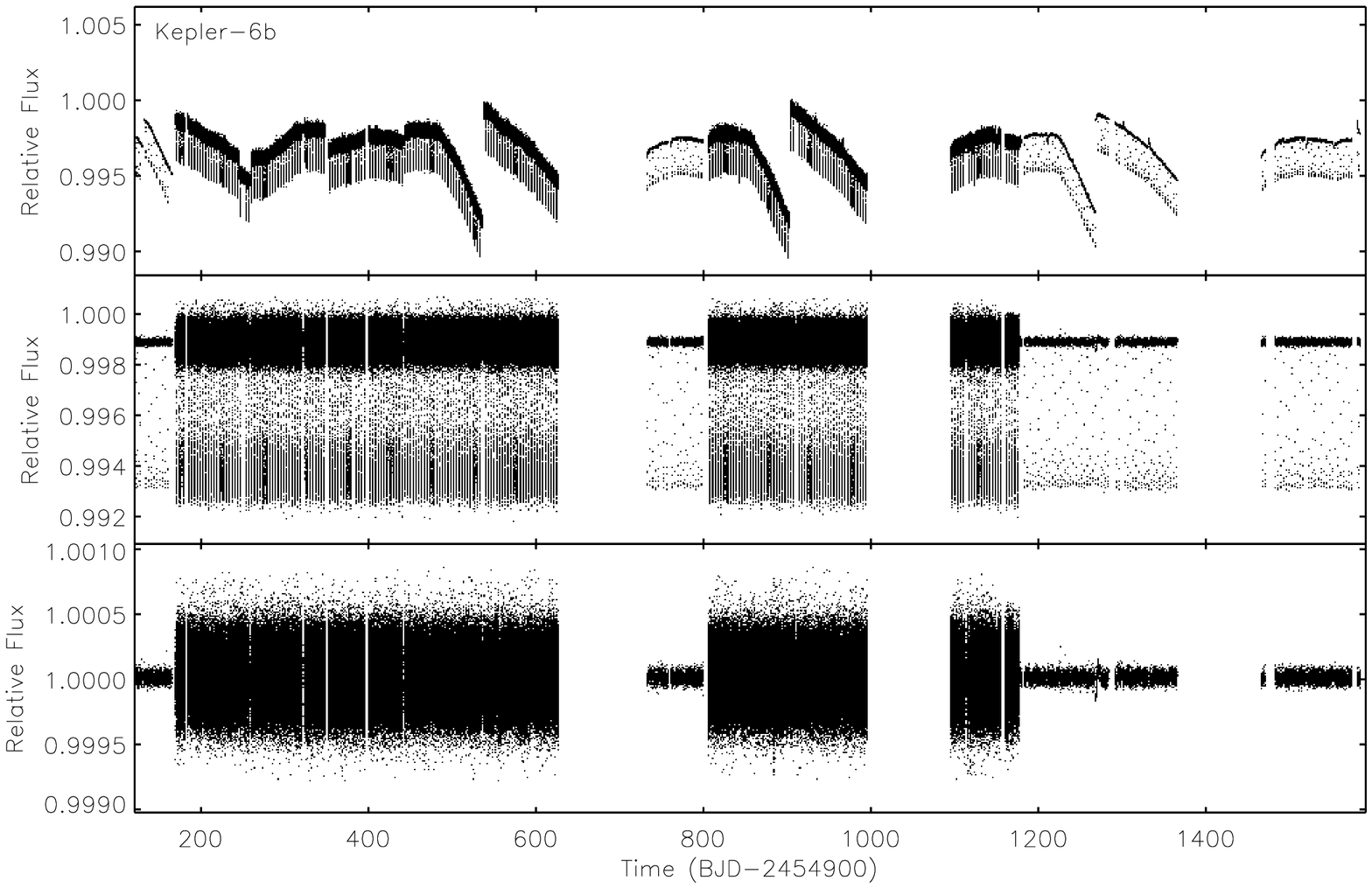}}
\end{center}
\caption{For Kepler-5b (top plot) and Kepler-6b (bottom plot), the top panel contains the raw {\it Kepler} simple aperature photometry light curve, the middle is after cotrending and the bottom is after cotrending and removing the transits and outliers.}
\label{fig:A1}
\end{figure*}
\begin{figure*}[ht]
\begin{center}
\scalebox{0.8}{\includegraphics{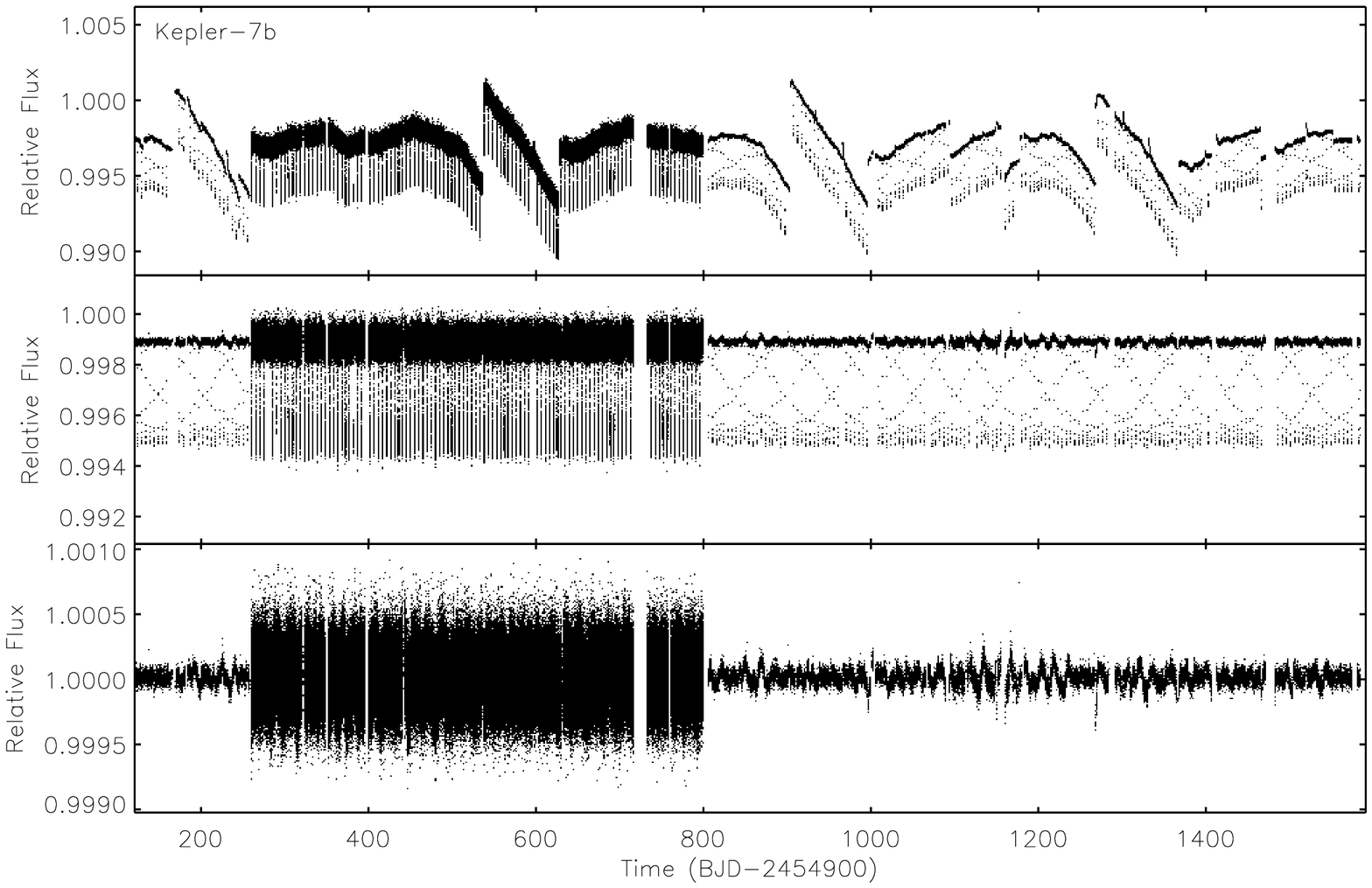}}
\scalebox{0.8}{\includegraphics{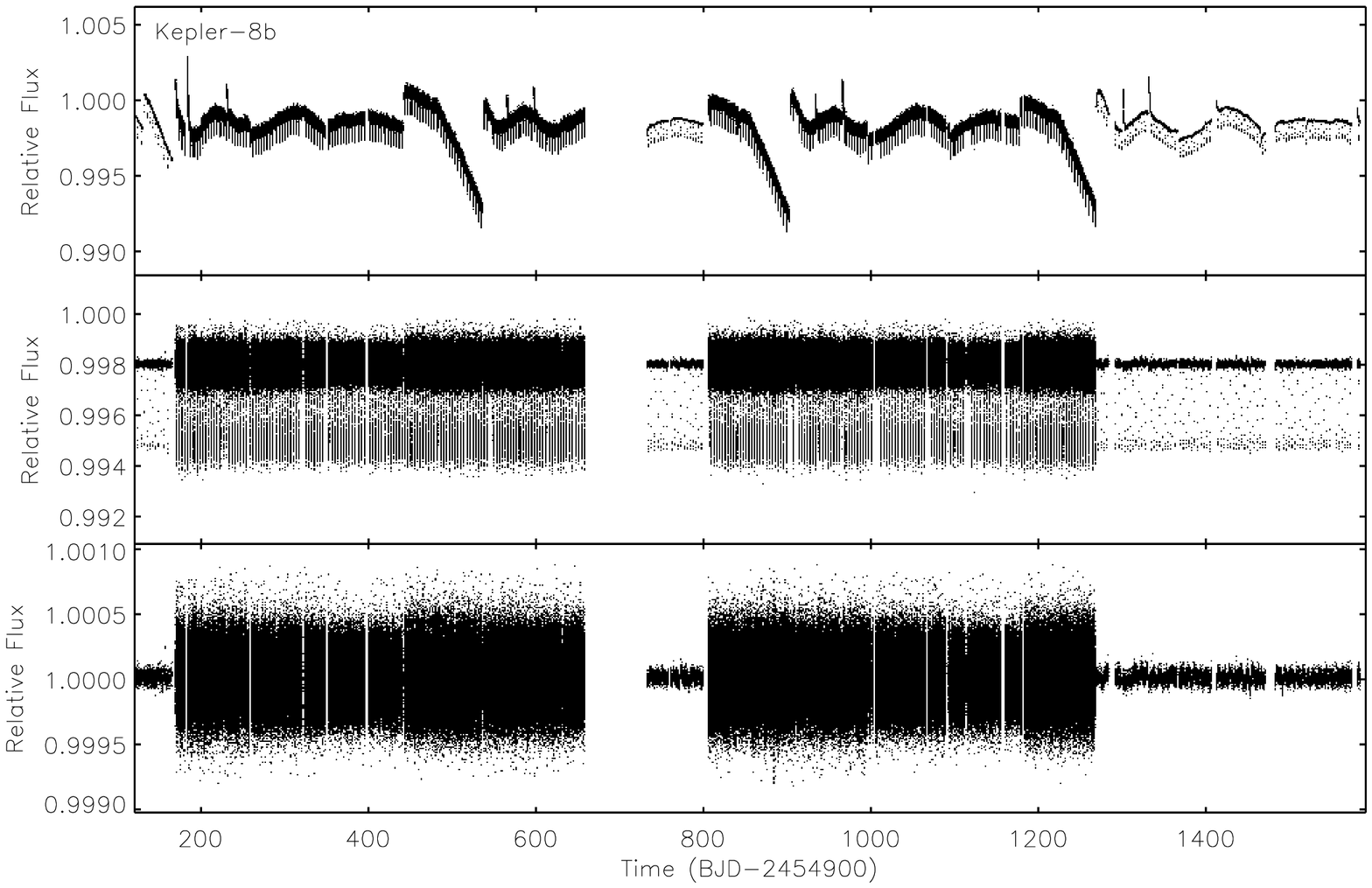}}
\end{center}
\caption{Same as Figure~\ref{fig:A1}, but for Kepler-7b (top plot) and Kepler-8b (bottom plot).}
\label{fig:A2}
\end{figure*}
\begin{figure*}[ht]
\begin{center}
\scalebox{0.8}{\includegraphics{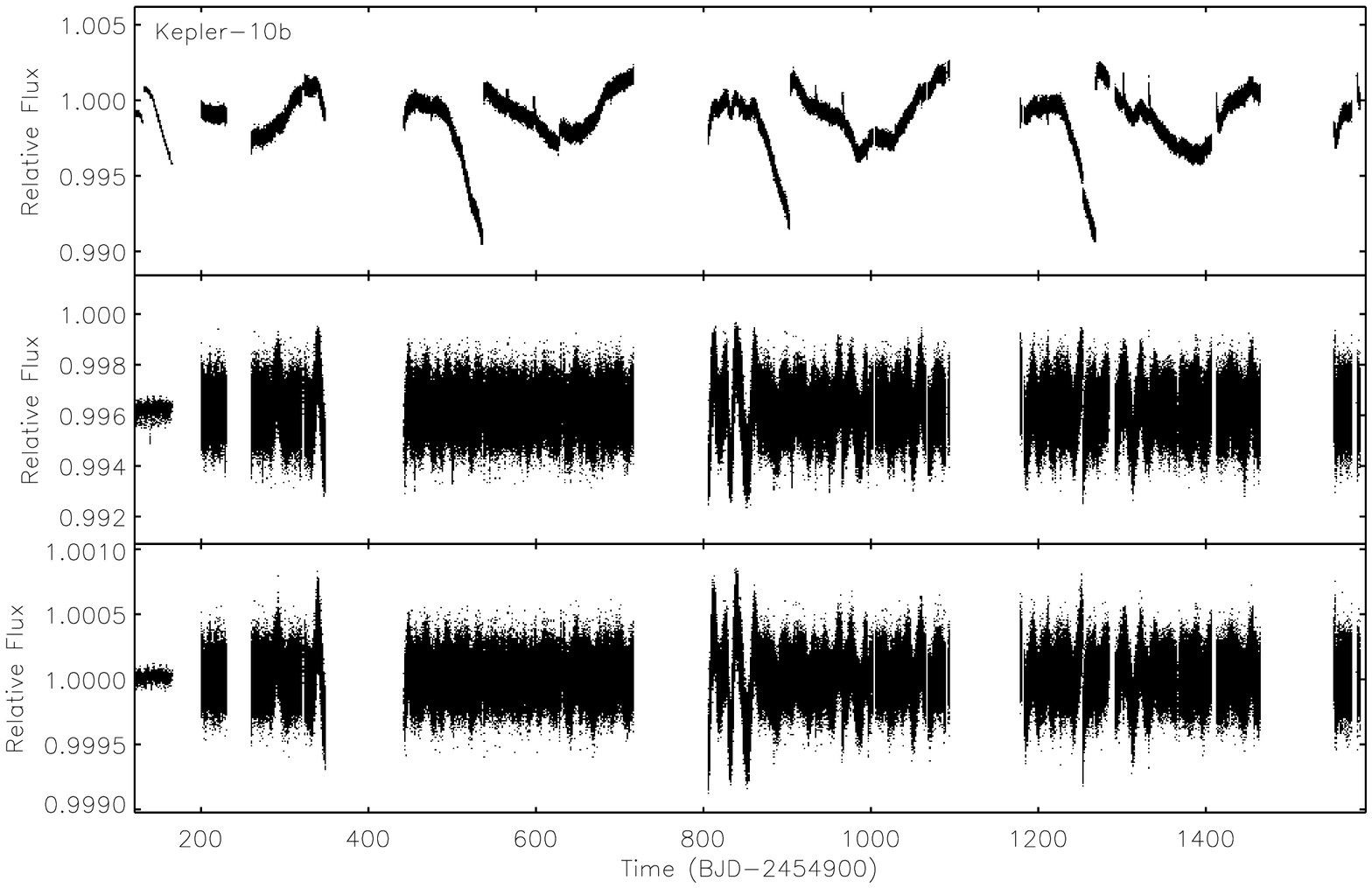}}
\scalebox{0.8}{\includegraphics{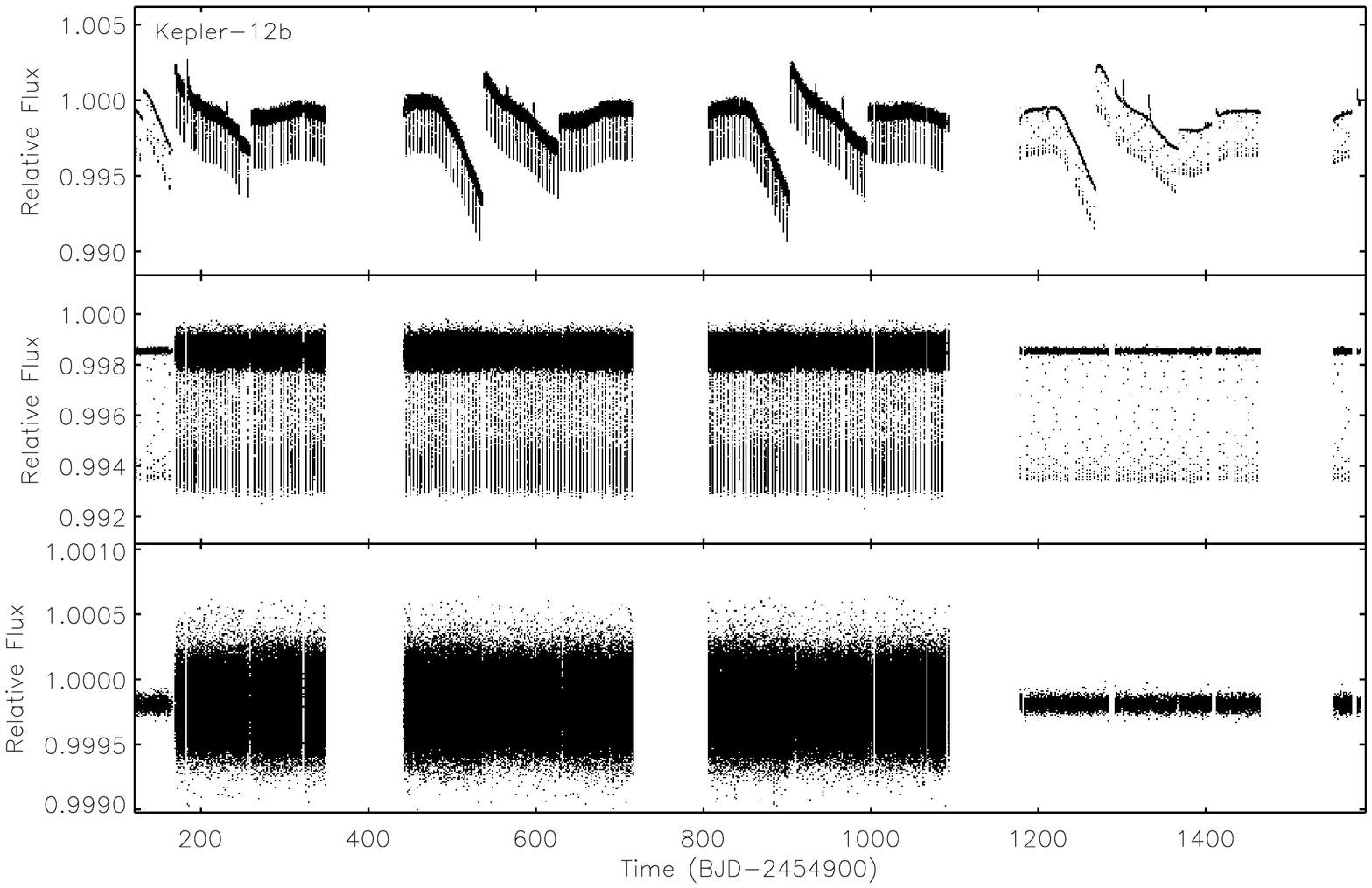}}
\end{center}
\caption{Same as Figure~\ref{fig:A1}, but for Kepler-10b (top plot) and Kepler-12b (bottom plot).}
\label{fig:A3}
\end{figure*}
\begin{figure*}[ht]
\begin{center}
\scalebox{0.8}{\includegraphics{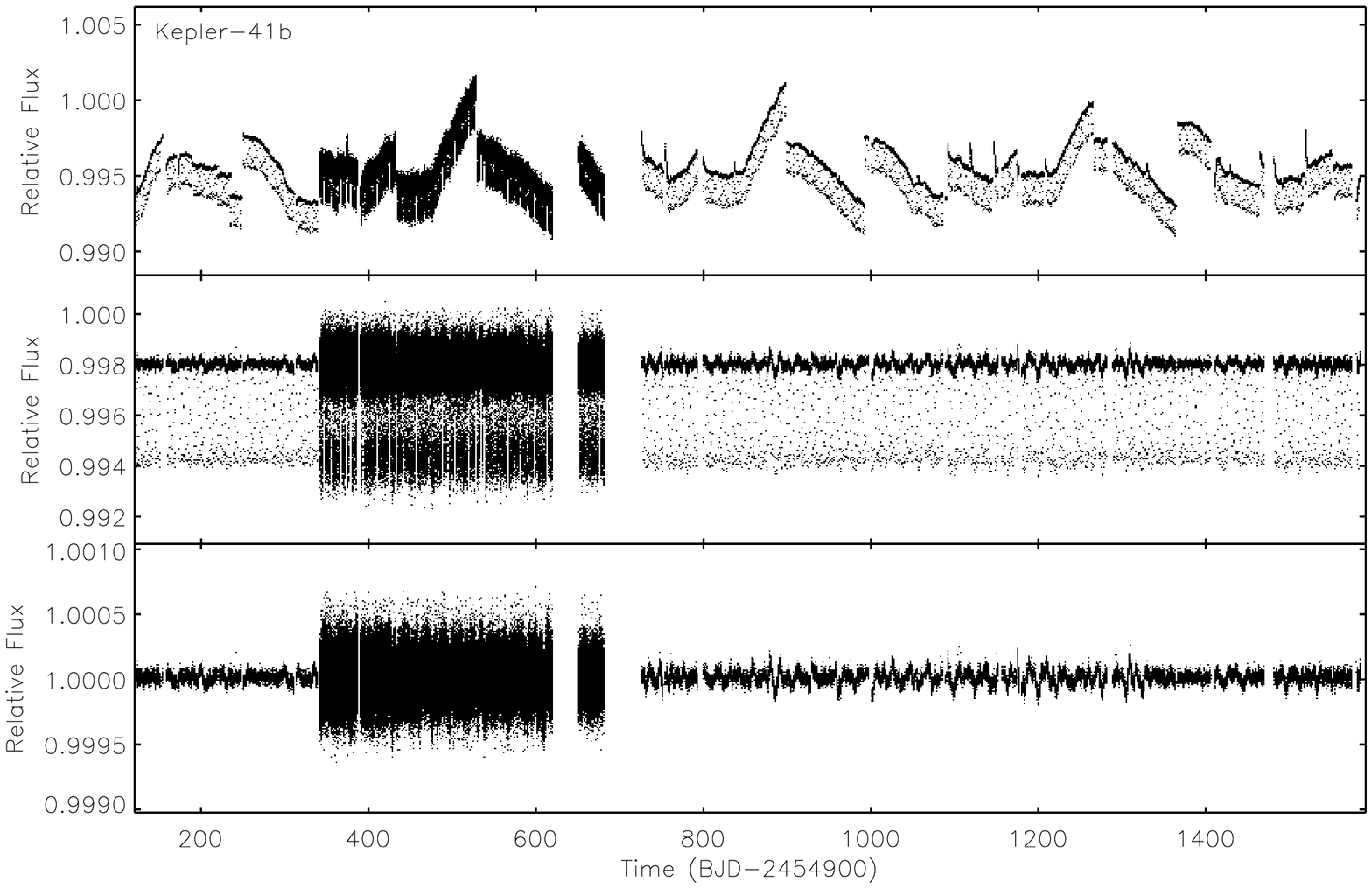}}
\scalebox{0.8}{\includegraphics{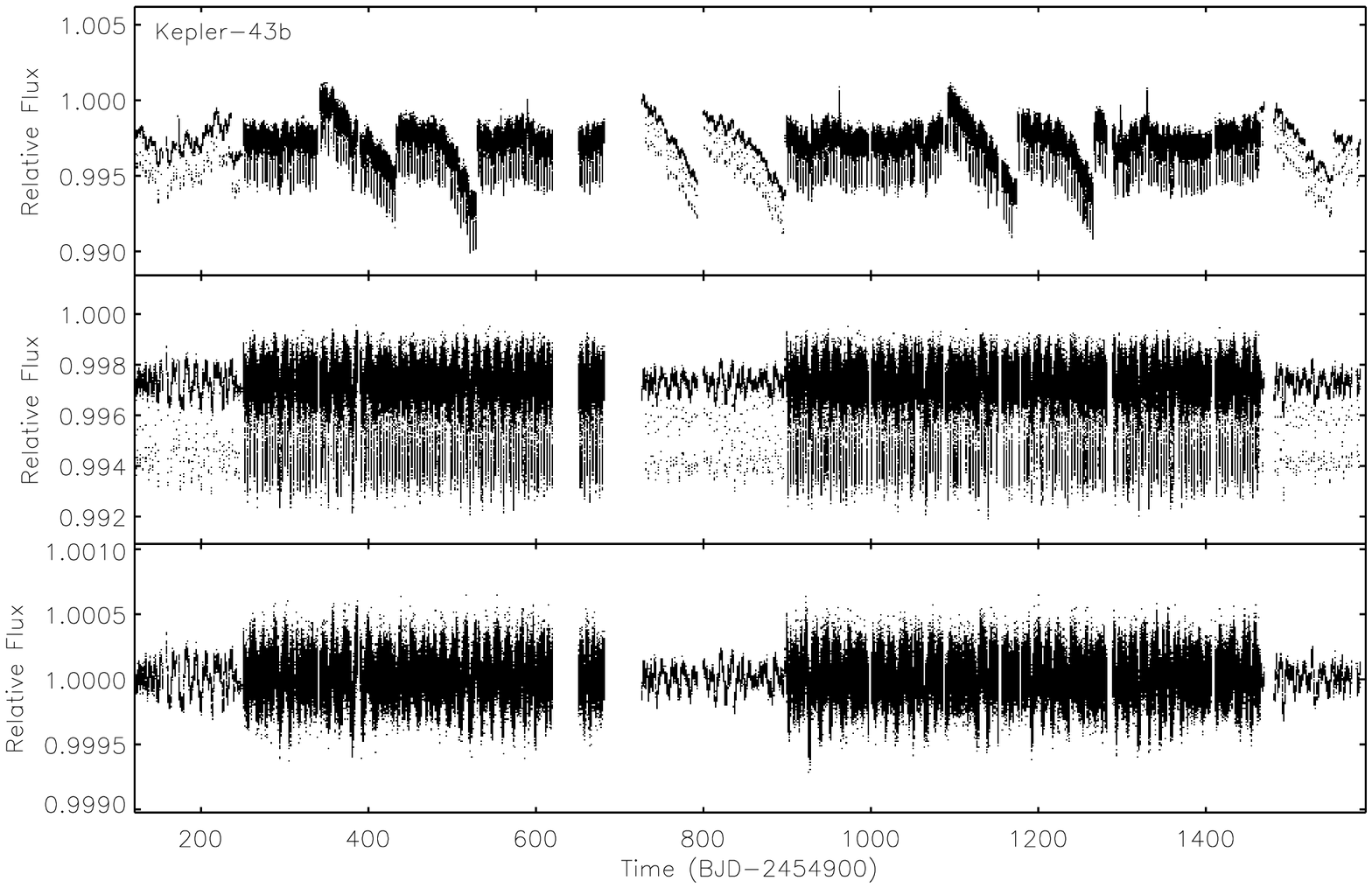}}
\end{center}
\caption{Same as Figure~\ref{fig:A1}, but for Kepler-41b (top plot) and Kepler-43b (bottom plot).}
\label{fig:A4}
\end{figure*}
\begin{figure*}[ht]
\begin{center}
\scalebox{0.8}{\includegraphics{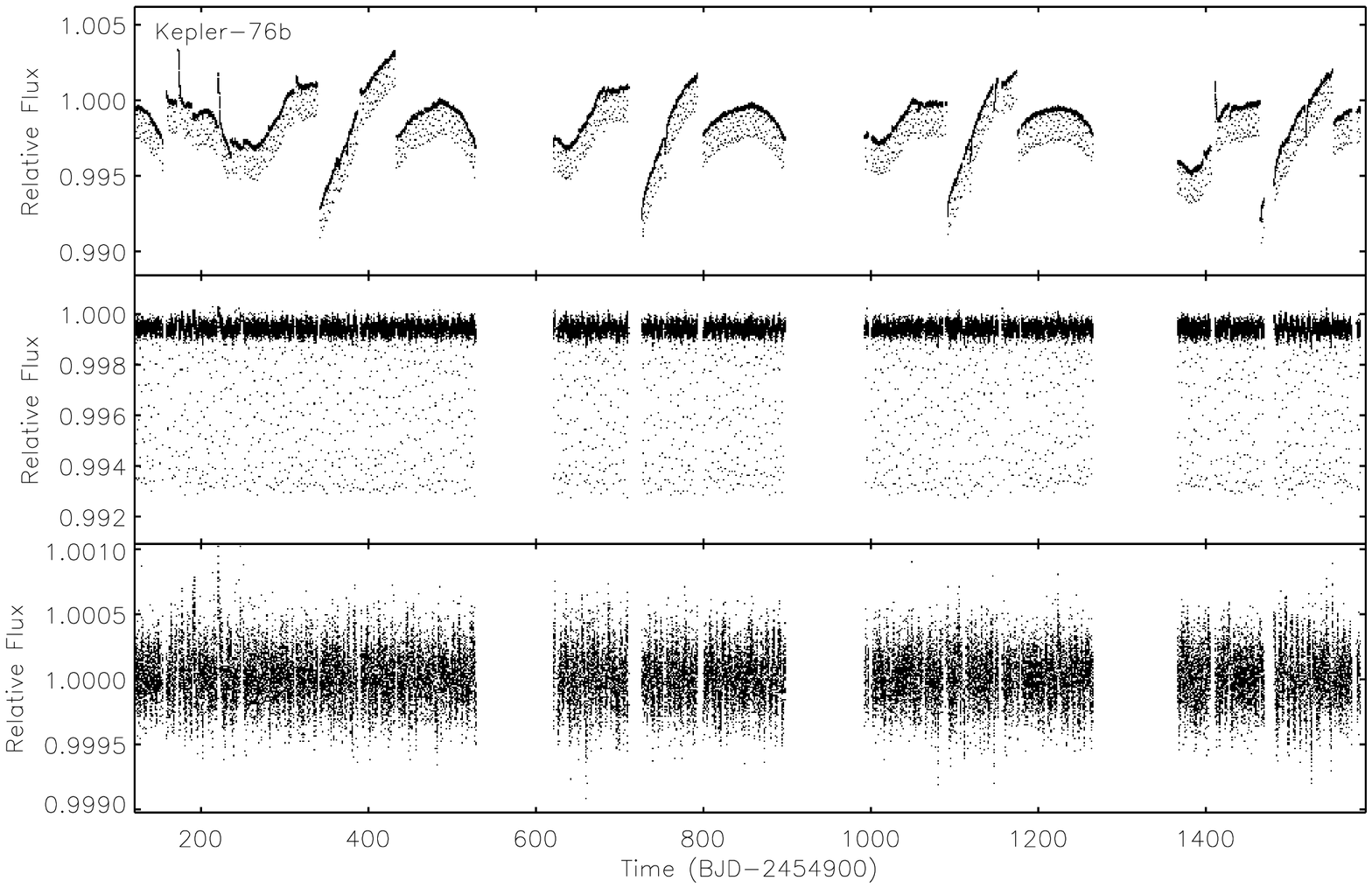}}
\scalebox{0.8}{\includegraphics{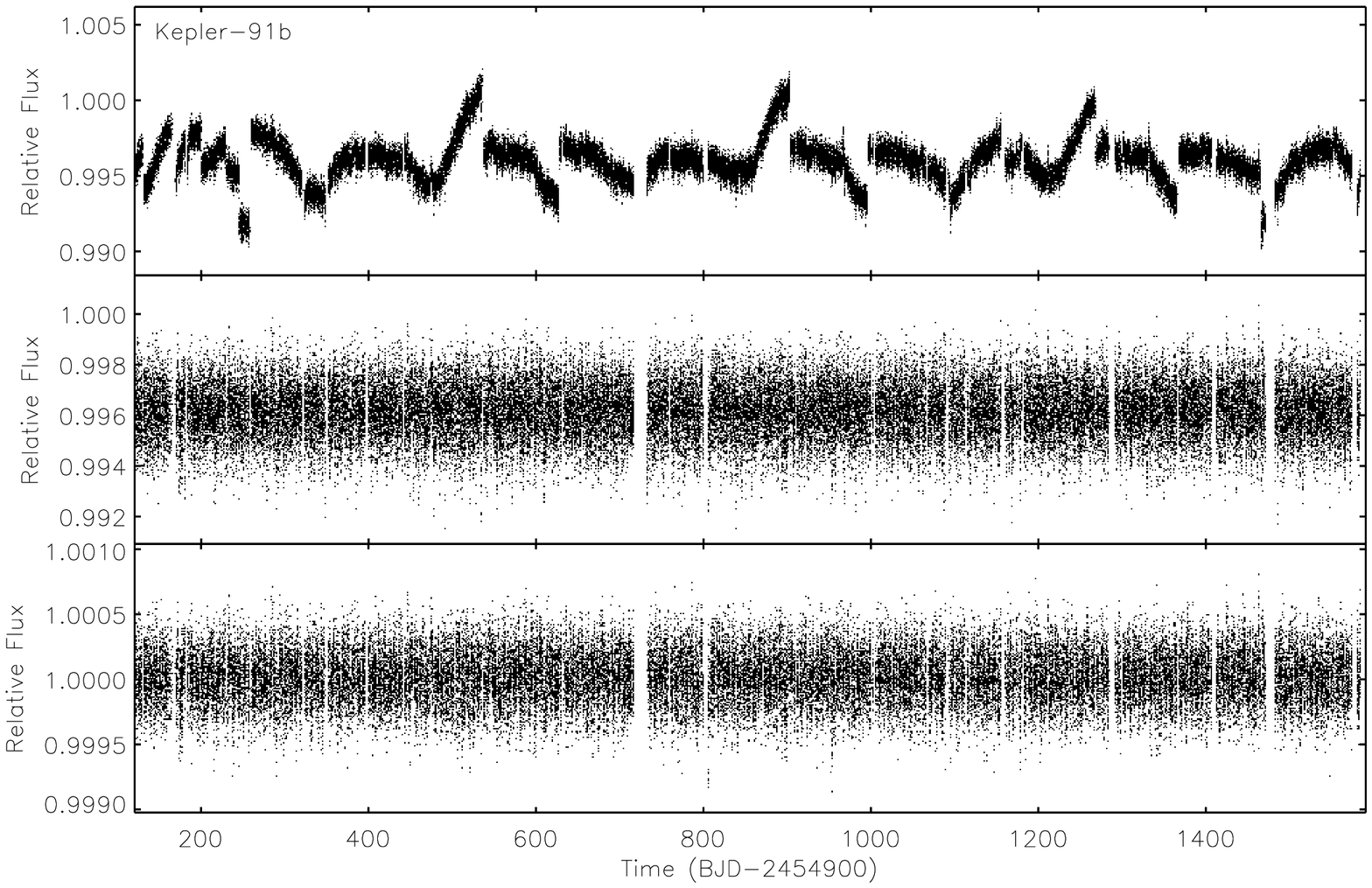}}
\end{center}
\caption{Same as Figure~\ref{fig:A1}, but for Kepler-76b (top plot) and Kepler-91b (bottom plot).}
\label{fig:A5}
\end{figure*}
\begin{figure*}[ht]
\begin{center}
\scalebox{0.8}{\includegraphics{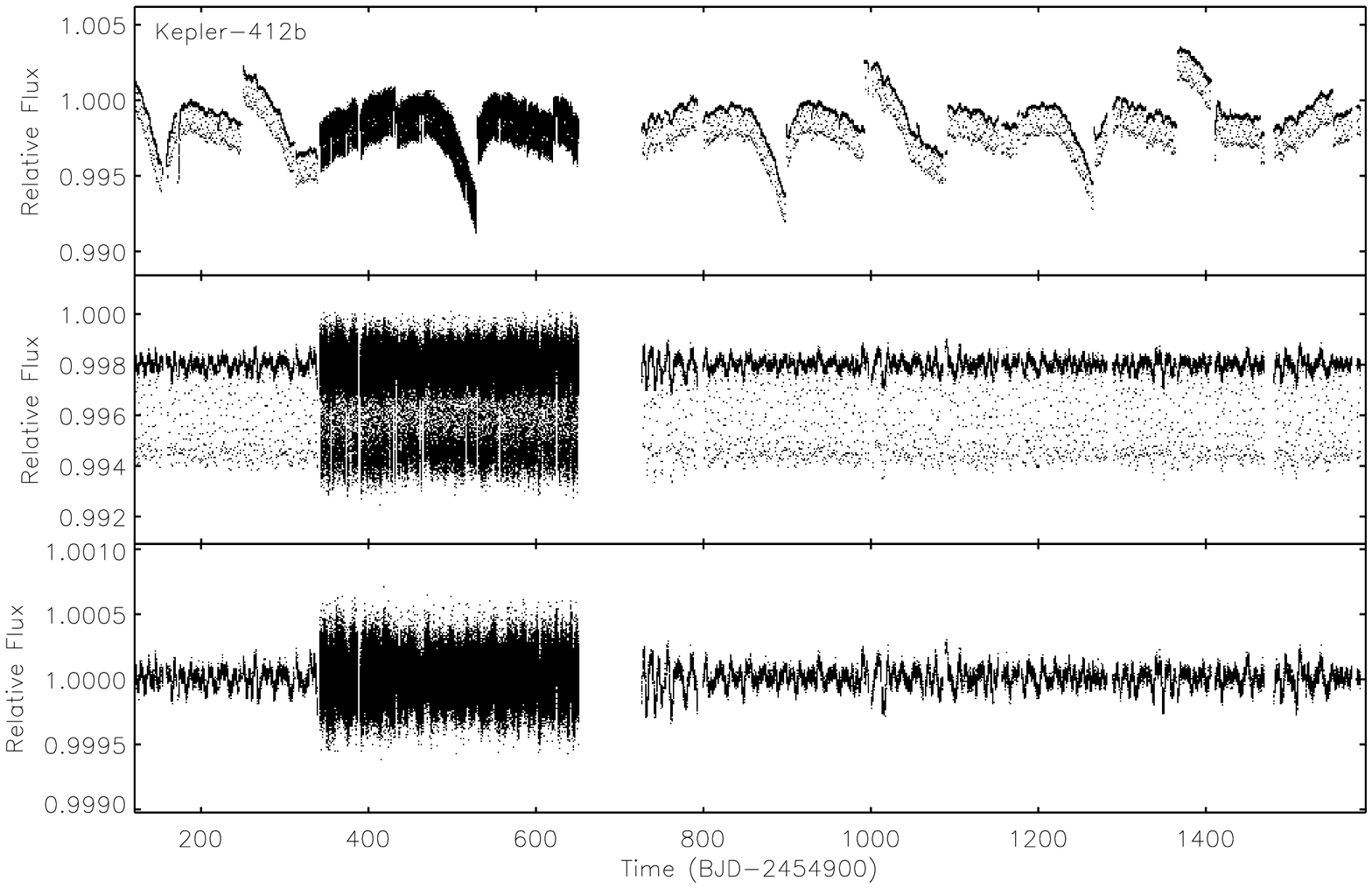}}
\scalebox{0.8}{\includegraphics{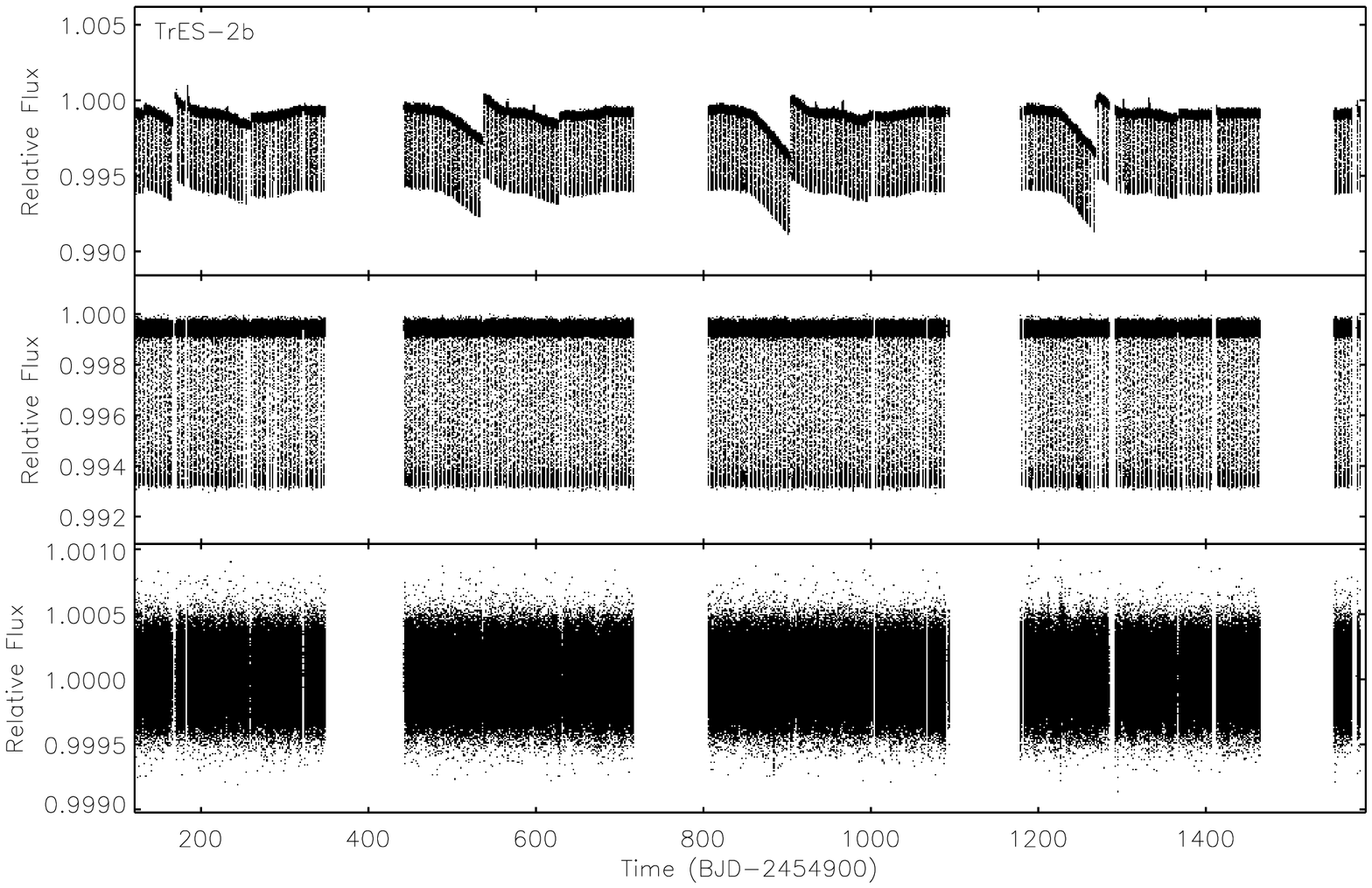}}
\end{center}
\caption{Same as Figure~\ref{fig:A1}, but for Kepler-412b (top plot) and TrES-2b (bottom plot).}
\label{fig:A6}
\end{figure*}
\begin{figure*}[ht]
\begin{center}
\scalebox{0.8}{\includegraphics{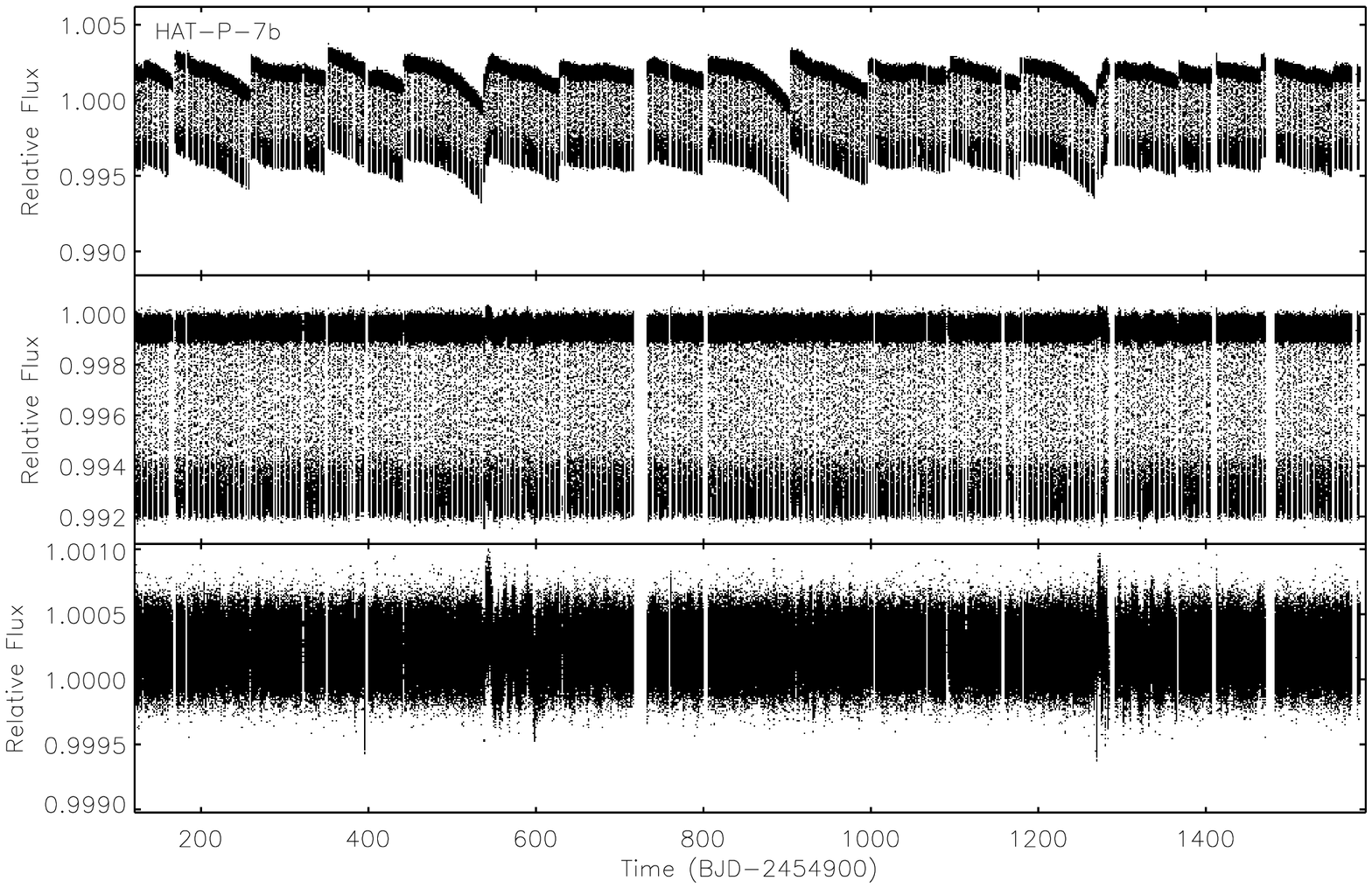}}
\scalebox{0.8}{\includegraphics{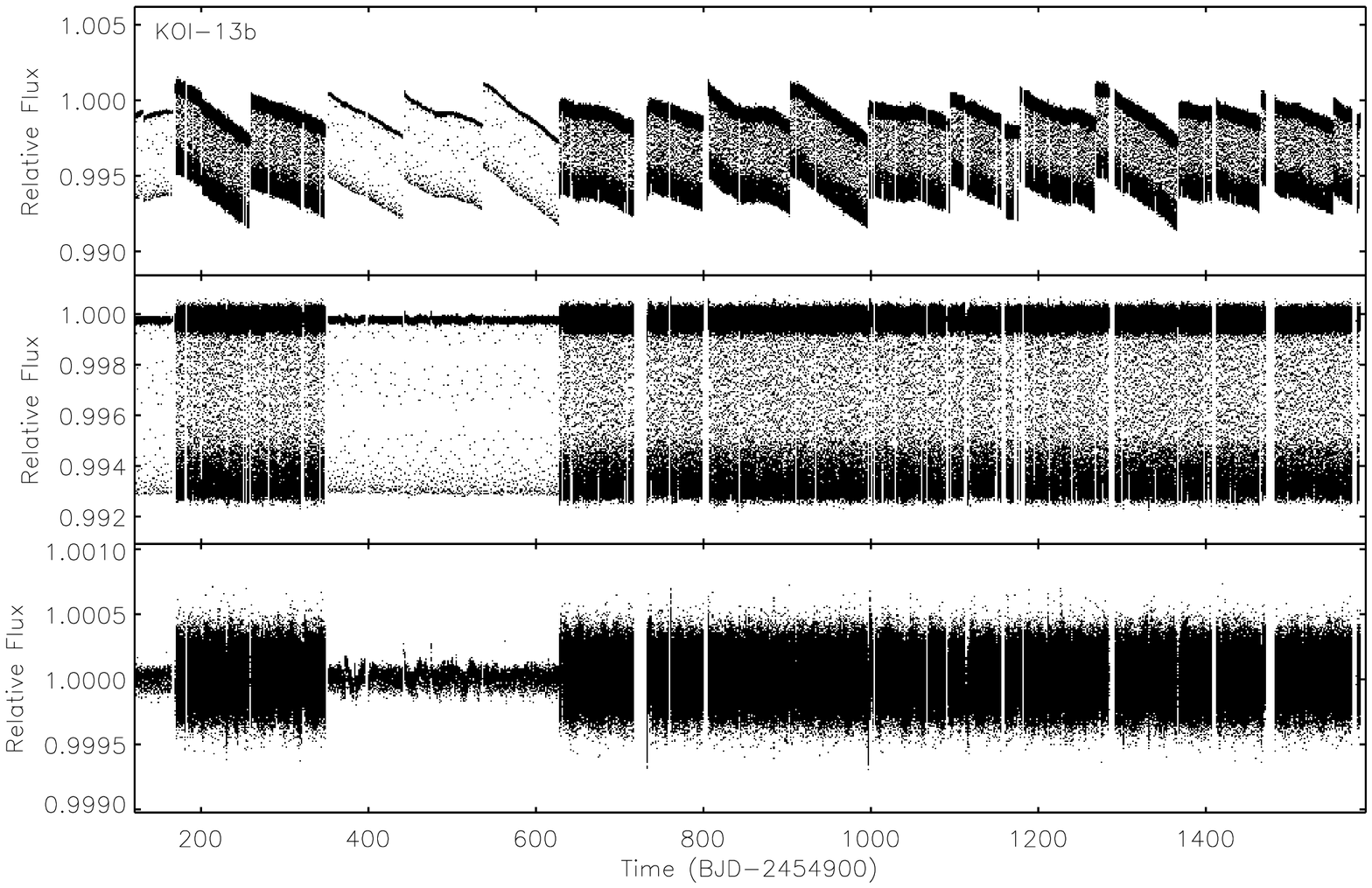}}
\end{center}
\caption{Same as Figure~\ref{fig:A1}, but for HAT-P-7b (top plot) and KOI-13b (bottom plot).}
\label{fig:A7}
\end{figure*}
\end{document}